\documentstyle[12pt]{article}
\def \unit {{  1 \kern -0.27em {\rm I}}\,}
\def \pom {{\scriptscriptstyle \kern -0.1em I \kern -0.25em P}}
\def \MSbar {\vbox{\hrule\kern 1pt\hbox{\rm MS}}}

\def\lesssim{\ \hbox{\raise 2pt \hbox{$<$} \kern -13pt
                     \lower 3pt \hbox{$\sim$}}\ }
\def\greatersim{\ \hbox{\raise 2pt \hbox{$>$} \kern -13pt
                     \lower 3pt \hbox{$\sim$}}\ }
\def\bk{{\mbox{\bf k}}}
\def\bq{{\mbox{\bf q}}}
\def\bs{{\mbox{\bf s}}}
\def\bsp{{\mbox{$\bf s^{'}$}}}
\def\bri1{{\mbox{${\bf r}^{(i)}_1$}}}
\def\brj2{{\mbox{${\bf r}^{(j)}_2$}}}
\def\bnul{{\mbox{\bf 0}}}

\input epsf.tex
\def\desepsf(#1 width #2){\epsfxsize=#2 \epsfbox{#1}}
\newcounter{sect}
\renewcommand{\theequation}{\thesect.\arabic{equation}}

\begin{document}
\begin{titlepage}
\renewcommand{\thefootnote}{\fnsymbol{footnote}}

\vspace*{5mm}
%\par \vskip 10mm

\begin{center}
{\Large \bf  Hard scattering factorization and light cone hamiltonian 
approach to diffractive processes} \\
%\end{center}
\vspace*{1cm}

        \par \vskip 5mm \noindent
        {\bf F.\ Hautmann}$^a$, {\bf Z.\ Kunszt}$^b$ and 
        {\bf D.E.\ Soper}$^{c,d}$\\
        \par \vskip 3mm \noindent
        $^a$ Department of Physics, Pennsylvania State University, 
        University Park PA 16802, USA\\ 
        \par \vskip 1mm \noindent 
        $^b$ Institute of Theoretical Physics, 
        ETH, CH-8093 Zurich, Switzerland\\     
        \par \vskip 1mm \noindent 
        $^c$ Institute of Theoretical Science, 
        University of Oregon, Eugene OR 97403, USA\\        
        \par \vskip 1mm \noindent 
        $^d$ Theory Division, 
        CERN, CH-1211 Geneva 23, Switzerland\\
         
\par \vskip 1cm

\end{center}
%\par \vskip 2mm
\vspace*{1cm}

\begin{center} {\large \bf Abstract} \end{center}
\begin{quote}
We describe  diffractive deeply inelastic scattering  in terms of 
diffractive parton distributions. We investigate   
these distributions in a hamiltonian formulation that 
emphasizes the spacetime picture of diffraction scattering. 
For hadronic systems with  small transverse size, 
diffraction occurs predominantly at short distances and 
the diffractive parton distributions can be studied 
by perturbative methods. For realistic, large-size systems 
we discuss the possibility that 
diffractive parton distributions are  controlled  essentially 
by semihard physics at a scale of nonperturbative origin of 
the order of a   GeV.  We find that this possibility accounts for 
two  important qualitative aspects of the diffractive data from HERA: 
the flat behavior in $\beta$ and the delay in the fall-off with $Q^2$. 
\end{quote}
\vspace*{2.3 cm}

\begin{flushleft}
     CERN-TH/99-154\\
     ETH-TH/99-09\\
     PSU-TH/207\\
     May 1999 \\
\end{flushleft}
\end{titlepage}

\renewcommand{\thefootnote}{\fnsymbol{footnote}}

\setcounter{equation}{0}
\setcounter{sect}{1}

\noindent  {\Large \bf 1. Introduction} 
\vskip .1 true cm

In hadron-hadron scattering, a substantial fraction of the events are
diffractive: one or both of the initial hadrons emerges with small
transverse momentum in the final state, having lost only a small
fraction of its energy. Assuming that quantum chromodynamics is the
theory of the strong interactions, one expects that diffractive
scattering is due to the exchange of gluons.  Since gluons are
pointlike objects, the gluon exchange picture suggests the possibility
of {\em hard} diffractive  scattering, in which exchanged gluons
moving in opposite directions participate in a hard process such as jet
production, with a transferred-momentum scale  much larger than
$\Lambda_{\rm {QCD}}$. Similarly, in lepton-hadron scattering it should
be possible to have a hard process, deeply inelastic scattering, in
which the incoming hadron is diffractively scattered. These
possibilities were suggested in 1985 by Ingelman and 
Schlein~\cite{ingelschl}. Experimental data from both 
hadron-hadron~\cite{hhdiffcern,hhdifffnal} 
and  lepton-hadron~\cite{disdiff1,h1scalviol,zeusf2d2}
colliders have confirmed the existence of hard diffractive scattering.

Since hard diffractive scattering contains a hard subprocess, one may
ask whether the QCD factorization theorem that holds for inclusive hard
scattering  works also for diffractive hard scattering.
Investigation of this question indicates that factorization does not
hold for diffractive hard processes when there are two hadrons in the
initial state. In {\it inclusive} hard scattering there are
contributions from particular final states that would  break
factorization, but these contributions cancel because of the unitarity
of the scattering matrix when one sums over all final states~\cite{css}. 
If only
diffractive final states are allowed, this cancellation is
spoiled~\cite{spoil,proo}. However, factorization {\it does} hold
for diffractive deeply inelastic scattering, in which there is only one
hadron in the initial state~\cite{proo,bere,fracture}. It is
diffractive deeply inelastic scattering that is of concern in this
paper.

In diffractive  deeply inelastic scattering, one can measure the
diffractive structure function 
$ d F_2^{\rm diff}(x , Q^2 , x_\pom , t) / [ d x_\pom \, dt]$, 
where $Q^2$ and $x$ are as usual the photon virtuality and the
Bjorken variable of deeply inelastic scattering, $x_\pom$ is the
fractional loss of longitudinal momentum by the diffracted hadron,
and $t$ is the  invariant momentum transfer from the diffracted hadron. 
This structure function is often called $ F_2^{D (4)}$.   
The factorization theorem allows us to write
\begin{equation}
\label{fact} 
{ { d F_2^{\rm diff} 
(x , Q^2 , x_\pom , t) } \over { dx_\pom\, dt}}\
= \sum_a \, \int  
d \xi   \, { df_{a/A}^{\rm diff}(\xi,x_\pom,t,\mu)
\over dx_\pom\, dt}
 {\hat F}_{ a} (x / \xi , Q^2 / \mu^2 ) 
\hspace*{0.3 cm} .
\end{equation}
The function ${\hat F}_{a} $ is the hard scattering function, 
calculable in perturbation theory. It is the {\it same} hard scattering
function as in inclusive deeply inelastic scattering. The function $d
f_{a/A}^{\rm {diff}}/ [dx_\pom\,dt]$ in  Eq.~(\ref{fact}) is the 
diffractive parton distribution, containing the long distance physics. 
It is interpreted as the probability to find a parton of type $a$ in a 
hadron of type $A$
carrying momentum fraction $\xi$ and, at the same time, to find that 
the hadron appears in the final state carrying a fraction $1 - x_\pom$
of its longitudinal momentum, having been scattered with an invariant
momentum  transfer $t$. Both the hard scattering function $\hat F$ and
the diffractive parton distribution functions $df_{a/A}^{\rm
diff}/[dx_\pom\, dt]$ depend on a factorization scale $\mu$.
The $\mu$ dependence of the distribution functions is given by the
usual renormalization group evolution equation,
\begin{equation} 
\label{eveq} 
{\partial \over {\partial \ln \mu^2} } \, 
{ df_{a/A}^{\rm diff}(\xi,x_\pom,t,\mu)
\over dx_\pom\, dt}
= \sum_a \, \int  d \xi^\prime \, K_{a a^{\prime} } (\xi^\prime,\mu) 
  \, { df_{a^{\prime}/A}^{\rm diff}(\xi / \xi^\prime,x_\pom,
t,\mu)\over dx_\pom\, dt}
\hspace*{0.3 cm} ,
\end{equation}
where the   kernel $K_{a a^{\prime}}$ has a perturbative expansion
in powers of $\alpha_s(\mu)$. The diffractive parton distributions are
defined as certain matrix elements of quark and gluon field operators,
analogously to the  definition of the ordinary (inclusive) parton
distributions~\cite{bere}. We will discuss these definitions later in
the paper.

Our purpose in this paper is to investigate the diffractive parton
distribution functions, expanding on the analysis reported in
Ref.~\cite{hkslett}. We are interested in the leading behavior of these
functions when $x_\pom \ll 1$. In the language of Regge theory, 
this corresponds to looking in the region where the pomeron is dominant 
over other Regge poles. Although the evolution equation for the diffractive
parton distribution functions is the same as that of the inclusive
parton distribution functions, their behavior at a fixed scale $\mu_0$
that serves as the starting point for evolution may be very different
from the behavior of the inclusive functions. The different
phenomenology that characterizes diffractive versus inclusive deeply
inelastic scattering depends entirely on this. 

Of course, the diffractive parton distributions in a proton  
at the scale $\mu_0$ are not perturbatively calculable. 
Notice that the problem lies with the large transverse size of the proton. 
Suppose one had a hadron of a size $1 / M$ that is small compared to 
$1/\Lambda_{\rm {QCD}}$. Then one could compute diffractive parton 
distributions as a perturbation expansion.
In this paper,  we first consider 
diffraction of small-size hadronic systems. 
We study this in detail. 
Then, we discuss how the  picture of diffraction changes as we let 
the size  increase. This involves 
nonperturbative dynamics. We 
explore whether one may extract 
(at least, qualitative) 
information on the diffractive parton  distributions for a 
large-size system 
by supplementing the computation at a much smaller size scale 
with a  hypothesis on 
the infrared behavior of the diffraction process. 

As a simple case of a small-size hadronic system, we consider a 
diquark system produced by a 
color-singlet current that couples only to heavy quarks of mass 
$ M \gg \Lambda_{\rm {QCD}}$. This system gets diffracted and 
acts as a color source with small radius of order $1 / M$. 
In this case, the perturbation expansion for the leading 
$1 / x_\pom \to \infty$ 
terms in the  diffractive parton distributions begins at order
$\alpha_s^4$. Although this is a rather high order of perturbation
theory, we find that the result has quite a simple structure and can be
expressed in terms of integrals that can be evaluated numerically.

One can view the problem that we address as being that of diffractive
deeply inelastic scattering at scale $Q$ from a hadron of size $1/M$
with $Q \greatersim M$ and $x_\pom \ll 1$. There are two main ingredients in
our analysis. 

The first ingredient has already been introduced: the factorization
formula (\ref{fact}).  Using
factorization, we are led to analyze the diffractive parton
distribution functions, which are simpler than 
$F_2^{\rm diff}$. This ingredient is especially
important because (as we will see) the diffractive gluon distribution
makes an important contribution to $F_2^{\rm diff}$, but
this contribution is not so easy to analyze 
systematically without the use of  the
factorization formula. 

The second ingredient is the physical picture that, in a suitable
reference frame, the partons that are ``measured'' in the process are
created by the measurement operator outside of the hadron and, much
later, interact with the hadron. This picture, called the ``aligned
jet'' model by Bjorken~\cite{alignedjet}, applies when $x_\pom \ll 1$. 
We will see that it  emerges most naturally 
when one works in configuration space using 
 light cone coordinates 
\begin{equation} 
\label{lccoord} 
 x^\pm = { { x^0 \pm x^3 } \over { \sqrt{2} } } \hspace*{0.3 cm} .  
\end{equation} 
The calculation becomes particularly transparent when one uses 
a hamiltonian formulation  
in which the theory is quantized on   planes of equal 
$x^-$.  We will see how this works in Sec.~3. 

The plan of the paper is as follows. In Sec.~2 we review the 
operator definitions for diffractive parton distributions and  
 describe their  structure at large $1 / x_\pom$.  
In Sec.~3 we show how to compute these distributions 
in the light cone hamiltonian formulation  of the theory.  
In Sec.~4  we present 
  general properties and  
 numerical results for  the distributions. 
In Sec.~5 we comment on their 
evolution and the structure of ultraviolet divergences.  
In  Sec.~6 we discuss the relation of the previous calculations 
with the phenomenology of diffractive deeply inelastic scattering 
at HERA. 
We give conclusions in Sec.~7. In Appendices~A-D we 
give calculational details 
on certain operator matrix elements,   we outline 
the main steps of the covariant 
formulation 
alternative to the one presented in the text, 
and we collect some  integrals.

\vskip 1.5 true cm

\setcounter{equation}{0}
\setcounter{sect}{2}

\noindent  {\Large \bf   2. Basic definitions and approximations } 
\vskip .1 true cm

In this section, we outline the parts of our analysis that bear on the
general structure of the result. We begin with the operator definitions of
the diffractive parton distributions. Then we describe how the diffractive
parton distributions break into a convolution of a part associated with
these operators and a part associated with the wave function of the 
incoming state. 

\vskip 0.7 true cm 

\noindent  {\large \bf 2.1 Operator definitions  } 
\vskip .1 true cm

Let us briefly recall the definition of the  diffractive parton
distributions  in terms of matrix elements of  bilocal field
operators~\cite{bere}. This is the same  definition~\cite{cs82,cfp}  as for 
inclusive parton distributions except that one requires that the final  state
include the diffractively scattered hadron.  
Let $(p_A , s_A )$ and 
$( p_{\!A^\prime}, s_{\!A^\prime} )$ denote the momentum and spin  
of the incident and the diffracted hadron. Let us 
adopt  the standard notation   $ \beta x_\pom$ 
for the hadron momentum fraction carried by the  parton. 
For gluons one has
\begin{eqnarray}
\label{opg}
&& {{d\, f_{g/A}^{\rm {diff}}(\beta x_\pom, x_\pom , t , \mu ) } \over
{dx_\pom\,dt}} = 
{1 \over {16 \, \pi^2}} \, 
{1 \over 2\pi \beta x_\pom p_{\!A}^+}{1 \over 2}\sum_{s_{\!A}} \int d y^-
e^{i\beta x_\pom p_{\!A}^+ y^-}
\sum_{X,s_{\!A^\prime}}
\nonumber\\
&& \hskip 1 cm  \times 
\langle p_{\!A},s_{\!A} |\widetilde F_a(0)^{+j}
| p_{\!A^\prime},s_{\!A^\prime}; X \rangle 
\langle p_{\!A^\prime},s_{\!A^\prime}; X|
\widetilde F_a(0,y^-,{\bf 0})^{+j}| p_{\!A},s_{\!A} \rangle 
\hspace*{0.2 cm} ,
\end{eqnarray}
where there is an implicit sum over $j= 1,2$ and where $\widetilde
F_a(0,y^-,{\bf 0})^{+\nu}$ is the field strength operator modified by
multiplication by an exponential of a line integral of the vector potential:
\begin{equation}
\label{Ftilde}
\widetilde F_a(y)^{+j}
=
E(y)_{ab}
F_b(y)^{+j} \hspace*{0.2 cm} ,  
\end{equation}
where\footnote{Here the sign in the
exponent is opposite to that in Refs.~\cite{hkslett,cs82}. The sign choice
depends on the convention for the sign of $g$. Here we choose the sign of $g$
so that $D_\mu = \partial_\mu + i g A_\mu^a t_a$.}
\begin{equation}
\label{eikdef}
E(y) = 
{\cal P}
\exp\left(
- i g \int_{y^-}^\infty d x^-\, A_c^+(y^+,x^-,{\bf y})\, t_c
\right) \hspace*{0.3 cm} .
\end{equation}
The symbol ${\cal P}$ denotes path ordering of the exponential. The matrices
$t_c$ in Eq.~(\ref{eikdef}) are the generators of the adjoint representation
of SU(3), $(t_c)_{bd} = - i f_{cbd}$.

The field operators in Eq.~(\ref{opg}) are evaluated at points 
separated by lightlike distances. There are 
ultraviolet divergences from the  operator products.  It is
understood that these are renormalized at the scale $\mu$ using the
$\overline{\rm MS}$ prescription.   

For purposes of computations, there is a more useful way \cite{cs82} to write
$\widetilde F^{+j}$. Starting with 
\begin{equation}
\widetilde F_a(y)^{+j}
=
E(y)_{ab}
\left[
\partial^+ A_b(y)^j - \partial^j A_b(y)^+
+ i g A_c(y)^+ A_d(y)^j\ (t_c)_{bd}
\right] \hspace*{0.3 cm} , 
\end{equation}
we have
\begin{eqnarray}
\widetilde F_a(y)^{+j}&=&
\partial^+ ( E(y)_{ab} A_b(y)^j )
- i g E(y)_{ab} A_c(y)^+ (t_c)_{bd}\ A_d(y)^j
\nonumber\\
&& 
- E(y)_{ab}\partial^j A_b(y)^+
 + i g E(y)_{ab} A_c(y)^+ A_d(y)^j\ (t_c)_{bd} \hspace*{0.3 cm} .
\end{eqnarray}
The second and fourth terms cancel. Furthermore, inside the integration in
Eq.~(\ref{opg}), $\partial^+ = \partial/\partial x^-$ becomes 
$- i \beta x_\pom
p_A^+$:
\begin{equation}
\label{newF}
\widetilde F_a(y)^{+j}\to
 E(y)_{ab}[ -i \beta x_\pom p_A^+\ A_b(y)^j 
- \partial^j A_b(y)^+]  \hspace*{0.3 cm} .
\end{equation}
This form is useful because it does not have an $A\times A$ term and it does
not have a $\partial/\partial x^-$. (In the complex conjugate matrix element
in Eq.~(\ref{opg}), the  $-i\beta x_\pom p_A^+$ becomes 
$+i\beta x_\pom p_A^+$.)

For  quarks of type $j$ one has 
\begin{eqnarray}
\label{opq}
&& {{d\, f_{j/A}^{\rm {diff}}(\beta x_\pom, x_\pom , t , \mu ) } \over
{dx_\pom\,dt}} = 
{1 \over {16 \, \pi^2}} \, 
{ 1 \over 2\pi\beta x_\pom p_A^+}  
{1 \over 2}\sum_{s_{\!A}}\int d y^- e^{i\beta x_\pom p_{\!A}^+ y^-}
\sum_{X,s_{\!A^\prime}} 
\nonumber\\
&& \hskip 1 cm  \times 
{\beta x_\pom p_A^+ \over 2}
\langle p_{\!A},s_{\!A} |\widetilde {\overline \Psi}(0)
| p_{\!A^\prime},s_{\!A^\prime}; X \rangle
\gamma^+ \langle p_{\!A^\prime},s_{\!A^\prime}; X|
{\widetilde \Psi}(0,y^-,{\bf 0})
| p_{\!A},s_{\!A} \rangle \hspace*{0.2 cm} 
, 
\end{eqnarray}
where
\begin{equation}
\label{qtilde}
\widetilde \Psi_j(y)
=
E(y) \Psi_j(y)  \hspace*{0.2 cm}   . 
\end{equation}
Here $E(y)$ is given by Eq.~(\ref{eikdef}) with, now, the matrices $t_c$ being
the generators of the fundamental representation of SU(3).

\vskip 0.7 true cm

\noindent  {\large \bf \boldmath 
2.2 General structure at large $1 / x_\pom$ } 
\vskip .1 true cm

As discussed in Sec.~1, 
   the scattering process  we consider  is initiated by a color-singlet 
current.  
  We  specialize the definitions of the previous subsection to the case 
 in which the incident hadron $A$ is a special 
photon that couples to a heavy quark-antiquark pair of mass $M$. This 
pair couples again to the diffractively scattered photon in the final state.  

In the discussion that follows, we use the
same method for the diffractive quark distribution and the diffractive gluon
distribution. For the sake of definiteness, we present the discussion in the
case of the diffractive gluon distribution.  
At the end of Sec.~3 we will give the 
extension of the results to the case of the quark.

We consider the limiting $1 / x_\pom \to \infty$ 
 behavior of graphs for the diffractive
gluon distribution. We shall find contributions proportional to $x_\pom^{-2}$
at fixed $\beta$. 
 At higher orders in perturbation theory, the $x_\pom^{-2}$ would be
supplemented by logarithms of $x_\pom$. In this paper we 
will limit ourselves to considering the graphs
at the lowest order of perturbation theory for which $x_\pom^{-2}$ behavior
emerges. 

\begin{figure}[htb]
\centerline{ \desepsf(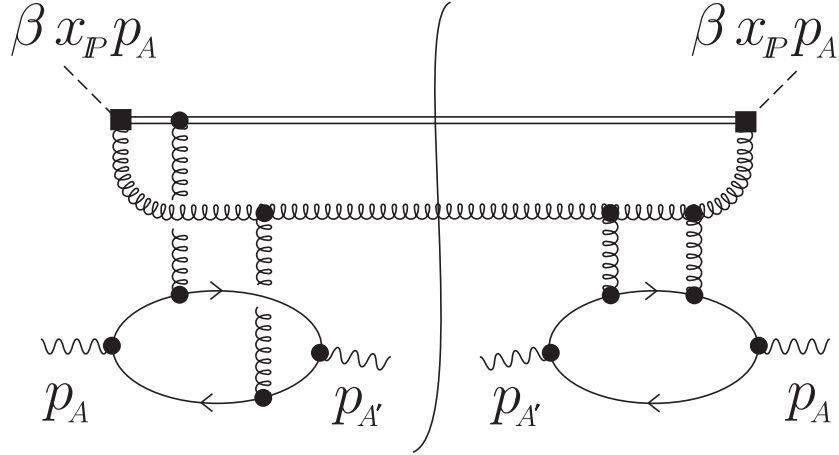 width 11 cm) }
\caption{A particular graph that contributes to the diffractive gluon
distribution.}
\label{fig:prlgraph}
\end{figure}

What are the lowest order graphs that can give $x_\pom^{-2}$ behavior? 
Note that there must be at least one gluon in the final state in order to
balance the color of the operator that measures the gluon distribution. We
find that  with exactly one gluon in the final state there are $x_\pom^{-2}$
contributions at order $\alpha_s^4$. An example of a contributing graph is
shown in Fig.~\ref{fig:prlgraph}. 
Note that there are graphs that contribute to the
diffractive gluon distribution at one lower order of perturbation theory,
but they yield fewer powers of $1/x_\pom$. Thus we consider 
$\alpha_s^4$ contributions.

We write the hadron momenta as
\begin{eqnarray}
p_A^\mu &=& (p_A^+,0,{\bf 0})
\nonumber\\
p_{A'}^\mu &=& 
((1- x_\pom) p_A^+,{{\bf q}^2\over 2(1- x_\pom) p_A^+},-{\bf q)} 
\hspace*{0.3 cm} .
\end{eqnarray}
Thus the momentum transfer is\footnote{Note that in our notation, $q^\mu$ is
not the virtual photon momentum associated with deeply inelastic scattering in
Eq.~(\ref{fact}), as is common in the literature. This should cause no
confusion since in this section there is no virtual photon.}
\begin{equation}
q^\mu \equiv p_A^\mu - p_{A'}^\mu =
( x_\pom p_A^+,-{{\bf q}^2\over 2(1- x_\pom) p_A^+},{\bf q)} 
\hspace*{0.3 cm} .
\end{equation}
The invariant momentum transfer is $t \equiv q_\mu q^\mu = - {\bf
q}^2/(1-x_\pom)$. Since we are interested in the $x_\pom \to 0$ limit, we use
the approximation $t \approx - {\bf q}^2$. We suppose that ${\bf q}^2 \lesssim
M^2$.

It is useful to work in a frame in which $p_A^+ \sim M$, so that the initial
hadron is approximately at rest.

We have a final state particle with momentum
\begin{equation}
k^\mu = ({ {\bf k}^2 \over 2 k^-}, k^-, {\bf k}) \hspace*{0.3 cm} . 
\label{gluonmomentum}
\end{equation}
Since plus momentum $q^+ =  x_\pom p_A^+$ is delivered from the scattered
hadron and plus momentum $\beta x_\pom p_A^+$ is removed by the measurement
operator, plus momentum 
\begin{equation}
{{\bf k}^2 \over 2 k^-} =
(1-\beta) x_\pom p_A^+
\label{gluonkminus}
\end{equation}
remains for the final state particle. Let us assume (as we will find,
self-consistently, in this calculation) 
that there are no important integration
regions with ${\bf k}^2 \ll M^2$ or with ${\bf k}^2 \gg M^2$. That is, ${\bf
k}^2 \sim M^2$ in the integration regions that give leading contributions.
Then $k^-$ must be large, $k^- \greatersim M/x_\pom$. The observation that the
final state parton has large minus momentum is crucial to the calculation.

Our analysis is simplified if we choose a physical gauge. Since the gluon in
the final state has large minus momentum, it is natural \cite{KogutSoper} to
use the null-plane gauge $A^- = 0$. (We make a few remarks on 
 the calculation in Feynman gauge in Appendix~C.)

We have seen that the final state hadron has a minus component of momentum of
order $M$ while the final state parton has minus momentum of order $M/x_\pom
\gg M$. For the virtual particles, we divide the integration over
minus momenta into regions $\ell^- \sim M$ and
$\ell^- \gg M$. In a general Feynman graph, it is far from easy to make this
division, but at order $\alpha_s^4$ the situation turns out to be quite simple.

In particular, the heavy quarks in our model hadron must have $\ell^- \sim M$.
In order to leave the hadron in a color singlet state, two gluons (at least)
must attach to the heavy quarks. In order that the intermediate heavy quark
lines not be far off shell, the minus momentum delivered by each of these
gluons must not be large. Finally, the only sink for the large minus momentum
carried by the final state gluon is the vertex representing the measurement
operator $\widetilde F^{\nu +}$, which can absorb large $\ell^-$ since it is
evaluated at a fixed value of plus position, $x^+ = 0$. With a little thought,
one realizes that all of the remaining internal propagator lines must carry
large minus momentum.

\begin{figure}[htb]
\centerline{ \desepsf(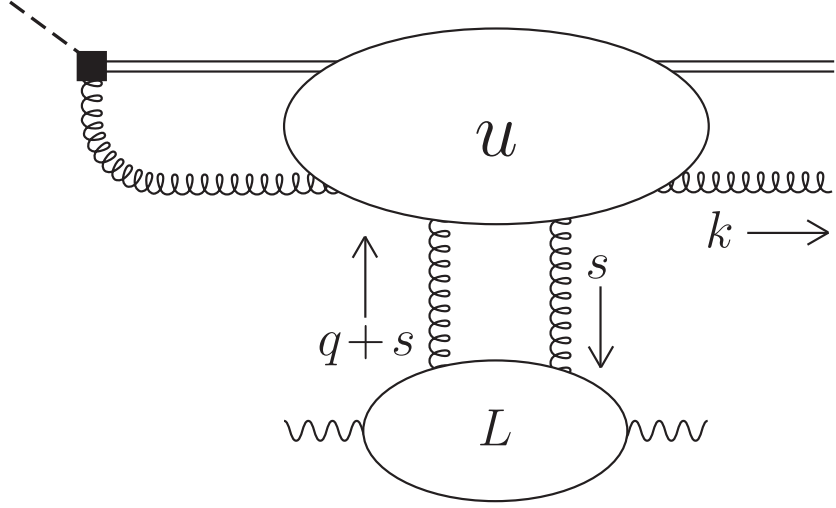 width 11 cm) }
\caption{Structure of the diffractive gluon distribution. In the amplitude,
two gluons are exchanged, one gluon is absorbed by the measurement operator,
and one gluon is emitted into the final state. The subgraphs $u$ and $L$ are
evaluated at lowest order of perturbation theory.}
\label{fig:structure}
\end{figure}

We are thus led to the picture shown in Fig.~\ref{fig:structure}. In the lower
subgraph, all loop momenta have $\ell^- \sim M$. In the upper subgraph, all
loop momenta have $\ell^- \gg M$. Two gluon lines with $\ell^- \sim M$
communicate between the two subgraphs. An example of a graph that contributes
to Fig.~\ref{fig:structure} is shown in Fig.~\ref{fig:prlgraph}. 

The definition (\ref{opg}) together with the structure represented by
Fig.~\ref{fig:structure} lead to the following expression for the diffractive
gluon distribution:
\begin{eqnarray}
\label{start} 
{{d f_{g/A}^{\rm {diff}} 
(\beta x_\pom,  x_\pom , {\mbox{\bf q}}^2 , M
) } \over
{dx_\pom\,dt}} &=& 
{1 \over {16 \pi^2}} \,  
{ 1 \over 2\pi\beta x_\pom p_A^+}\,
{1 \over 2}\, \sum_{\varepsilon\,\varepsilon^\prime} \, 
\int\! {d^4 s \over (2\pi)^4 }\,
\int\! {d^4 s' \over (2\pi)^4 }\,
\int\! {d^2 {\bf k}\,dk^+ \over (2\pi)^3 2k^+}\,
\nonumber\\ 
&&  \times 
{\rm Tr}\left\{
u_0^\dagger(k,q,s')^{\alpha\beta}_{ab}\
u_0(k,q,s)^{\gamma\delta}_{cd}
\right\}
(2\pi)\delta(k^+ - (1-\beta)x_\pom p_A^+)
\nonumber\\ 
&&  \times 
{1\over 4}\
{-iD(s')_{\alpha\mu}\over s^{\prime 2} - i\epsilon}\
{-iD(q+s')_{\beta\nu}\over (q + s')^2 - i\epsilon}\
{iD(s)_{\gamma\rho}\over s^2 + i\epsilon}\
{iD(q+s)_{\delta\sigma}\over (q+s)^2 + i\epsilon}\
\nonumber\\ 
&&  \times 
L_0(q,s',\varepsilon,\varepsilon^\prime)^{\mu\nu}_{ab}\
L_0(q,s,\varepsilon,\varepsilon^\prime)^{\rho\sigma}_{cd}\ 
\hspace*{0.3 cm} .
\end{eqnarray}
Here the $(2\pi)\delta(k^+ - (1-\beta)x_\pom p_A^+)$ results from performing
the integration over $y^-$ in Eq.~(\ref{opg}). We note immediately that the
integration over the plus momentum of the final state gluon gives
\begin{equation}
\int {dk^+ \over (2\pi)2k^+}\ (2\pi)\delta(k^+ - (1-\beta)x_\pom p_A^+)
= {1 \over 2 (1-\beta) x_\pom p_A^+} \hspace*{0.3 cm} . 
\label{longint}
\end{equation}
The functions $u_0$ and $L_0$ are the amputated Green functions represented in
Fig.~\ref{fig:structure}. The subscripts 0 distinguish these functions from
simpler functions $u$ and $L$ that are defined below in terms of $u_0$ and
$L_0$ and appear in the final formula (\ref{structure}) for the diffractive
gluon distribution. The function $u_0$ carries two transverse vector indices
that are not shown. One is the index $j$ carried by the operator $F_a^{+j}$ in
Eq.~(\ref{opg}), the other is the transverse polarization index of the final
state gluon. It also carries two color indices: the color index $a$ of
$F_a^{+j}$ and the color index of the final state gluon. The notation ${\rm
Tr}\{u_0^\dagger u_0\}$ denotes a summation over these polarization and color
indices. The indices carried by $u_0$ that are displayed are the color and
polarization indices for the exchanged gluons. The counting factor 1/4
associated with the gluon exchange accounts for the two ways for attaching the
labels $s'$ and $q+s'$ to the two gluons to the right of the final state cut
and the two ways for attaching the labels $s$ and $q+s$ to the two gluons to
the left of the final state cut. The notation  $iD(\ell)_{\mu\nu}/[\ell^2 +
i\epsilon]$ denotes the propagator of a gluon in null-plane gauge. The Green
functions $L_0$ depend on the transverse polarization vectors $\varepsilon$ and
$\varepsilon'$ of the initial and final state photons, respectively.
There is a sum over the two choices for each of these polarizations.

We can now make a number of approximations that simplify Eq.~(\ref{start}).
These approximations will be discussed again in the following section, but we
outline them here in order to exhibit their effect on the overall structure
of Eq.~(\ref{start}).

First, consider the momenta $s$, $q-s$, $s'$, and $q+s'$ of the exchanged
gluons.  
We have taken $|{\bf q}| \lesssim M$. Let us assume that (as we will find,
 self-consistently, in this calculation) that $|{\bf s}|$ and $|{\bf s}'|$
 are of order $M$ in the integration region that gives the leading
 contributions. 
Now $s^+$ must be of order $M^2/k^-$, where $k^-$ is the minus momentum
of the final state gluon, in order that the gluons in the upper subgraph are
not too far off shell. As we have just seen, $k^- \sim M/x_\pom$, so $s^+ \sim
x_\pom M$. On the other hand, $s^-$ must be of order $M^2/p_A^+ \sim M$ in
order that the partons in the lower subgraph are not too far off shell. Thus
$s^+s^- \sim x_\pom M^2 \ll {\bf s}^2$. For this reason, the propagator for
the exchanged gluon with momentum $s$ is $iD_{\gamma\rho}/s_\mu s^\mu \approx
-iD_{\gamma\rho}/{\bf s}^2$. A similar power counting shows that only the
transverse part of the momentum in each of the other exchanged gluon
propagators contributes in the $1 / x_\pom \to \infty$ limit.

Second, $s^+$, $s^{\prime +}$ and $q^+$, being of order $x_\pom M$, are
negligible compared to the plus momenta in the subgraphs $L_0$. Thus we
replace $s^+$, $s^{\prime +}$ and $q^+$ by 0 in the subgraphs $L_0$.
Then the integrations over $s^+$ and $s^{\prime +}$ can be associated with the
upper subgraphs $u_0$. Similarly,  $s^-$, $s^{\prime -}$, and $q^-$, being of
order $M$, are negligible compared to the minus momenta in the subgraphs $u_0$,
which are of order $M/x_\pom$. Thus we replace $s^-$, $s^{\prime -}$, and $q^-$
by 0 in the subgraphs $u$. Then the integrations over $s^-$ and $s^{\prime -}$
can be associated with the lower subgraphs $L_0$.

Third, the lower subgraphs in Fig.~\ref{fig:structure} are proportional to
color matrices ${\rm Tr}[t_a t_b] = {\scriptstyle{1\over 2}}\delta_{ab}$. Thus
we can replace
\begin{equation}
L_0(q,s,\varepsilon,\varepsilon^\prime)^{\rho\sigma}_{cd}
\to 
\delta_{cd}\
{L_0(q,s,\varepsilon,\varepsilon^\prime)^{\rho\sigma}_{ee}
\over N_c^2 - 1} \hspace*{0.3 cm} .
\label{colorsinglet}
\end{equation}

Fourth, we will find that in the end only the $g_{\mu\nu}$ parts of the
propagators of the exchanged gluons count. Then the Lorentz index structure
in Fig.~\ref{fig:structure} is $u_0^{\gamma\delta} L_{0\gamma\delta}$. Since
the partons in the upper subgraphs have very large minus momenta, this becomes
approximately $u_0^{--} L_{0}^{++}$.

With these changes, Eq.~(\ref{start}) becomes
\begin{eqnarray}
\label{mid1} 
{{d f_{g/A}^{\rm {diff}} 
(\beta x_\pom,  x_\pom , {\mbox{\bf q}}^2 , M
) } \over
{dx_\pom\,dt}} &=& 
{1 \over {64 \pi^2}} \,  
{ 1 \over 4\pi\beta(1-\beta) x_\pom^2 (p_A^+)^2}\,
{1 \over 2}\, \sum_{\varepsilon\,\varepsilon^\prime} \, 
\int\! {d^2 {\bf s} \over (2\pi)^2 }\,
\int\! {d^2 {\bf s}' \over (2\pi)^2 }\,
\int\! {d^2 {\bf k} \over (2\pi)^2 }\,
\nonumber\\ 
&&  \times 
{\rm Tr}\left\{
\left(\int\! {d s^{\prime +} \over 2\pi}\,u_0^\dagger(k,q,s')^{--}_{aa}\right)
\left(\int\! {d s^+ \over 2\pi}u_0(k,q,s)^{--}_{cc}\right)
\right\}
\nonumber\\ 
&&  \times 
{1\over {\bf s}^2}\
{1\over ({\bf q}+{\bf s})^2}\
{1\over {\bf s}^{\prime 2}}\
{1\over ({\bf q}+{\bf s}')^2}\
\nonumber\\ 
&&  \times 
\left(\int\! {d s^{\prime -} \over 2\pi}\,
{ L_0(q,s',\varepsilon,\varepsilon^\prime)_{bb}^{++} \over N_c^2 - 1}
\right)
\left(\int\! {d s^- \over 2\pi}\,
{ L_0(q,s,\varepsilon,\varepsilon^\prime)_{dd}^{++} \over N_c^2 - 1}
\right).
\end{eqnarray}
Let us define simplified functions $L$ and $u$ by
\begin{equation}
\int\! {d s^- \over 2\pi}\,
{ L_0(q,s,\varepsilon,\varepsilon^\prime)_{bb}^{++} \over N_c^2 - 1}
=  p_A^+\ L({\bf q},{\bf s},\varepsilon,\varepsilon^\prime)
\label{Ldef}
\end{equation}
and
\begin{equation}
\int\! {d s^+ \over 2\pi}\ u_0(k,q,s)^{--}_{aa}
= {\unit} \,C_A\,g_s^2\,
u(\beta,{\bf k},{\bf q},{\bf s}) \hspace*{0.3 cm} ,
\label{udef}
\end{equation}
where $\unit$ is a unit matrix in color space. We insert these definitions
into Eq.~(\ref{mid1}) and use 
\begin{equation}
{\rm Tr}\,\unit = N_c^2 -  1 \hspace*{0.3 cm} .
\end{equation}
This leaves a factor ${\rm Tr}\{u^\dagger u\}$ in which the trace is now over
transverse polarization indices but not color.

These changes give
\begin{eqnarray}
\label{structure} 
{{d f_{g/A}^{\rm {diff}} 
(\beta x_\pom,  x_\pom , {\mbox{\bf q}}^2 , M
) } \over
{dx_\pom\,dt}} &=& 
{1 \over {64 \pi^2}} \,  
{ C_A^2 (N_c^2 - 1) g_s^4 \over 4\pi\beta(1-\beta) x_\pom^2 }\,
{1 \over 2}\, \sum_{\varepsilon\,\varepsilon^\prime} \, 
\int\! {d^2 {\bf s} \over (2\pi)^2 }\,
\int\! {d^2 {\bf s}' \over (2\pi)^2 }\,
\int\! {d^2 {\bf k} \over (2\pi)^2 }\,
\nonumber\\ 
&&  \times 
{\rm Tr}\left\{
u^\dagger(\beta,{\bf k},{\bf q},{\bf s}')\,
u(\beta,{\bf k},{\bf q},{\bf s})
\right\}
\nonumber\\ 
&&  \times 
{1\over {\bf s}^2}\
{1\over ({\bf q} + {\bf s})^2}\
{1\over {\bf s}^{\prime 2}}\
{1\over ({\bf q} + {\bf s}')^2}\
\nonumber\\ 
&&  \times 
 L({\bf q},{\bf s}',\varepsilon,\varepsilon^\prime)\,
 L({\bf q},{\bf s},\varepsilon,\varepsilon^\prime) \hspace*{0.3 cm} .
\end{eqnarray}
This formula gives the basic structure of the answer for the 
matrix element (\ref{opg}). We will now examine this in detail.

\vskip 1.5 true cm

\setcounter{equation}{0}
\setcounter{sect}{3}

\noindent  {\Large \bf  3. Diffractive parton distributions and 
 null-plane field theory} 
\vskip .1 true cm

In this section, we use the formulation of QCD 
quantized on planes of equal light cone coordinates 
to analyze the structure depicted  in 
Fig.~\ref{fig:structure}. 
In doing so, we will derive explicit expressions for the 
subgraphs $u$ and $L$. 
This style of analysis is perhaps 
less familiar than the approach using covariant 
Feynman graphs in Feynman gauge, but it 
 expresses the physics of the process 
in configuration space   in a more transparent
fashion. For those readers who prefer a standard covariant calculation, we
present some of the essential steps in such a calculation in Appendix~C.
We carry out the calculation of this section for the specific case of the
diffractive gluon distribution. Then, in Sec.~3.4, we assemble the
complete result for both the gluon and quark distributions.

\vskip 0.7 true cm

\noindent  {\large \bf 3.1 The upper subgraph} 
\vskip .1 true cm

Consider the function $u(\beta,{\bf k},{\bf q},{\bf s})$ that appears 
in Eq.~(\ref{structure}) and is represented by the upper subgraph 
in Fig.~\ref{fig:structure}.  The partons in $u$ move with very
large minus momentum through the gluon field that accompanies the heavy quark
state that is approximately at rest. Our analysis is designed to draw the
consequences of this, concentrating on the
development of the states in space and time.

\vskip 0.2 true cm 

\noindent  {\em External field to represent the exchanged gluons} 
\vskip .1 true cm

It is convenient to replace the gluons coming from the lower subgraph by an
external color field ${\cal A}^\mu(x)$. We thus consider the matrix element
\begin{equation}
{\cal M} = \int dy^- e^{i\beta x_\pom p_A^+ y^-}
\langle k, s| \widetilde F_a(0,y^-,{\bf 0})^{+j}|0\rangle_{\cal A} 
\hspace*{0.2 cm} .
\label{calMdef}
\end{equation}
Here $\widetilde F^{+j}$ is the measurement operator (\ref{newF}) for the
gluon distribution function and $k,s$ are the momentum and spin of the final
state gluon. The matrix element is evaluated in the presence of an external
color field $\cal A$. We expand ${\cal M}$ in perturbation theory and 
extract the term ${\cal M}_{0,2}$ proportional to two powers of $g_s{\cal A}$
and zero additional powers of $g_s$. The coefficient in this term is the Green
function $u_0$ in Eq.~(\ref{start}):
\begin{equation}
{\cal M}_{0,2} = {1 \over 2} 
\int\! { d^4 s \over (2\pi)^4}
\int\! { d^4 q \over (2\pi)^4}\
\widetilde{\cal A}_\gamma^c(-s)\,
\widetilde{\cal A}_\delta^d(q+s)\,
2\pi\delta(q^+ - x_\pom p_A^+)\,
u_0(k,q,s)^{\gamma\delta}_{cd} \hspace*{0.2 cm} .
\end{equation}
Here $\widetilde{\cal A}$ is the Fourier transform of $\cal A$:
\begin{equation}
\tilde{\cal A}_a^\mu(k)
= \int dx\ e^{i k\cdot x}{\cal A}_a^\mu(x) \hspace*{0.2 cm} .
\end{equation}

Since $\cal A$ represents the gluons exchanged with the lower state and these
gluons are in a color singlet state ({\it cf.} Eq.~(\ref{colorsinglet})), we
will replace
\begin{equation}
\widetilde{\cal A}_\gamma^c(-s)\,
\widetilde{\cal A}_\delta^d(q+s)\,
\to 
\delta_{cd}\ { 1 \over N_C^2 - 1}\,
\widetilde{\cal A}_\gamma^e(-s)\,
\widetilde{\cal A}_\delta^e(q+s)\, \hspace*{0.2 cm} . 
\label{colorsinglet2}
\end{equation}
After making this replacement, the arguments that led to  
Eq.~(\ref{mid1}) lead us to anticipate that
${\cal M}_{0,2}$ becomes much simpler in the $1 / x_\pom \to \infty$ limit, 
taking the
form
\begin{eqnarray}
{\cal M}_{0,2} &\approx& {1 \over 2} 
\int\! { d^2 {\bf s} \over (2\pi)^2}
\int\! { d^2 {\bf q} \over (2\pi)^2}
\int\! { ds^- \over 2\pi}
\int\! { dq^- \over 2\pi}
%\nonumber\\
%&&\times
{1 \over N_c^2 -1}\
\widetilde{\cal A}^{c+}(0,-s^-,-{\bf s})\,
\widetilde{\cal A}^{c+}(0,q^- + s^-,{\bf q} + {\bf s}) \,
\nonumber\\
&&\times
\int\!{ds^+ \over 2\pi}\
u_0(k; x_\pom p_A^+,0,{\bf q};s^+,0,{\bf s})^{--}_{aa} 
\hspace*{0.2 cm} .
\end{eqnarray}
Using Eq.~(\ref{udef}), we identify the function $u$ that appears in the final
formula (\ref{structure}) for the diffractive gluon distribution.  Thus
\begin{eqnarray}
{\cal M}_{0,2} &\approx& {1 \over 2} 
\int\! { d^2 {\bf s} \over (2\pi)^2}
\int\! { d^2 {\bf q} \over (2\pi)^2}
\int\! { ds^- \over 2\pi}
\int\! { dq^- \over 2\pi}
%\nonumber\\
%&&\times
{1 \over N_c^2 -1}\
\widetilde{\cal A}^{c+}(0,-s^-,-{\bf s})\,
\widetilde{\cal A}^{c+}(0,q^- + s^-,{\bf q} + {\bf s}) \,
\nonumber\\
&&\times
{\unit} \,C_A\,g_s^2\,
u(\beta,{\bf k},{\bf q},{\bf s}) \hspace*{0.2 cm} .
\label{extractu}
\end{eqnarray}
where $\unit$ is a unit matrix in color space. Our aim in this section is to
calculate ${\cal M}_{0,2}$ in the $1 / x_\pom \to \infty$ limit, then to use
Eq.~(\ref{extractu}) to extract $u(\beta,{\bf k},{\bf q},{\bf s})$.

In order to better illustrate the physical principles involved and to give
some indication of how the present calculation would work at higher orders, it
is useful to generalize the problem that we attack. Let us therefore consider
a matrix element
\begin{equation}
{\cal M} = \int dy^- e^{i\beta x_\pom p_A^+ y^-}
\langle k_1,s_1,\dots,k_N,s_N | 
\widetilde F_a(0,y^-,{\bf 0})^{+j}|0\rangle_{\cal A} \hspace*{0.2 cm} .
\label{Mgeneralized}
\end{equation}
Here $k_i,s_i$ are the momenta and spins of one or possibly more final state
partons, with $k_i^- \sim M/x_\pom$. The matrix element is evaluated in full
QCD in the presence of an external color field $\cal A$ considered at all
orders of perturbation theory. As $x_\pom \to 0$, the momenta $k_i^-$ become
large. On the other hand, the external field $\cal A$ stays fixed. That is to
say, the minus momenta of the quantum particles become large while the
minus momenta of the gluons in the field produced by the diffracted hadron
stay of order $M$.

\vskip 0.2 true cm 

\noindent  {\em The eikonal line} 
\vskip .1 true cm

The operator $\widetilde F^{+j}$ in Eq.~(\ref{newF}) contains the exponential
of a line integral of the color vector potential, which now includes both the
quantum potential $A$ and the external potential ${\cal A}$,
\begin{equation}
E(0,y^-,{\bf 0}) =
\exp\left(
-ig\int_{y^-}^\infty dx^- [A_c^+(0,x^-,{\bf 0}) + {\cal A}_c^+(0,x^-,{\bf 0})]
t_c 
\right) \hspace*{0.2 cm} .
\label{eikonal}
\end{equation}
This eikonal line operator produces the same effect as if there were a special
color octet particle, ${\cal E}$, with a propagator
\begin{equation}
{ i \over \ell^+ + i\epsilon}
\end{equation}
and an interaction vertex with the color field
\begin{equation}
- i g t_c n^\mu
\end{equation}
with $n\cdot \epsilon = \epsilon^+$. We can build such a particle into the
theory. Let the particle be created with an operator $Q_a^\dagger(y^-)$  and
destroyed with an operator $Q_a(y^-)$. The commutation relation is
$[Q_a(y^-),Q_b^\dagger(y^-)] = \delta_{ab}$. The action
\begin{equation}
\int_{-\infty}^{+\infty}\! d x^-\ 
Q_b^\dagger(x^-)
\left[
i { d \over dx^-} \delta_{ba}
 - g[A_c^+(0,x^-,{\bf 0}) + {\cal A}_c^+(0,x^-,{\bf 0})]
(t_c)_{ba}  
\right]
Q_a(x^-)
\label{eikonalaction}
\end{equation}
will produce the desired propagator and vertices. We need a notation for the
states. We use $Q_b^\dagger(0) |0\rangle = |{\cal E},b\rangle$ and  
\begin{equation}
Q_b^\dagger(0) |
k_1,s_1,\dots,k_N,s_N \rangle = |k_1,s_1,\dots,k_N,s_N,{\cal E},b\rangle 
\hspace*{0.2 cm} .
\end{equation}
Thus we can replace the operator $\widetilde F^{+j}$ by
\begin{equation}
Q^\dagger(y^-)_a {\cal O}_a^j(0,y^-,{\bf 0}) \hspace*{0.2 cm} , 
\end{equation}
where, using Eq.~(\ref{newF}),
\begin{equation}
\label{calOdef}
{\cal O}_a^j(y) = 
 -i\beta x_\pom p_A^+\ A_a(y)^j 
- \partial^j A_a(y)^+ \hspace*{0.2 cm} .
\end{equation}
When making this replacement, we include the ${\cal E}$ particle in the final
state and add the extra action (\ref{eikonalaction}) to the action for QCD in
an external color field.

In subsequent equations, we do not explicitly indicate the color index, $b$,
for the special eikonal particle and we write $Q^\dagger(y^-) {\cal
O}^j(0,y^-,{\bf 0})$ for $Q^\dagger(y^-)_a {\cal O}_a^j(0,y^-,{\bf 0})$.
Indeed, the color indices for all of the partons are also left implicit.

Note that treating the eikonal factor as being produced by a quantum particle
with special properties is more than just a technical trick. In the
experimental determination of the gluon distribution, there is a short
distance interaction that scatters a gluon constituent of the
hadron and produces a system of jets with very large $\ell^-$. (This applies
for either the inclusive or the diffractive $F_2$, for $x_\pom \sim 1$ or
$x_\pom \ll 1$.) If the color field of the hadron is too soft to resolve the
internal structure of the jet system, then the jet system looks like a color
octet particle with  infinite $\ell^-$. We approximate
\begin{equation}
{ i \over 2 \ell^+ \ell^- - \ell_T^2 + i\epsilon}\
(- ig) t_c (2 \ell^- n^\mu + \cdots)
\to 
{ i \over \ell^+ + i\epsilon}\
(-ig) t_c  n^\mu 
\end{equation}
and arrive at the interactions of the special eikonal particle. This
idealization is incorporated into the definition \cite{bere,cs82} of the
\MSbar\ gluon distribution function. Deviations from the idealization are
accounted for in the perturbative calculation of the hard scattering matrix
elements for the physical process. 

\vskip 0.2 true cm 

\noindent  {\em Taking the high energy limit} 
\vskip .1 true cm

Our problem is now to find the limiting behavior of
\begin{equation}
{\cal M} = \int dy^- e^{i\beta x_\pom p_A^+ y^-}
\langle k_1,s_1,\dots,k_N,s_N,{\cal E}| Q^\dagger(y^-)
{\cal O}^j(0,y^-,{\bf 0})
|0\rangle_{\cal A} 
\label{Mdef1}
\end{equation}
when $1/x_\pom \to \infty$ and all of the $k_i^-$ tend to infinity like
$M/x_\pom$. We analyze this problem using approximations suggested by the
``aligned jet'' picture of small $x$ deeply inelastic 
scattering~\cite{alignedjet}.
Similar problems have been addressed in many papers over the past few 
years (see, for instance, Refs.~\cite{nikoplus},~\cite{buch} and 
~\cite{wuest}); 
the treatment in \cite{buch} is especially close
to that given below. Here we note that essentially the same problem was solved
in Ref.~\cite{bks}, in which the authors addressed deeply inelastic scattering
producing a $\mu\bar\mu$ pair in an external $U(1)$ field in the high energy
limit $k_i^- \to \infty$. We simply adapt the treatment of \cite{bks} into the
problem at hand.

Our analysis of ${\cal M}$ is based on null-plane-quantized field theory, as in
Ref.~\cite{bks}. (However, we use the theory quantized on planes of equal $x^-
\equiv (x^0 - x^3)/\sqrt 2$ instead of planes of equal $(x^0 + x^3)/\sqrt 2$
used in Ref.~\cite{bks}. This is appropriate to a system with large 
minus momentum.) In this formulation of the theory, the role of the
hamiltonian is played by $P^+$, which is the generator of translations in
$x^-$. We refer to this operator as $H$. In the problem at hand, $H$ is
the generator of $x^-$ translations in full QCD in the presence of the external
color field $\cal A$. To start, let us change to the interaction picture based
on full QCD without the external field as the base hamiltonian $H_0$ and the
interaction with the external field as the perturbation $V$. In this picture,
we write ${\cal M}$ as
\begin{eqnarray}
{\cal M} &=& \int dy^- e^{i\beta x_\pom p_A^+ y^-}
\langle k_1,s_1,\dots,k_N,s_N,{\cal E}|
\nonumber\\
&&\times
U(\infty,y^-)
[Q^\dagger(y^-) {\cal O}^j(0,y^-,{\bf 0})]_I
U(y^-,-\infty)
|0\rangle \hspace*{0.2 cm} .
\label{Mdef2}
\end{eqnarray}
Here a subscript $I$ on an operator denotes the operator in the interaction
picture specified above. The evolution operator $U$ is
\begin{equation}
U(\tau_2,\tau_1) = T\ \exp\left(
- i \int_{\tau_1}^{\tau_2} dz^- V_I(z^-)
\right) \hspace*{0.2 cm} .
\end{equation}
Now, within the approximations used here, the interaction $V$ does not produce
soft parton pairs from the vacuum and thus $V|0\rangle \approx 0$. (We discuss
the approximations and their validity at the end of this subsection.) Thus, we
can replace $U(y^-,-\infty)$ by $1$:
\begin{eqnarray}
{\cal M} &=& \int dy^- e^{i\beta x_\pom p_A^+ y^-}
\langle k_1,s_1,\dots,k_N,s_N,{\cal E}|
\nonumber\\
&&\times
U(\infty,y^-)
[Q^\dagger(y^-) {\cal O}^j(0,y^-,{\bf 0})]_I
|0\rangle \hspace*{0.2 cm} .
\label{Mdef3}
\end{eqnarray}

Since $V_I(z^-)$ vanishes when ${\cal A} = 0$, it is substantially non-zero
only for $z^- \sim 1/M$. (That is, the exchanged gluons carry plus momenta of
order $M$.) On the other hand, the integral in (\ref{Mdef3}) extends over a
much larger range, $|y^-|\sim 1/(x_\pom M)$. Thus most of the contribution to
${\cal M}$ comes from the regions $y^- \gg 1/M$ and $y^- \ll - 1/M$. In the
region $y^- \ll - 1/M$, we can approximate $U(\infty,y^-)$ by
$U(\infty,-\infty)$. In the region $y^- \gg 1/M$, we can approximate
$U(\infty,y^-)$ by $1$. Then
\begin{eqnarray}
{\cal M} &\approx& \int_{-\infty}^0 dy^- e^{i\beta x_\pom p_A^+ y^-}
\langle k_1,s_1,\dots,k_N,s_N,{\cal E}|
\nonumber\\
&&\times
U(\infty,-\infty)
[Q^\dagger(y^-) {\cal O}^j(0,y^-,{\bf 0})]_I
|0\rangle
\nonumber\\
&& + \int_{0}^\infty dy^- e^{i\beta x_\pom p_A^+ y^-}
\langle k_1,s_1,\dots,k_N,s_N,{\cal E}|
\nonumber\\
&&\times
[Q^\dagger(y^-) {\cal O}^j(0,y^-,{\bf 0})]_I
|0\rangle \hspace*{0.2 cm} ,
\end{eqnarray}
where we are allowed to set the integration endpoints to zero instead of, say,
$\pm 1/M$ because the difference is of order $x_\pom$ compared to the
integral. Now, adding and subtracting a term in the integral over $y^- < 0$,
we obtain
\begin{eqnarray}
{\cal M} &\approx& \int_{-\infty}^0 dy^- e^{i\beta x_\pom p_A^+ y^-}
\langle k_1,s_1,\dots,k_N,s_N,{\cal E}|
\nonumber\\
&&\times
[U(\infty,-\infty) - 1]
[Q^\dagger(y^-) {\cal O}^j(0,y^-,{\bf 0})]_I
|0\rangle
\nonumber\\
&& + \int_{-\infty}^\infty dy^- e^{i\beta x_\pom p_A^+ y^-}
\langle k_1,s_1,\dots,k_N,s_N,{\cal E}|
\nonumber\\
&&\times
[Q^\dagger(y^-) {\cal O}^j(0,y^-,{\bf 0})]_I
|0\rangle \hspace*{0.2 cm} .
\end{eqnarray}
The second term here is proportional to
\begin{equation}
\delta(\beta x_\pom p_A^+ + \sum k_i^+) \hspace*{0.2 cm} ,
\end{equation}
which vanishes because all of the terms in the argument of the delta function
are positive. Thus
\begin{eqnarray}
{\cal M} &=& \int_{-\infty}^0 dy^- e^{i\beta x_\pom p_A^+ y^-}
\langle k_1,s_1,\dots,k_N,s_N,{\cal E}|
\nonumber\\
&&\times
[U(\infty,-\infty) - 1]
[Q^\dagger(y^-) {\cal O}^j(0,y^-,{\bf 0})]_I
|0\rangle
\times \left(1 + {\cal O}(x_\pom)\right) \hspace*{0.2 cm} .
\label{usecond}
\end{eqnarray}

We can understand Eq.~(\ref{usecond}) as follows. First, the operator
$Q^\dagger {\cal O}$ creates a gluon and one of the special eikonal particles.
Then this state evolves according to QCD, possibly evolving into a system with
more partons. Since it has very large momentum in the minus direction, its
evolution in $y^-$ is slow (except for the inevitable ultraviolet
renormalizations). At $y^- \approx 0$, this system of quarks and gluons passes
through the external field. After that, it continues its slow evolution. 

With a straightforward derivation (which is given in Ref.~\cite{bks} in the
case of abelian gauge theory), one finds that in the high energy limit the
interaction with the external field becomes a very simple operator,
which, following \cite{bks}, we denote by $\bf F$:
\begin{equation}
U(\infty,-\infty) \approx {\bf F} \hspace*{0.2 cm} .
\label{UtoF}
\end{equation}
The action of $\bf F$ is simply to produce an eikonal phase for each
parton while leaving its minus momentum and its transverse position unchanged.
If the parton is at transverse position $\bf b$ when it passes through the
external field, then the phase is
\begin{equation}
F({\bf b}) \equiv
{\cal P}\ \exp\left\{
-ig\int_{-\infty}^{+\infty} d z^- {\cal A}_a^+(0,z^-,{\bf b})\, t_a
\right\} \hspace*{0.2 cm} .
\label{Fdef}
\end{equation}
Here the color matrices $t_a$ are the generators of SU(3) in the
representation appropriate to the color of the parton and the $\cal P$
indicates path ordering of the color matrices. In the case of the special
eikonal particle, ${\bf b} = 0$. Then Eq.~(\ref{usecond}) becomes
\begin{eqnarray}
{\cal M} &=& \int_{-\infty}^0 dy^- e^{i\beta x_\pom p_A^+ y^-}
\langle k_1,s_1,\dots,k_N,s_N,{\cal E}|
\nonumber\\
&&\times
[{\bf F}- 1]
[Q^\dagger(y^-) {\cal O}^j(0,y^-,{\bf 0})]_I
|0\rangle
\times \left(1 + {\cal O}(x_\pom)\right) \hspace*{0.2 cm} .
\label{uthird}
\end{eqnarray}

The approximation (\ref{UtoF}) is, in essence, very simple. For a scalar
parton with large minus momentum $k^-$ that absorbs a soft gluon with
momentum $q$, the approximation is
\begin{equation}
{ i \over (k + q)^2 + i\epsilon}\,(-ig\,2k^\mu \varepsilon_\mu)
\approx
{ i \over 2 k^- q^+ + i\epsilon}\,(-ig\,2k^- \varepsilon^+)
=
{ i \over  q^+ + i\epsilon}\,(-ig\, \varepsilon^+) \hspace*{0.2 cm} .
\end{equation}
For partons with spin 1/2 and 1, the diagrammatic derivation is similar, but
with a little work required to deal with the numerator structure. There is,
however, a substantial difficulty that is related to the question of which
gluons are soft and which are large $k^-$ partons. For us, the external field
represents the soft gluons and the partons all have large $k^-$. However,
really a soft external gluon can produce multiple soft quantum gluons with an
interaction that does not have the eikonal form. The soft quantum gluons can
interact somewhere else with the large $k^-$ partons (with interactions that
do have the eikonal form). Thus the present derivation works only when
such interactions among soft gluons are neglected. The present derivation also
neglects interactions with large transverse momentum quanta that affect
ultraviolet renormalization. Fortunately, in our application we extract a
result at the lowest nontrivial order of perturbation theory, where none of
these complications arise. 

\vskip 0.2 true cm 

\noindent  {\em The operator ${\bf F}$} 
\vskip .1 true cm

Let us specify in more detail the matrix elements of the
operator ${\bf F}$ between parton states. First of all, there are separate
factors for each parton:
\begin{eqnarray}
\lefteqn{
\langle k_1,s_1,\dots,k_N,s_N,{\cal E}|
{\bf F}
| p_1,s'_1,\dots,p_N,s'_N,{\cal E}\rangle
=}
\nonumber\\
&&
\langle k_1,s_1|{\bf F}| p_1,s'_1\rangle
\cdots
\langle k_N,s_N|{\bf F}| p_N,s'_N\rangle
\langle {\cal E}|{\bf F}|{\cal E}\rangle
+{\rm permutations}  \hspace*{0.2 cm} ,
\end{eqnarray}
where the notation `` + permutations'' indicates that we should match identical
partons in all possible ways. For a single parton state specified at time $x^-
= 0$ by its momentum $k = (k^-,{\bf k})$ and null-plane helicity $s$, we have
\begin{equation}
\langle k,s|{\bf F}| p,s'\rangle =
\delta_{s s'}\ (2 \pi)\,2 k^- \delta( k^{-} - p^-)
\widetilde F({\bf p} - {\bf k}) \hspace*{0.2 cm} ,
\label{partonF}
\end{equation}
where $\widetilde F({\bf p} - {\bf k})$ is the Fourier transform of $F({\bf
p})$ defined in Eq.~(\ref{Fdef}):
\begin{equation}
\widetilde F({\bf p} - {\bf k}) =
\int d{\bf b}\
e^{i ({\bf p} - {\bf k})\cdot {\bf b}}
F({\bf b}) \hspace*{0.2 cm} .
\end{equation}
For the special particle $\cal E$ we have
\begin{equation}
\langle {\cal E}|{\bf F}| {\cal E}\rangle =
F({\bf 0}) \hspace*{0.2 cm} ,
\label{eikonalF}
\end{equation}
thus giving us back the eikonal phase factor that was part of the definition
of the measurement operator, with the lower limit on the $y^-$ integration
approximated by $-\infty$.

\vskip 0.2 true cm

\noindent  {\em The Born approximation} 
\vskip .1 true cm

We now revert to the lowest order of perturbation theory at which a leading 
$1 / x_\pom \to \infty$ contribution is obtained. 
We take the order $g^2$ contribution to ${\bf F} -1$. 
For the evolution of the partonic state between time $y^-$ and time zero, 
and then from time zero to the final state at $x^- \to \infty$
we take order zero of QCD perturbation theory. Then there is but one gluon in
the final state and we have
\begin{eqnarray}
{\cal M}_{0,2} &=& \int_{-\infty}^0 dy^- e^{i\beta x_\pom p_A^+ y^-}
(2\pi)^{-3}\int { d p^- \over 2 p^-}\int d^2 {\bf p}\sum_{s'}\
\langle k,s,{\cal E}|[{\bf F}- 1]_{g^2}| p,s',{\cal E}\rangle
\nonumber\\
&&\times
\exp(i p^+ y^-)
\langle p,s',{\cal E}|
Q^\dagger(0) {\cal O}^j(0,0,{\bf 0})
|0\rangle \hspace*{0.2 cm} ,
\label{Mfourth}
\end{eqnarray}
where the subscript $g^2$ reminds us that we are to use the $g^2$ term in the
expansion of $[{\bf F} - 1]$ and where $p^+$ is the free particle
plus momentum,
\begin{equation}
p^+ = { {\bf p}^2 \over 2 p^-} \hspace*{0.2 cm} .
\end{equation}
For the matrix element of $[{\bf F}-1]$ we have, using Eqs.~(\ref{partonF}) and
(\ref{eikonalF}), 
\begin{equation}
\langle k,s,{\cal E}|[{\bf F}- 1]_{g^2}| p,s',{\cal E}\rangle =
\delta_{ss'}\,(2\pi)\,2p^- \delta(k^- - p^-)
\left[
\widetilde F({\bf k}-{\bf p})\
F(0)
\right]_{g^2} \hspace*{0.2 cm} .
\end{equation}
Also, we perform the $y^-$ integration to produce an energy denominator. Then
\begin{equation}
{\cal M}_{0,2} =
\int {d^2 {\bf p}\over (2\pi)^2}\
\left[
\widetilde F({\bf k}-{\bf p})\
F(0)
\right]_{g^2}
{ -i \over \beta x_\pom p_A^+ +  {\bf p}^2/( 2 k^-)}\
\langle k^-,{\bf p},s|
{\cal O}^j(0,0,{\bf 0})
|0\rangle \hspace*{0.2 cm} .
\label{ufifth}
\end{equation}
It will be useful at this point to adopt the notation
\begin{equation}
\label{wavefctndef}
\psi^{ji}({\bf k},{\bf p}) = 
i\ { \langle k^-,{\bf p};i|
{\cal O}^j(0)
|0\rangle 
\over \beta x_\pom p_A^+ + {\bf p}^2/(2k^-)} \hspace*{0.3 cm} .
\end{equation}
Here $\psi$ represents the wave function of the gluon state just before it
interacts with the external field. In $\psi$, ${\bf p}$ is 
the transverse momentum
of the gluon in the intermediate state and $i$ is its transverse polarization
($i = (1,2)$). Since the minus momentum of this gluon has been set to 
the minus momentum of the final state gluon, we have
\begin{equation}
\label{kminus}
k^- = {{\bf k}^2 \over 2(1-\beta)x_\pom p_A^+} \hspace*{0.3 cm} .
\end{equation}

Thus 
\begin{equation}
{\cal M}_{0,2} =
-\int {d^2 {\bf p}\over (2\pi)^2}\
\left[
\widetilde F({\bf k}-{\bf p})\
F(0)
\right]_{g^2}
\psi^{js}({\bf k},{\bf p}) \hspace*{0.2 cm} .
\label{usepsi}
\end{equation}

\vskip 0.2 true cm 

\noindent  {\em Evaluation of $\psi$} 
\vskip .1 true cm

With the replacement in Eq.~(\ref{kminus}), Eq.~(\ref{wavefctndef}) reads
\begin{equation}
\label{wavefctnmid}
\psi^{js}({\bf k},{\bf p}) = 
i\ { {\bf k}^2 \over x_\pom p_A^+}\
{ \langle k^-,{\bf p};s|{\cal O}^j(0)|0\rangle 
\over \beta {\bf k}^2 + (1-\beta){\bf p}^2} \hspace*{0.3 cm} .
\end{equation}
To complete the evaluation of $\psi$, we need to evaluate the operator matrix
element
\begin{equation}
\langle k^-,{\bf p},s,b|
{\cal O}_a^j(0)
|0\rangle \hspace*{0.2 cm} ,
\end{equation}
where, here, we have made the color indices $a,b$ explicit. Using
Eq.~(\ref{calOdef}),
${\cal O}^j(y) = 
 -i\beta x_\pom p_A^+\ A_a(y)^j 
- \partial^j A_a(y)^+$,
we obtain
\begin{equation}
\langle k^-,{\bf p},s,b|
{\cal O}_a^j(0)
|0\rangle
 =
-i\,\delta_{ab}\left(
\beta x_\pom p_A^+\
\epsilon^j(k^-,{\bf p},s)
+  p^j 
\epsilon^+(k^-,{\bf p},s)
\right) \hspace*{0.2 cm} .
\end{equation}
The polarization vectors, for transverse polarization $s = 1,2$, are those
appropriate to $A^- = 0$ gauge:
\begin{equation}
\epsilon^j(k^-,{\bf p},s) = \delta^{js} \;\; , \hskip 1 cm
\epsilon^+(k^-,{\bf p},s) = { 1 \over k^-}\ p^s
= { 2 (1-\beta) x_\pom p_A^+ \over {\bf k}^2}\ p^s  \hspace*{0.2 cm} . 
\end{equation}
Thus
\begin{eqnarray}
\langle k^-,{\bf p},s|
{\cal O}^j(0)
|0\rangle
 &=&
-i\,\left(
\beta x_\pom p_A^+\
\delta^{js}
+  { 2 (1-\beta) x_\pom p_A^+ \over {\bf k}^2}\  p^j p^s 
\right)
\nonumber\\
 &=&
-i\,
{ x_\pom p_A^+ \over {\bf k}^2}
\left(
\beta{\bf k}^2 \delta^{js}
+  2(1-\beta) p^j p^s  
\right) \hspace*{0.2 cm} . 
\end{eqnarray}
Inserting this result into Eq.~(\ref{wavefctnmid}), we have
\begin{equation}
\label{wavefctnfinal}
\psi^{js}({\bf k},{\bf p}) = 
{\beta {\bf k}^2 \delta^{js} +  2(1-\beta) p^j p^s
\over \beta {\bf k}^2 + (1-\beta){\bf p}^2} \hspace*{0.2 cm} .
\end{equation}

\vskip 0.2 true cm 

\noindent  {\em Expansion in powers of the external field} 
\vskip .1 true cm

We now expand the eikonal phase factor $\widetilde F({\bf k}-{\bf p})\ F(0)$
in Eq.~(\ref{usepsi}), picking out the contribution proportional to two powers
of the external field. We choose to call the momentum labels of the fields $q
+ s$ and $-s$, and we symmetrize over which label belongs to which field. This
gives
\begin{eqnarray}
{\cal M}_{0,2}&=&{ g^2 \over 2}
\int\!{ d^4 q \over (2\pi)^4}\int\!{ d^4 s \over (2\pi)^4}\
\widetilde {\cal A}^+_b(q + s)\,\widetilde {\cal A}^+_a(- s)
\nonumber\\
&&\times\biggl\{
\psi_{js}({\bf k},{\bf k})
\left[
{ i t_a t_b\over s^+ + i\epsilon}
+ { i t_b t_a\over - s^+ + i\epsilon}
\right] 2\pi \delta(q^+)
\nonumber\\
&&\ \ \ +
\psi_{js}({\bf k},{\bf k}-{\bf q})
\left[
{ i t_b t_a\over s^+ + i\epsilon}
+ { i t_a t_b\over - s^+ + i\epsilon}
\right] 2\pi \delta(q^+)
\nonumber\\
&&\ \ \ -
\psi_{js}({\bf k},{\bf k}+{\bf s})\
t_b t_a\
2\pi \delta(q^+ + s^+) 2\pi \delta(s^+)
\nonumber\\
&&\ \ \ -
\psi_{js}({\bf k},{\bf k}-{\bf q}-{\bf s})\
t_a t_b\
2\pi \delta(q^+ + s^+) 2\pi \delta(s^+)
\biggr\} \hspace*{0.2 cm} .
\label{toppart1}
\end{eqnarray}
Note the denominators $1/[\pm s^+ + i\epsilon]$, which arise from the path
ordering instruction in Eq.~(\ref{Fdef}). For instance,
\begin{equation}
\int\! dx^- \int\! dy^-\, \theta(y^- > x^-) \
e^{-i(q^+ + s^+) x^-}\,e^{i s^+ y^-}
= 2\pi \delta(q^+)\,{ i \over s^+ + i\epsilon} \hspace*{0.3 cm} .
\end{equation}

\vskip 0.2 true cm 

\noindent  {\em Color singlet simplification} 
\vskip .1 true cm

We can simplify the result in Eq.~(\ref{toppart1}) if we recall that in our
problem we are to replace the classical field product $\widetilde{\cal
A}^+_b(q + s)\,\widetilde {\cal A}^+_a(- s)$ by the corresponding matrix
element of the quantum fields. In this matrix element, the color field is in a
color singlet configuration, since it results from the scattering of a meson
that starts in a color singlet state and ends in a color singlet state. Thus
we are entitled to make the replacement (\ref{colorsinglet2}),
\begin{equation}
\widetilde{\cal A}^+_b\,\widetilde {\cal A}^+_a
\to 
\delta_{ba}\ { 1 \over N_C^2 - 1}\,\widetilde{\cal A}^+_c\,
\widetilde {\cal A}^+_c \hspace*{0.2 cm} .
\label{colproj}
\end{equation}
Then we can evaluate
\begin{equation}
t_a t_b\, \delta_{ba} = C_A \equiv N_C 
\end{equation}
since the generator matrices $t_a$ are in the adjoint representation of SU(3).
This provides a great simplification because
\begin{equation}
\left[
{ i \over s^+ + i\epsilon}
+ { i \over - s^+ + i\epsilon}
\right]
= 2\pi \delta(s^+) \hspace*{0.2 cm} .
\end{equation}
Thus
\begin{eqnarray}
{\cal M}_{0,2}&=&{g_s^2 \over 2}{ C_A \over N_C^2 - 1}
\int\!{ dq^- d^2 {\bf q} \over (2\pi)^3}
\int\!{ ds^- d^2 {\bf s} \over (2\pi)^3}\
\widetilde {\cal A}^+_c(0,q^- + s^-,{\bf q} + {\bf s})\,
\widetilde {\cal A}^+_c(0,- s^-, - {\bf s})
\nonumber\\
&&\times\biggl\{
\psi_{js}({\bf k},{\bf k})
 +
\psi_{js}({\bf k},{\bf k}-{\bf q})
\nonumber\\
&&\ \ \ \ 
 -
\psi_{js}({\bf k},{\bf k}+{\bf s})\
 -
\psi_{js}({\bf k},{\bf k}-{\bf q}-{\bf s})\
\biggr\} \hspace*{0.2 cm} .
\label{toppart2}
\end{eqnarray}

This result has the form anticipated in Eq.~(\ref{extractu}). We are thus able
to extract the function $u(\beta,{\bf k},{\bf q},{\bf s})$. We have
\begin{equation}
u(\beta,{\bf k},{\bf q},{\bf s}) =
\psi({\bf k},{\bf k})
 +
\psi({\bf k},{\bf k}-{\bf q})
 -
\psi({\bf k},{\bf k}+{\bf s})\
 -
\psi({\bf k},{\bf k}-{\bf q}-{\bf s}) \hspace*{0.2 cm} .
\label{ucomb1}
\end{equation}
Here $\psi$ is given in Eq.~(\ref{wavefctnfinal}). Our evaluation of 
$u(\beta,{\bf k},{\bf q},{\bf s})$ is thus complete.

\vskip 0.7 true cm 

\noindent  {\large \bf 3.2 The exchanged gluons } 
\vskip .1 true cm

Each of the exchanged gluons appears between the upper subgraph $u$, in
which the partons have large momentum in the minus direction, and  the
subgraph $L$, in which the  momentum components in the plus
direction and in the minus direction are of the same order. The
factor for one of the gluons is, in an obvious notation,
\begin{equation}
{\cal U}_\mu\ 
{ i \over 2 l^+ l^- - {\bf l}^2}
\left(
-g^\mu_\nu + { l^\mu v_\nu +  v^\mu 
l_\nu\over l\cdot v}
\right)
{\cal L}^\nu \hspace*{0.2 cm} ,
\label{gluon}
\end{equation}
where we have denoted the gluon momentum, either $q^\mu + s^\mu$ or $-
s^\mu$, by $l^\mu$. Here $v^\mu = (1,0,{\bf 0})$ so that $v\cdot l = l^-$. The
vector $v^\mu$ appears because we are using $A^- = 0$ gauge.

In going to the high energy limit, we have already made some replacements that
simplify Eq.~(\ref{gluon}). First, according to Eq.~(\ref{toppart2}), only the
plus component of the external field appears. Thus the factor (\ref{gluon})
becomes
\begin{equation}
{\cal U}^- 
{ i \over 2 l^+ l^- - {\bf l}^2}
\left(
-g^+_\nu + { l^+ v_\nu +  v^+ l_\nu\over l\cdot v}
\right)
{\cal L}^\nu \hspace*{0.2 cm} .
\label{gluon2}
\end{equation}
Second, in Eq.~(\ref{toppart2}), our exchanged gluon fields are
evaluated with $l^+ = 0$. Thus Eq.~(\ref{gluon}) becomes
\begin{equation}
{\cal U}^- 
{ i \over  - {\bf l}^2}
\left(
-g^+_\nu + { v^+ l_\nu\over l\cdot v}
\right)
{\cal L}^\nu \hspace*{0.2 cm} .
\label{gluon3}
\end{equation}
We can make one more simplification. Because of gauge invariance, when we sum
over all ways of attaching our gluon to the quarks in the lower subgraph, we
can drop the term proportional to $l_\nu{\cal L}^\nu$. (Here we use the fact
that the two gluons must be in a net color singlet state, so that we
effectively have abelian Ward identities.) Thus, our factor (\ref{gluon})
finally becomes 
\begin{equation}
{\cal U}^- 
{ i  \over  {\bf l}^2}
{\cal L}^+  \hspace*{0.2 cm} . 
\label{gluon4}
\end{equation}
This result is built into Eq.~(\ref{structure}).

\vskip 0.7 true cm 

\noindent  {\large \bf 3.3 The lower subgraph} 
\vskip .1 true cm

We now turn to the lower subgraph. Consider the matrix element
\begin{equation}
{\cal M} =
\int\!{dp_{\!A^\prime}^+\over 2\pi} \int\! d^4y \int\! d^4x\
e^{i p_{\!A^\prime}\cdot y - i p_{\!A}\cdot x}\
\langle 0|T{\cal O}'(y) {\cal O}(x) |0\rangle_{\cal A} \hspace*{0.2 cm} .
\end{equation}
The operator ${\cal O}$ creates the initial heavy quark-antiquark state and
then the operator ${\cal O}'$ destroys the final heavy quark state:
\begin{eqnarray}
{\cal O}(x) &=& e\, e_Q\
\varepsilon_\mu\overline\Psi(x)\gamma^\mu\Psi(x)
\nonumber\\
{\cal O}'(y) &=& e\, e_Q\
\varepsilon'_\mu\overline\Psi(y)\gamma^\mu\Psi(y) \hspace*{0.2 cm} .
\label{lowerops}
\end{eqnarray}
Here $\varepsilon_\mu$ and $\varepsilon'_\mu$ are the polarization vectors for 
the initial and final state photons and $e\, e_Q$ is the coupling of 
the heavy quark to the photon. The initial and final momenta of the 
photon are
\begin{eqnarray}
p_{\! A}^\mu &=&(p_{\! A}^+,0,{\bf 0}),
\nonumber\\
p_{\! A^\prime}^\mu &=&
(p_{\! A^\prime}^+,{ {\bf q}^2 \over 2 p_{\! A'}^+},-{\bf q)} 
\hspace*{0.2 cm}.
\end{eqnarray}
We expand ${\cal M}$ in perturbation theory 
and extract the term ${\cal M}_{0,2}$
proportional to two powers of $g_s {\cal A}$ and zero additional powers of
$g_s$. The coefficient of this term is the Green function $L_0$ of
Eq.~(\ref{start}):
\begin{eqnarray}
{\cal M}_{0,2}&=& {1 \over 2} 
\int\!{dp_{\!A^\prime}^+\over 2\pi}
\int\!{d^4 s\over (2\pi)^4}
\int\!{d^4 w\over (2\pi)^4}\
\widetilde A(s)^c_\rho\,
\widetilde A(w)^d_\sigma
\nonumber\\
&&\times
(2\pi)^4 \delta^4(p_A + s + w - p_{A'})\
L_0(q,s,\varepsilon,\varepsilon')_{cd}^{\rho\sigma} \hspace*{0.2 cm} . 
\end{eqnarray}
When we use the color singlet nature of $L_0$ and make the simplifications
that result in the $1 / x_\pom \to  \infty$ limit, Eq.~(\ref{mid1}) leads us to
anticipate ${\cal M}_{0,2}$ takes the form
\begin{eqnarray}
{\cal M}_{0,2}&=& {1 \over 2} 
\int\!{d^2 {\bf s}\over (2\pi)^2}
\int\!{d s^+\over 2\pi}\int\!{d w^+\over 2\pi}\
\widetilde A(s^+,0,{\bf s})_c^-\,
\widetilde A(w^+,0,- {\bf q} - {\bf s})_c^-
\nonumber\\
&&\times
\int\!{d s^-\over 2\pi}\
{ 1 \over N_c^2 -1}\,
L_0({\bf q};0,s^-,{\bf s},\varepsilon,\varepsilon')_{ee}^{++} 
\hspace*{0.2 cm} . 
\end{eqnarray}
Using Eq.~(\ref{Ldef}), we identify the function $L$ that appears in the 
formula (\ref{structure}) for the diffractive gluon distribution. Thus
\begin{equation}
{\cal M}_{0,2}= {1 \over 2} 
\int\!{d^2 {\bf s}\over (2\pi)^2}
\int\!{d s^+\over 2\pi}\int\!{d w^+\over 2\pi}\
\widetilde A(s^+,0,{\bf s})_c^-\,
\widetilde A(w^+,0,- {\bf q} - {\bf s})_c^-\
p_A^+
L({\bf q};{\bf s},\varepsilon,\varepsilon') \hspace*{0.2 cm} .
\label{extractL}
\end{equation}
Our aim in this section is to calculate ${\cal M}_{0,2}$ in the 
$1 / x_\pom \to \infty$
limit and then to use Eq.~(\ref{extractL}) to extract $L({\bf q};{\bf
s},\varepsilon,\varepsilon')$.

The matrix element ${\cal M}$ is evaluated in the presence of an external gluon
field ${\cal A}$, which we think of as created by the scattering in the upper
subgraph that we analyzed earlier. In this subsection, it is convenient to use
a reference frame in which the minus momenta of the partons in the upper
subgraph, $k^- \sim {\bf k}^2/(x_\pom p_{\! A}^+)$, are fixed to be of order
$M$ as $x_\pom \to 0$, so that the external field can be regarded as remaining
fixed in the $x_\pom \to 0$ limit. Then $p_{\! A}^+$ is large:
\begin{equation}
p_{\! A}^+ \sim M/x_\pom \hspace*{0.2 cm} .
\end{equation}

We are now ready for the evaluation of $\cal M$. We use the same
methods that we used for the upper subgraph. As we found for the
upper subgraph, the interaction with the external field can be
approximated by the eikonal operator ${\bf F}$ that supplies for a
parton at transverse position ${\bf b}$ a phase
\begin{equation}
F({\bf b}) \equiv
{\cal P}\exp\left\{ - i g 
\int_{-\infty}^\infty dy^+ {\cal A}_a^-(y^+,0,{\bf b}) t_a
\right\} \hspace*{0.2 cm} .
\end{equation}
Note that here the indices $+$ and $-$ are interchanged compared to what 
they were in our evaluation of the upper subgraph. We do not, however, make 
this change explicit by introducing a new name for the operator ${\bf F}$, 
the phase function $F(b)$, or its Fourier transform $\widetilde F({\bf q})$,
\begin{equation}
\widetilde F({\bf q}) =
\int d{\bf b}\
e^{i {\bf q}\cdot {\bf b}}
F({\bf b}) \hspace*{0.2 cm} .
\end{equation}
In ${\cal M}_{0,2}$, we have the order $g^2$ term in the perturbative expansion
of $\bf F$, with everything else evaluated at order zero in QCD interactions:
\begin{eqnarray}
{\cal M}_{0,2} &\approx & 
\int\! { dp_{\!A^\prime}^+ \over (2\pi)^2}
\int_0^{\infty}\! dy^+
\int\! dy^- \int\! d{\bf y}\ 
\int_{-\infty}^0\! dx^+ 
\int\! dx^- \int\! d{\bf x}\
e^{i p_{\!A^\prime}\cdot y - i p_{\!A}\cdot x}
\nonumber\\
&&\times
\langle 0|{\cal O}'(y)
[{\bf F}]_{g^2}
 {\cal O}(x) |0\rangle \hspace*{0.2 cm} .
\label{lower0}
\end{eqnarray}

Equation (\ref{lower0}) is actually the main result. The rest of the derivation
amounts to some straightforward manipulations. We first insert intermediate
states into the expression:
\begin{eqnarray}
{\cal M}_{0,2} &\approx& 
\int\! {dp_{\!A^\prime}^+ \over 2\pi}\ 
\int_0^{\infty}\! dy^+
\int\! dy^- \int\! d{\bf y}\ 
\int_{-\infty}^0\! dx^+ 
\int\! dx^- \int\! d{\bf x}\
e^{i p_{\!A^\prime}\cdot y - i p_{\!A}\cdot x}
\nonumber\\
&&\times
(2\pi)^{-12}\int{ dz_1 \over 2z_1}\int d{\bf r}_1
\int{ dz_2 \over 2z_2}\int d{\bf r}_2
\int{ dz'_1 \over 2z'_1}\int d{\bf r}'_1
\int{ dz'_2 \over 2z'_2}\int d{\bf r}'_2
\nonumber\\
&&\times
\langle 0|{\cal O}'(y)|
z'_1 p_{A'}^+,{\bf r}'_1,s'_1;z'_2 p_{A'}^+,{\bf r}'_2,s'_2
\rangle
\nonumber\\
&&\times
\langle
z'_1 p_{A'}^+,{\bf r}'_1,s'_1;z'_2 p_{A'}^+,{\bf r}'_2,s'_2
|[{\bf F}]_{g^2}|
z_1 p_A^+,{\bf r}_1,s_1;z_2 p_A^+,{\bf r}_2,s_2
\rangle
\nonumber\\
&&\times
\langle 
z_1 p_A^+,{\bf r}_1,s_1;z_2 p_A^+,{\bf r}_2,s_2
|{\cal O}(x) |0\rangle \hspace*{0.2 cm} .
\end{eqnarray}
Here the state created by the operator $\cal O$ consists of an antiquark with
momentum $(z_1 p_A^+,{\bf r}_1)$ and spin $s_1$ and a quark with momentum
$(z_2 p_A^+,{\bf r}_2)$ and spin $s_2$. After the operator ${\bf F}$
acts, we have a similar state with the momenta and spins denoted with primes.
Next, we perform the space integrals. The integrations over $y^+$ and $x^+$
give $k^-$ denominators. The other space integrals give delta functions that
can be used to eliminate some of the momentum integrations:
\begin{eqnarray}
\lefteqn{{\cal M}_{0,2} \approx }
\nonumber\\
&&
- { 1 \over 4 (2\pi)^{7}}
\int\! dp_{\!A^\prime}^+\ 
\int{ dz \over z (1-z)}\int d{\bf r}
\int{ dz' \over z'(1-z')}\int d{\bf r}'
\nonumber\\
&&\times
\left[
{\bf q}^2 - {{\bf r}^{\prime 2} + M^2\over z'}
- {({\bf q}+{\bf r}')^2 + M^2 \over (1-z')}
+ i\epsilon
\right]^{-1}
\left[ 
- {{\bf r}^2 + M^2 \over z}
- {{\bf r}^2 + M^2 \over (1-z)}
+ i\epsilon
\right]^{-1}
\nonumber\\
&&\times
\langle 0|{\cal O}'(0)|
z' p_{A'}^+,{\bf r}',s'_1;(1-z')p_{A'}^+,-{\bf q}-{\bf r}',s'_2
\rangle
\nonumber\\
&&\times
\langle
z' p_{A'}^+,{\bf r}',s'_1;(1-z')p_{A'}^+,-{\bf q}-{\bf r}',s'_2
|[{\bf F}]_{g^2}|
z p_A^+,{\bf r},s_1;(1-z) p_A^+,-{\bf r},s_2
\rangle.
\nonumber\\
&&\times
\langle
z p_A^+,{\bf r},s_1;(1-z) p_A^+,-{\bf r},s_2
|{\cal O}(0) |0\rangle \hspace*{0.2 cm} .
\end{eqnarray}
We introduce here a wave function $\Phi$ defined in Appendix B, 
so that we can make the replacements 
\begin{eqnarray}
\lefteqn{
{\langle
z p_A^+,{\bf r},s_1;(1-z) p_A^+,-{\bf r},s_2
|{\cal O}(0) |0\rangle
\over
- [{\bf r}^2 + M^2]/ z
- [{\bf r}^2 + M^2]/ (1-z) }
=}
\nonumber\\
&&\hskip 1cm
 e\, e_Q\ \sqrt{z(1-z)}\
\Phi(z,{\bf r},M,\varepsilon)_{s_1 s_2} \hspace*{0.2 cm} ,
\label{Phiuse1}
\end{eqnarray}
and
\begin{eqnarray}
\lefteqn{
{\langle 0|{\cal O}'(0)|
z' p_{A'}^+,{\bf r}',s'_1;(1-z')p_{A'}^+,-{\bf q}-{\bf r}',s'_2
\rangle
\over
{\bf q}^2
- [{\bf r}^{\prime 2} + M^2]/ z'
- [({\bf q}+{\bf r}')^2 + M^2 ]/ (1-z') }
=}
\nonumber\\
&&\hskip 1cm
e\, e_Q\ \sqrt{z'(1-z')}\
\Phi^\dagger
(z',{\bf r}' + z' {\bf q},M,\varepsilon')_{s'_2 s'_1} \hspace*{0.2 cm} .
\label{Phiuse2}
\end{eqnarray}
For the matrix element of ${\bf F}$ we can write
\begin{eqnarray}
\lefteqn{
\langle
z' p_{A'}^+,{\bf r}',s'_1;(1-z')p_{A'}^+,-{\bf q}-{\bf r}',s'_2
|[{\bf F}]_{g^2}|
z p_A^+,{\bf r},s_1;(1-z) p_A^+,-{\bf r},s_2
\rangle
=}
\nonumber\\
&&
\delta_{s'_1 s_1}\delta_{s'_2 s_2}\,
(2\pi) 2 z p_A^+ \delta(z'p_{A'}^+ - z p_A^+)\,
(2\pi) 2 (1-z) p_A^+ \delta((1-z')p_{A'}^+ - (1-z) p_A^+)\,
\nonumber\\
&&\times
[\widetilde F({\bf r}' - {\bf r})\,
\widetilde F({\bf r} - {\bf q} -{\bf r}' )]_{g^2} \hspace*{0.2 cm} .
\end{eqnarray}

With these substitutions we have
\begin{eqnarray}
\lefteqn{{\cal M}_{0,2} \approx }
\nonumber\\
&&
- {  e^2\, e^2_Q\,p_A^+\over  (2\pi)^{5}}
\int dz \int d{\bf r}
\int d{\bf r}'
\nonumber\\
&&\times
\Phi^\dagger
(z,{\bf r}' + z {\bf q},M,\varepsilon')_{s_2 s_1}
[\widetilde F({\bf r}' - {\bf r})\,
\widetilde F({\bf r} - {\bf q} -{\bf r}' )]_{g^2}
\Phi(z,{\bf r},M,\varepsilon)_{s_1 s_2} \hspace*{0.2 cm} .
\end{eqnarray}
In this equation, we expand 
$\widetilde F({\bf r}' - {\bf r})\,  
\widetilde F({\bf r} - {\bf q} -{\bf r}' )$ 
in powers of ${g\cal A}$ and keep the order $g^2$ terms. There are four
contributions. Using Eq.~(\ref{extractL}), we  extract $L$: 
\begin{eqnarray}
\label{genl}
\lefteqn{
L({\mbox{\bf q}}, {\mbox{\bf s}},  M, 
\varepsilon, 
\varepsilon^\prime) = {{e^2_Q \, e^2 \, g_s^2  } 
\over {4 \, \pi} } \, 
\int \, { {d^2 {\mbox{\bf r}}} \over {(2 \, \pi)^2} } 
\int_0^1 \, d z }
\\
&& \times {\mbox{Tr}} \left\{ 
\left[ -
\Phi^\dagger (z, {\mbox{\bf r}} + {\mbox{\bf s}}
+ z \, {\mbox{\bf q}}, M,\varepsilon^\prime )  
 - 
\Phi^\dagger (z, {\mbox{\bf r}} - {\mbox{\bf s}}
- (1-z) \, {\mbox{\bf q}}, M,\varepsilon^\prime) 
\right. \right. 
\nonumber\\
&& + \left. \left. 
\Phi^\dagger (z, {\mbox{\bf r}} + z \, {\mbox{\bf q}}, 
M,\varepsilon^\prime)  
 + 
\Phi^\dagger (z, {\mbox{\bf r}} - (1-z) \, {\mbox{\bf q}}, 
M, \varepsilon^\prime)  \right] \, 
\Phi (z, {\mbox{\bf r}} , M, \varepsilon) \right\} \hspace*{0.3 cm} .     
\nonumber
\end{eqnarray}
We find in  Appendix B that
\begin{equation}
\label{spinphi} 
\Phi (z, {\mbox{\bf k}} ,M, \varepsilon) 
=  { 1 \over 
{  \left(  {\mbox{\bf k}}^2 + M^2 \right) } } \, 
 \left[  
(1 -  z) \, \varepsilon \cdot \sigma \, {\mbox{\bf k}} \cdot \sigma 
- z \, {\mbox{\bf k}} \cdot \sigma \, \varepsilon \cdot \sigma 
 + i \, M \, \varepsilon \cdot  \sigma   \right]
\hspace*{0.3 cm} .
\end{equation}
This completes our evaluation of $L({\bf q}, {\bf s},  M, \varepsilon,
\varepsilon')$.

\vskip 0.7 true cm 

\noindent  {\large \bf 3.4 The gluon and quark distributions} 
\vskip .1 true cm

We now recap the result for the diffractive gluon 
 distribution  and  give its extension 
 to the case of the diffractive quark distribution. 
 Let us introduce a parton index $a$,  with $a = g , q$.  
 Let us define functions $U_a$ in terms of the functions 
$u$ of Eq.~(\ref{udef}) as follows: 
\begin{equation}
\label{capu}
U_a( x_\pom , \beta , {\mbox{\bf q}},{\mbox{\bf s}},
{\mbox{\bf s}}^\prime   ) = 
{ { g_s^4 \, c_a } \over { 4 \, \pi \,  \beta \, 
(1- \beta) \, 
x_{\pom}^2 }} \, \int \,{{ d^2 {\mbox{\bf k}} } \over 
{( 2 \, \pi )^2}} \, {\mbox{Tr}} \left( u_{a}^{\dagger} 
( \beta , {\mbox{\bf k}} , {\mbox{\bf q}},{\mbox{\bf s}}^\prime )
\,  
u_{a} ( \beta , {\mbox{\bf k}} , {\mbox{\bf q}},{\mbox{\bf s}} ) 
\right) \hspace*{0.2 cm} ,  
\end{equation}  
with $u_a$  given by the linear combination 
(\ref{ucomb1}) of wave functions, 
\begin{equation}
\label{comb}
u_{a} ( \beta , {\mbox{\bf k}} , {\mbox{\bf q}},{\mbox{\bf s}} ) = 
\psi_a ( {\mbox{\bf k}} , {\mbox{\bf k}} ) 
- \psi_a ( {\mbox{\bf k}} , {\mbox{\bf k}} + 
{\mbox{\bf s}} ) 
+ \psi_a ( {\mbox{\bf k}} , {\mbox{\bf k}} - 
{\mbox{\bf q}}) - \psi_a ( {\mbox{\bf k}} , {\mbox{\bf k}} - 
{\mbox{\bf q}} - {\mbox{\bf s}} ) \hspace*{0.2 cm} .  
\end{equation} 
The difference between the gluon and the quark cases is 
in the expressions for the color factors $c_a$ 
and the wave functions $\psi_a$. 
The color factor for gluons may be read from Eq.~(\ref{structure}),   
\begin{equation}
\label{colg} 
  c_g = C_A^2 \, (N_c^2 - 1) 
\hspace*{0.2 cm} . 
\end{equation} 
A similar calculation yields the color factor for quarks:  
\begin{equation}
\label{colq} 
 c_q = C_F^2 \, N_c 
\hspace*{0.2 cm} . 
\end{equation} 
The wave function for gluons is given in Eq.~(\ref{wavefctnfinal}), 
\begin{equation}
\label{psig}
 \psi_g^{ i j} ( {\mbox{\bf k}} , 
 {\mbox{\bf p}} ) = { 
 {  \beta \, {\mbox{\bf k}}^2  \, \delta^{i j }   + 2 \, ( 1 - \beta) 
 {\mbox{\bf p}}^i \, {\mbox{\bf p}}^j } 
\over { \beta \, {\mbox{\bf k}}^2 + (1-\beta) \, {\mbox{\bf p}}^2 }}  
\hspace*{0.2 cm} . 
\end{equation} 
We find in Appendix A that 
the wave function for quarks is 
\begin{equation}
\label{psiq}
 \psi_q( {\mbox{\bf k}} , 
 {\mbox{\bf p}} ) = { 
{\sqrt{ \beta \, (1-\beta) \, {\mbox{\bf k}}^2}   } 
\over { \beta \, {\mbox{\bf k}}^2 + (1-\beta) \, {\mbox{\bf p}}^2 }}  
\, {\mbox{\bf p}} \cdot \sigma 
\hspace*{0.2 cm} .  
\end{equation}
 
Then  we may rewrite the overall structure (\ref{structure}) of the result 
in the following form, for both the gluon and the quark distributions: 
\begin{eqnarray} 
\label{conv} 
&& {{d f_{a/A}^{\rm {diff}} 
(\beta x_\pom,  x_\pom , {\mbox{\bf q}}^2 , M
) } \over
{dx_\pom\,dt}} = 
{1 \over {64 \, \pi^2}} \,  
{1 \over 2} \, \sum_{\varepsilon} \, \sum_{\varepsilon^\prime} \, 
\int \, { {d^2  {\mbox{\bf s}}}  \over {(2 \, \pi)^2} }\,
 {1 \over { {{\mbox{\bf s}}}^2 \, 
 ( {\mbox{\bf q}}+{\mbox{\bf s}})^2 } } \,  
\int \, { {d^2  {\mbox{\bf s}}^{\prime}}  \over {(2 \, \pi)^2} }\,
 {1 \over { {{\mbox{\bf s}}}^{\prime \, 2} 
( {{\mbox{\bf q}}}+{{\mbox{\bf s}}}^{\prime })^2} } \, 
\nonumber\\ 
&& \hskip 4 cm \times 
L({\mbox{\bf q}}, {\mbox{\bf s}}, M
, \varepsilon, \varepsilon^\prime
) \, 
U_a(x_\pom ,  \beta, {\mbox{\bf q}},
{\mbox{\bf s}},{\mbox{\bf s}}^\prime) \, 
L( {\mbox{\bf q}}, {\mbox{\bf s}}^{\prime }, M
, \varepsilon, \varepsilon^\prime
) \;\;\;\;,       
\end{eqnarray} 
with $U_a$ given in Eq.~(\ref{capu}) and $L$ in Eq.~(\ref{genl}). 

\vskip 1.5 true cm 

\setcounter{equation}{0}
\setcounter{sect}{4}

\noindent  {\Large \bf 
4. Behavior of the diffractive Green functions} 
\vskip .1 true cm 

The result (\ref{conv}) for the diffractive parton distributions 
is given in terms of two quantities: the functions $U$ and 
the functions $L$. 
The latter contain the dependence on the 
specific  diffracted system. 
The functions $U$, on the other hand, are universal 
 Green functions.  They   
control the process of diffractive deeply inelastic scattering
 for any small-size 
hadronic system. In this section we examine some of their properties 
and present results from the numerical integration of Eq.~(\ref{conv}).

\vskip 0.7 true cm 

\newpage

\noindent  {\large \bf 4.1 Ultraviolet and infrared finiteness} 
\vskip .1 true cm

 Observe first, in Eq.~(\ref{conv}),  the  factors $1 / 
{{\mbox{\bf s}}}^2 $, $1 / 
 ( {\mbox{\bf q}}+{\mbox{\bf s}})^2$ 
   from the propagators for the exchanged gluons 
  (and analogous factors 
 with ${\mbox{\bf s}} \to {\mbox{\bf s}}^{\prime} $).       
 The   
 poles at ${\mbox{\bf s}} = 0$ and ${\mbox{\bf s}} = - {\mbox{\bf q}}$ 
are cancelled partly by  $U$ and partly by $L$. 
 From Eqs.~(\ref{comb}) and (\ref{genl}) 
we see that $U_a \propto |{\mbox{\bf s}}|$, 
$L \propto |{\mbox{\bf s}}|$ as $ {\mbox{\bf s}} \to 0$. 
Analogous behavior is observed for the other poles in  
${\mbox{\bf s}}$ and $  {\mbox{\bf s}}^{\prime} $.

The Green functions $U_a$ are constructed from the linear 
combinations of wave functions (\ref{comb}) by integrating 
over the $s$-channel transverse momentum ${\mbox{\bf k}}$. 
Note that each of the terms in Eq.~(\ref{comb}) would give rise to 
an ultraviolet-divergent integration over ${\mbox{\bf k}}$ 
in Eq.~(\ref{capu}), but that the bad behavior cancels among 
the terms.  This can be seen 
by expanding Eq.~(\ref{comb}) for ${\bf k}^2 \to \infty$.  
 Both the leading and   next-to-leading 
terms in the expansion vanish. The first nonvanishing contribution 
to $u_a$ is proportional to the second derivative of the wave 
function $\psi$. This goes like $ 1 / {\mbox{\bf k}}^2$ at large 
${\mbox{\bf k}}^2$.  
 The  net  contribution 
to Eq.~(\ref{capu}) 
from the  large 
${\mbox{\bf k}}^2$ region 
is therefore 
of the type 
$d |{\mbox{\bf k}}| / |{\mbox{\bf k}}|^3$.  
The physical reason for this cancellation lies with 
the partons being at the
same transverse position in the limit  
 ${\bf k}^2 \to \infty$.  Since the net color of the state is zero,
the coupling of the state to gluons vanishes in this limit.  
Similar power counting shows that all of the integrations are convergent 
in the ultraviolet. 

Since the integrals are convergent both in the ultraviolet 
and in the infrared, we conclude that the transverse momenta dominating 
the integrals are all of order $M$.

\vskip 0.7 true cm 

\noindent  {\large \bf \boldmath 4.2 Asymptotic behaviors of $U_a$} 
\vskip .1 true cm

The functions $U_a$ contain the dependence 
on the  longitudinal momentum fractions $x_\pom$, $\beta$.  
 The dependence on $x_\pom$ is 
given simply by the overall factor $1 / x_\pom^2$ that 
defines the leading  
$1 / x_\pom \to \infty$ power (at which level we are working). 
The dependence on $\beta$ is in contrast nontrivial.  
 Detailed plots of the $\beta$ dependence 
 that we find by numerically integrating 
 Eqs.~(\ref{conv}) and (\ref{capu}) 
will be given  in the next subsection.  
The limiting behaviors of 
$U_g$ and $U_q$ as $\beta \to 0$ and $\beta \to 1$, on the 
other hand, can be obtained analytically. 
Consider $\beta \to 0$. We may take the limit  inside  
 the integral in Eq.~(\ref{capu}) and evaluate the 
 wave 
 functions $\psi_a$ in Eqs.~(\ref{psig}),(\ref{psiq}) 
 for $\beta \to 0$. We get   
\begin{equation}
\label{psib0}
 \psi_g^{ i j} ( {\mbox{\bf k}} , 
 {\mbox{\bf p}} ) = { 
 {   2 \, 
 {\mbox{\bf p}}^i \, {\mbox{\bf p}}^j } 
\over { {\mbox{\bf p}}^2 }} \, \left[ 1 + {\cal O} (\beta) \right]   
\hspace*{0.2 cm} , \hspace*{0.4 cm}
\psi_q( {\mbox{\bf k}} , 
 {\mbox{\bf p}} ) = { 
{\sqrt{ \beta \, {\mbox{\bf k}}^2}   } 
\over {  {\mbox{\bf p}}^2 }}  
\, {\mbox{\bf p}} \cdot \sigma 
\, \left[ 1 + {\cal O} (\beta) \right] 
\hspace*{0.2 cm} , \hspace*{0.6 cm} \beta \ll 1 
\hspace*{0.2 cm}  .  
\end{equation}   
By using the expressions (\ref{psib0}) in 
Eqs.~(\ref{capu}),(\ref{comb})  
it is straightforward to check that the integral in 
$ d^2 {\mbox{\bf k}} $ is convergent. 
From Eq.~(\ref{capu}) and Eq.~(\ref{psib0})
we conclude that 
\begin{equation}
\label{Ub0}
U_g \propto \beta^{-1} \hspace*{0.1 cm} ,  
\hspace*{0.3 cm} U_q \propto \beta^{0} 
\hspace*{0.2 cm} , \hspace*{0.6 cm} 
\beta \ll 1 \hspace*{0.2 cm} . 
\end{equation}

The behavior (\ref{Ub0}) 
can be understood 
on general grounds. The emission of a soft vector quantum  
has a $1 / \beta$ spectrum, while the emission of 
a soft fermion does not. 
The behavior (\ref{Ub0}) 
holds for any value of the transferred  
momentum $ {\mbox{\bf q}}$. In the case $ {\mbox{\bf q}} = 0$, 
in particular,  
the trace and the integral in Eq.~(\ref{capu}) can be evaluated 
analytically in a simple way. 
The coefficients of the leading $\beta \to 0$ terms 
take a fairly simple form as functions of the 
 transverse momenta ${\mbox{\bf s}}$,${\mbox{\bf s}}^\prime$:   
\begin{eqnarray}
\label{gb0}
&& U_g(x_\pom, \beta ,  {\mbox{\bf q}} = 0, {\mbox{\bf s}},
{\mbox{\bf s}}^\prime     ) = 
{ {  g_s^4 \,  C_A^2 \, (N_c^2 - 1)  } \over {  4 \, \pi^2 
 \, 
x_{P}^2  }}
\, {1 \over \beta} \, 
 \left\{  
2 \, {\bf s} \cdot {\bf s}^\prime \, 
\ln \left(  { { ({\bf s} +   {\bf s}^\prime)^2
} \over 
 {  ({\bf s} -   {\bf s}^\prime)^2 } } \right) 
 + {\bf s}^2 \, \left[ 
\ln \left(  { { ({\bf s} +   {\bf s}^\prime)^2
} \over 
 {  {\bf s}^2 } } \right) \right. 
\right. 
\nonumber\\
&+& \left. \left.  
\ln \left(  { { ({\bf s} -   {\bf s}^\prime)^2
} \over 
 {  {\bf s}^2 } } \right)  \right] 
+  {\bf s}^{\prime \, 2} \, 
\left[ 
\ln \left(  { { ({\bf s} +   {\bf s}^\prime)^2
} \over 
 {  {\bf s}^{\prime \, 2} } } \right) + 
\ln \left(  { { ({\bf s} -   {\bf s}^\prime)^2
} \over 
 {  {\bf s}^{\prime \, 2} } } \right)  \right]  
 \right\} \, 
\left[ 1 + {\cal O} (\beta) \right] 
\hspace*{0.1 cm} , \hspace*{0.2 cm} \beta \ll 1 
\hspace*{0.1 cm}      
\end{eqnarray} 
and 
\begin{equation}
\label{qb0} 
 U_q(  x_\pom, \beta ,  {\mbox{\bf q}} = 0, {\mbox{\bf s}},
{\mbox{\bf s}}^\prime   ) =   
{ {  g_s^4 \, C_F^2 \, N_c } \over { 8 \, \pi^2 \, 
x_{P}^2 }} \, 
2 \, {\bf s} \cdot {\bf s}^\prime \, 
\ln \left(  { { ({\bf s} +   {\bf s}^\prime)^2
} \over 
 {  ({\bf s} -   {\bf s}^\prime)^2 } } \right) \,  
\left[ 1 + {\cal O} (\beta) \right] 
\hspace*{0.2 cm} , \hspace*{0.5 cm} \beta \ll 1 
\hspace*{0.2 cm}   .     
\end{equation} 
(See Appendix D for calculational details and 
general expressions for the functions $U_a$.)

Consider now $\beta \to 1$. It is convenient 
to  switch to a new integration variable  
${\mbox{\bf v}}$ in Eq.~(\ref{capu}) by setting 
\begin{equation}
\label{switch}
    {\mbox{\bf k}} = \lambda \, {\mbox{\bf v}} 
    \hspace*{0.2 cm} , \hspace*{0.4 cm} 
    \lambda \equiv \sqrt{ { {1 - \beta} \over \beta } } 
    \hspace*{0.2 cm} . 
\end{equation} 
 The functions $U_a$ are rewritten in 
the form 
\begin{equation}
\label{intv}
U_a = 
{ { g_s^4 \, c_a } \over { 4 \, \pi \,  \beta^2 \,  
x_{\pom}^2 }} \, \int \,{{ d^2 {\mbox{\bf v}} } \over 
{( 2 \, \pi )^2}} \, {\mbox{Tr}} \left( u_{a}^{\dagger} 
\,  
u_{a} 
\right) \hspace*{0.2 cm} . 
\end{equation}  
The wave functions $\psi_a$ of 
Eqs.~(\ref{psig}),(\ref{psiq}) are rewritten as 
\begin{equation}
\label{psiv}
 \psi_g^{ i j}  = { 
 {   {\mbox{\bf v}}^2  \, \delta^{i j }   + 2 \, 
 {\mbox{\bf p}}^i \, {\mbox{\bf p}}^j } 
\over {  {\mbox{\bf v}}^2 +  {\mbox{\bf p}}^2 }}  
\hspace*{0.2 cm} , \hspace*{0.4 cm}
 \psi_q= { 
{\sqrt{  {\mbox{\bf v}}^2}   } 
\over {  {\mbox{\bf v}}^2 +  {\mbox{\bf p}}^2 }}  
\, {\mbox{\bf p}} \cdot \sigma 
\hspace*{0.2 cm} .    
\end{equation}

The limit $\beta \to 1$ may be obtained by taking 
$\lambda \to 0$ inside the integral (\ref{intv}). 
Consider the leading term in the expansion 
of the functions $u_a$ (Eq.~(\ref{comb}))  
about $\lambda = 0$. By using the expressions 
 (\ref{psiv})  we get 
\begin{equation}
\label{combg} 
u_g \to \left( \delta^{i j }  - { 
 {   {\mbox{\bf v}}^2  \, \delta^{i j }   + 2 \, 
 {\mbox{\bf s}}^i \, {\mbox{\bf s}}^j } 
\over {  {\mbox{\bf v}}^2 +  {\mbox{\bf s}}^2 }}  
+ { 
 {   {\mbox{\bf v}}^2  \, \delta^{i j }   + 2 \, 
 {\mbox{\bf q}}^i \, {\mbox{\bf q}}^j } 
\over {  {\mbox{\bf v}}^2 +  {\mbox{\bf q}}^2 }}  
- { 
 {   {\mbox{\bf v}}^2  \, \delta^{i j }   + 2 \, 
 ({\mbox{\bf q}}^i + {\mbox{\bf s}}^i) 
 \, ( {\mbox{\bf q}}^j + {\mbox{\bf s}}^j) } 
\over {  {\mbox{\bf v}}^2 +  
({\mbox{\bf q}}+{\mbox{\bf s}} )^2 }}  \right) 
\, \left[ 1 + {\cal O} (\lambda) \right] 
\hspace*{0.2 cm}   
\end{equation} 
and 
\begin{equation}
\label{combq} 
u_q  \to \left( 
- { {\sqrt{  {\mbox{\bf v}}^2}   } 
\over {  {\mbox{\bf v}}^2 +  {\mbox{\bf s}}^2 }}  
\, {\mbox{\bf s}} \cdot \sigma 
- { {\sqrt{  {\mbox{\bf v}}^2}   } 
\over {  {\mbox{\bf v}}^2 +  {\mbox{\bf q}}^2 }}  
\, {\mbox{\bf q}} \cdot \sigma 
+ { {\sqrt{  {\mbox{\bf v}}^2}   } 
\over {  {\mbox{\bf v}}^2 +  
({\mbox{\bf q}}+{\mbox{\bf s}})^2 }}  
\,  ({\mbox{\bf q}}+{\mbox{\bf s}}) \cdot \sigma \right)
\, \left[ 1 + {\cal O} (\lambda) \right] 
\hspace*{0.2 cm} .   
\end{equation} 
We notice that the 
$\beta \to 1$ behavior of the functions $U_a$ 
is     different 
according to whether the transferred 
momentum ${\mbox{\bf q}}$ is zero or finite. 
Consider   $ {\mbox{\bf q}} = 0$. Then 
observe that in this case the 
terms of order $\lambda^0$ in $u_g$, $u_q$ 
vanish. 
(In particular, for the gluon this term vanishes 
after integrating over the angle of ${\mbox{\bf s}}$, so that 
$ 2 \,   {\mbox{\bf s}}^i \, {\mbox{\bf s}}^j \to 
{\mbox{\bf s}}^2 \, \delta^{i j }$.) By expanding further in 
powers of $\lambda$, for the quark we find 
\begin{equation}
\label{furtherq} 
u_q  \to \lambda \, 
 { {\sqrt{  {\mbox{\bf v}}^2}   } 
\over {  {\mbox{\bf v}}^2 +  {\mbox{\bf s}}^2 }}  
\left( { { 4 \, {\mbox{\bf v}} \cdot {\mbox{\bf s}}
 } \over 
{  {\mbox{\bf v}}^2 +  {\mbox{\bf s}}^2 }} 
\, {\mbox{\bf s}} \cdot \sigma + 
{ { 2 \, {\mbox{\bf s}}^2 
 } \over 
{  {\mbox{\bf v}}^2  }} \, {\mbox{\bf v}} \cdot \sigma 
\right)
 + {\cal O} (\lambda^2) 
\hspace*{0.2 cm} , \hspace*{0.4 cm} ( {\mbox{\bf q}} = 0 ) 
\hspace*{0.2 cm} .   
\end{equation} 
For the gluon, the $ {\mbox{\bf q}} = 0$ 
case has  an additional cancellation 
 at order $\lambda^1$, so that 
the first surviving term is 
proportional to $\lambda^2$: 
\begin{eqnarray}
\label{furtherg} 
&& u_g  \to 2 \, \lambda^2 \, 
\left\{    
{ { 2 \, {\mbox{\bf v}}^i \, {\mbox{\bf v}}^j } 
\over {  {\mbox{\bf v}}^2  }}  
- \delta^{i j }       -  
{ { 2 \, {\mbox{\bf v}}^i \, {\mbox{\bf v}}^j } 
\over {  {\mbox{\bf v}}^2 +  {\mbox{\bf s}}^2 }}  
+      { { 4 \, {\mbox{\bf v}} \cdot {\mbox{\bf s}} \, 
({\mbox{\bf v}}^i \, {\mbox{\bf s}}^j + 
{\mbox{\bf s}}^i \, {\mbox{\bf v}}^j) } 
\over {  ({\mbox{\bf v}}^2 +  {\mbox{\bf s}}^2 )^2 }}  
\right. 
\nonumber\\
&& \left. 
    -         { { 
[4 \, ( {\mbox{\bf v}} \cdot {\mbox{\bf s}} )^2 - 
{\mbox{\bf v}}^2 \, {\mbox{\bf s}}^2 - {\mbox{\bf v}}^4 
 ] \, ( {\mbox{\bf v}}^2  \, \delta^{i j }   + 2 \, 
 {\mbox{\bf s}}^i \, {\mbox{\bf s}}^j ) } \over 
{  ({\mbox{\bf v}}^2 +  {\mbox{\bf s}}^2 )^3 }}  
\right\} + {\cal O} (\lambda^3) 
\hspace*{0.2 cm} , \hspace*{0.4 cm} ( {\mbox{\bf q}} = 0 ) 
\hspace*{0.2 cm} .   
\end{eqnarray}  
By doing the $\beta \to 1$ power counting  from  
Eqs.~(\ref{switch}),(\ref{intv}),(\ref{furtherq}),(\ref{furtherg}) 
we obtain the  behaviors 
\begin{equation}
\label{Ub1q0}
U_g \propto (1-\beta)^{2} \hspace*{0.1 cm} ,  
\hspace*{0.3 cm} U_q \propto (1-\beta)^{1} 
\hspace*{0.1 cm} , \hspace*{0.3 cm} 
 1 - \beta \ll 1  \hspace*{0.3 cm} 
 ( {\mbox{\bf q}} = 0 ) \hspace*{0.2 cm} . 
\end{equation}

In the  $ {\mbox{\bf q}} = 0$ case 
one can also determine   the 
first nonzero  $\beta \to 1$   
coefficients  in a simple way 
by using the expressions (\ref{furtherq}),(\ref{furtherg}) 
in Eq.~(\ref{intv}) and performing the  trace and the 
integral in $ d^2 {\mbox{\bf v}} $.  This gives  
\begin{eqnarray}
\label{gb1av}
&&  U_g(x_\pom, \beta ,  {\mbox{\bf q}} = 0, {\mbox{\bf s}}^2,
{\mbox{\bf s}}^{\prime  \, 2}   ) = 
{ { g_s^4 \,  C_A^2 \, (N_c^2 - 1)  } \over {  4 \, 
\pi^2 \, 
x_{P}^2
 }}
\,  \left( 1-\beta \right)^2 \,  
\nonumber\\
&& \times \,  \left\{
{ {  2 \, {\bf s}^2 \,  {\bf s}^{\prime \, 2} 
\left[ 2 \, \left(  {\bf s}^2 \right)^3 
- 5 \, \left( {\bf s}^2 \right)^2 \, 
{\bf s}^{\prime \, 2} - 5 \, {\bf s}^2 \, 
\left( {\bf s}^{\prime \, 2} \right)^2 
+ 2 \,  \left( {\bf s}^{\prime \, 2} \right)^3 \right] }  \over 
{ \left( {\bf s}^2 -  {\bf s}^{\prime \, 2} \right)^4 } }
\right. 
\nonumber\\
&& + \left. 
{ { 6 \, {\bf s}^2 \,  {\bf s}^{\prime \, 2} \, 
\left[  \left( {\bf s}^2 \right)^4 
-5 \, \left( {\bf s}^2 \right)^3 \, {\bf s}^{\prime \, 2} 
+ 10 \, 
\left( {\bf s}^2 \right)^2 \, \left( {\bf s}^{\prime \, 2} \right)^2 
-5 \, {\bf s}^2 \, \left( {\bf s}^{\prime \, 2} \right)^3 
 +  
\left( {\bf s}^{\prime \, 2} \right)^4 \right] 
 }  \over 
{ \left( {\bf s}^2 -  {\bf s}^{\prime \, 2} \right)^5 } }
\, \ln \left(  { {\bf s}^2 \over  {\bf s}^{\prime \, 2} } \right) 
 \right\} 
\nonumber\\
&&
+  
{\cal O} \left( 1-\beta \right)^{3}
\hspace*{0.3 cm} ,  \hspace*{0.7 cm}  1- \beta \, \ll \, 1  
\hspace*{0.3 cm}  
\end{eqnarray}
and 
\begin{eqnarray}
\label{qb1} 
&&  U_q(x_\pom, \beta ,  {\mbox{\bf q}} = 0, {\mbox{\bf s}},
{\mbox{\bf s}}^\prime     ) =   
{ { g_s^4 \, C_F^2 \, N_c } \over {  \pi^2 \, 
x_{P}^2 }} \, 
  (1-\beta) \, 
  \left\{ 
{ {  \left( {\bf s} \cdot {\bf s}^\prime \right)^2 
\,  \left( {\bf s}^2 +  {\bf s}^{\prime \, 2} \right) }  \over 
{ \left( {\bf s}^2 -  {\bf s}^{\prime \, 2} \right)^2 } }
\right. 
\nonumber\\   
&& \left. + 
{ {   
   {\bf s}^2 \,  {\bf s}^{\prime \, 2} }  \over 
{  {\bf s}^2 -  {\bf s}^{\prime \, 2}  } } \, 
\left[ 1 - 
{ {  2 \, \left( {\bf s} \cdot {\bf s}^\prime \right)^2 
 }  \over 
{ \left( {\bf s}^2 -  {\bf s}^{\prime \, 2} \right)^2 } }
\right] \, 
\ln \left(  { {\bf s}^2 \over  {\bf s}^{\prime \, 2} } \right) 
\right\}
+ {\cal O} (1-\beta)^2   
\hspace*{0.5 cm} ,  \hspace*{0.8 cm}  1- \beta \, \ll \, 1  
\hspace*{0.2 cm} .   
\end{eqnarray}
(Again, we refer the reader to Appendix D for more details.) 

In the case $ {\mbox{\bf q}} \neq 0 $, on the other hand, 
there are no such cancellations as in 
Eqs.~(\ref{furtherq}),(\ref{furtherg}).  
The leading  $\lambda^0$ terms in 
Eqs.~(\ref{combg}),(\ref{combq}) do not vanish. The 
$\beta \to 1$ power counting 
 thus yields a constant behavior for the functions $U_a$: 
\begin{equation}
\label{Ub1} 
  U_g ,  U_q 
 \propto (1 - \beta)^0 
 \hspace*{0.1 cm} , \hspace*{0.3 cm} 
 1 - \beta \ll 1  \hspace*{0.3 cm} 
 ( {\mbox{\bf q}} \neq 0 ) \hspace*{0.2 cm} .  
\end{equation} 
It appears that near $\beta = 1$ the diffractive distributions 
will have a nontrivial ${\mbox{\bf q}}^2$-dependence and will 
be largest at nonzero ${\mbox{\bf q}}^2$.

\vskip 0.7 true cm 

\noindent  {\large \bf  4.3 Results from  Monte Carlo integration} 
\vskip .1 true cm

We are now in a position to determine numerical results for 
the diffractive parton distributions.  
 To  this end, we set 
up a Monte Carlo integration to evaluate the  
integrals in Eqs.~(\ref{conv}),(\ref{capu}),(\ref{genl}).

It is convenient to define the rescaled diffractive 
distributions $h_a$ 
\begin{equation}
\label{hresc}
h_a (\beta , {\mbox{\bf q}}^2 / M^2) = 
{ {  x_\pom^2 \, M^2 } \over { \alpha^2 \, e_Q^4 \, \alpha_s^4}  }\, 
 {{d f_{a/A}^{\rm {diff}} 
 } \over
{dx_\pom\,dt}}
\hspace*{0.2 cm} . 
\end{equation} 
\begin{figure}[htb]
\vspace{115mm}
\includegraphics{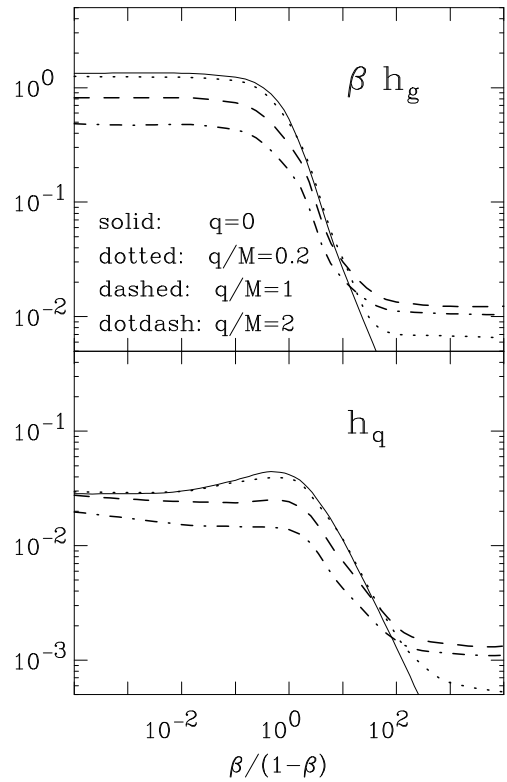}
\caption{ The $\beta$ dependence 
of the  gluon (above) and  quark (below) diffractive 
distributions for different values of 
${\mbox{\bf q}}^2 \simeq | t |$. 
 The rescaled distributions $h_a$ are defined 
 in Eq.~(\ref{hresc}). }
\label{fighlogbeta}
\end{figure}
In Fig.~\ref{fighlogbeta} we plot results for the $\beta$ dependence 
of these distributions at different values of ${\mbox{\bf q}}$.
To emphasize the behavior in the regions of small $\beta $ and 
large $\beta $ 
we make a logarithmic plot in the variable $\beta/(1-\beta)$. 
This behavior 
reflects the properties of the functions 
$U_a$ discussed in the previous subsection.  In particular, 
for small $\beta$ 
we have 
\begin{equation}
\label{hb0}
h_g \propto \beta^{-1} \hspace*{0.1 cm} ,  
\hspace*{0.3 cm} h_q \propto \beta^{0} \hspace*{0.5 cm} 
(\beta \to 0) \hspace*{0.2 cm} . 
\end{equation} 
For large $\beta$, 
both  the   
gluon and quark 
 distributions evaluated at any finite ${\mbox{\bf q}}$ 
 have a constant behavior: 
\begin{equation}
\label{hb1} 
  h_g ,  h_q 
 \propto (1 - \beta)^0 \hspace*{0.5 cm} 
(\beta \to 1 , \hspace*{0.1 cm} 
 {\mbox{\bf q}} \neq 0 ) \hspace*{0.2 cm} . 
\end{equation} 
The distributions at ${\mbox{\bf q}} = 0$, on the other hand, 
vanish in the $\beta \to 1$ limit, because of the 
cancellations in the 
leading $\beta \to 1$ coefficients of the 
functions 
$U_g$, $U_q$ 
observed in the previous subsection:  
\begin{equation}
\label{hb1q0}
h_g \propto (1-\beta)^{2} \hspace*{0.1 cm} ,  
\hspace*{0.3 cm} h_q \propto (1-\beta)^{1} \hspace*{0.5 cm} 
(\beta \to 1 , \hspace*{0.1 cm} 
 {\mbox{\bf q}} = 0 ) \hspace*{0.2 cm} . 
\end{equation}

\begin{figure}[htb]
\vspace{115mm}
\includegraphics{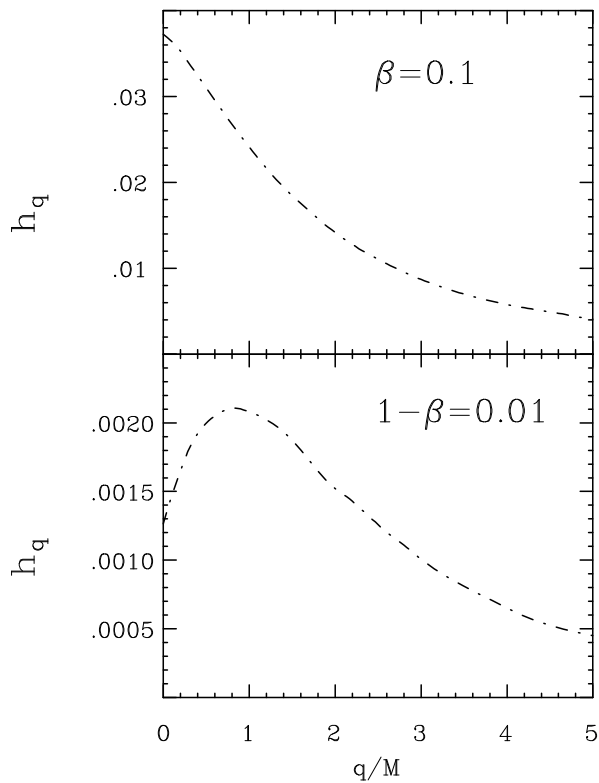}
\caption{ The ${\mbox{\bf q}}$ dependence 
of the    quark  diffractive 
distributions for small or intermediate $\beta$ (above) 
and for large $\beta$ (below).  }
\label{figtq}
\end{figure}

 The quantity that is perhaps  most interesting for 
experiment at present is the diffractive distribution integrated 
over $ | t | \simeq  {\mbox{\bf q}}^2 $ from $0$ up to a value 
 of the order of the squared heavy-quark 
mass, $M^2$. Therefore in the following 
we will also consider
 the integrated  gluon and flavor-singlet quark 
distributions, defined as follows 
\begin{equation}
\label{intt}
G  = {\cal N} \, \int_0^{M^2} \, d {\mbox{\bf q}}^2 \, 
 {{d f_{g/A}^{\rm {diff}} 
 } \over
{dx_P\,dt}} \hspace*{0.2 cm} , \hspace*{0.3 cm} 
\Sigma  = {\cal N} \, 
\sum_j \int_0^{M^2} \, d {\mbox{\bf q}}^2 \, 
 {{d f_{j/A}^{\rm {diff}} 
 } \over
{dx_P\,dt}}
\hspace*{0.3 cm} ,  
\end{equation} 
where  
$ {\cal N} = x_\pom^2 / ( \alpha^2 \, e_Q^4 \, \alpha_s^4 )$ 
and the sum in $\Sigma$ runs over
$j =\{u,\bar u,d,\bar d, s,\bar s\}$.

Fig.~\ref{fighlogbeta} shows that the   
asymptotic large-$\beta$ behavior is reached for 
rather small values of $1- \beta$, roughly of order $10^{-2}$.  
Furthermore, it shows that the asymptotic constants are 
numerically small compared to the values of the distributions 
at intermediate $\beta$. Correspondingly,   
the diffractive distributions fall off 
as one approaches the small $(1-\beta)$  region.  
The fall-off region  may be  most relevant phenomenologically, 
because  likely  it is to this range of $\beta$ values 
(rather than to the asymptotic region)  
that  current experiments on  diffraction  are most sensitive.

The results in Fig.~\ref{fighlogbeta} indicate that the 
 form of the ${\mbox{\bf q}}$-dependence 
of the distributions 
is different at small or intermediate $\beta$ and at large $\beta$. 
We illustrate this in Fig.~\ref{figtq} 
for the quark distribution.   
Qualitatively, the behavior is the same for the gluon.  
While for small and intermediate 
$\beta$ one has a monotonic fall-off 
with ${\mbox{\bf q}}$, at large $\beta$ the distribution peaks 
at a finite value of ${\mbox{\bf q}}$. The decrease seen as 
${\mbox{\bf q}} \to 0$ when $ 1 - \beta \ll 1$ arises because 
of the cancellations in the functions $U_g$ and $U_q$.

\begin{figure}[htb]
\vspace{95mm}
\includegraphics{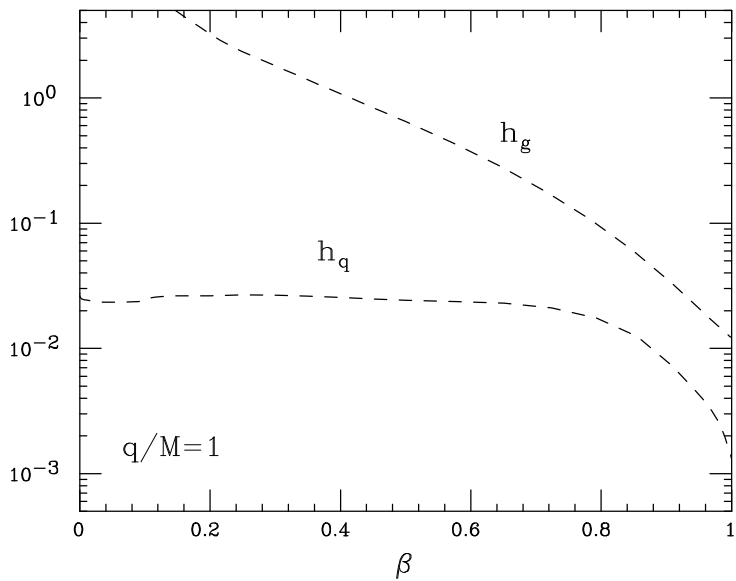} 
\caption{ The gluon ($ h_g $) and quark ($  h_q $) diffractive 
distributions versus $\beta$ (linear scale).  }
\label{fighlinbeta}
\end{figure}

We observe from Fig.~\ref{fighlogbeta} that the 
gluon distribution is much larger than the quark 
distribution.  The different order of magnitude 
at  intermediate values of $\beta$  
(say, about $\beta \approx 1 / 2$) 
is roughly   accounted for 
 by the color factors in Eqs.~(\ref{colg}),(\ref{colq}), 
 $c_g / c_q = 27/2$. 
To facilitate the comparison, in Fig.~\ref{fighlinbeta} we display  
$ h_g $ and $  h_q $ in the same graph. Here we use a linear scale 
for $\beta$. We see that the gluon remains large compared to the 
quark even at large values of $\beta$. 

\begin{figure}[htb]
\vspace{95mm}
\includegraphics{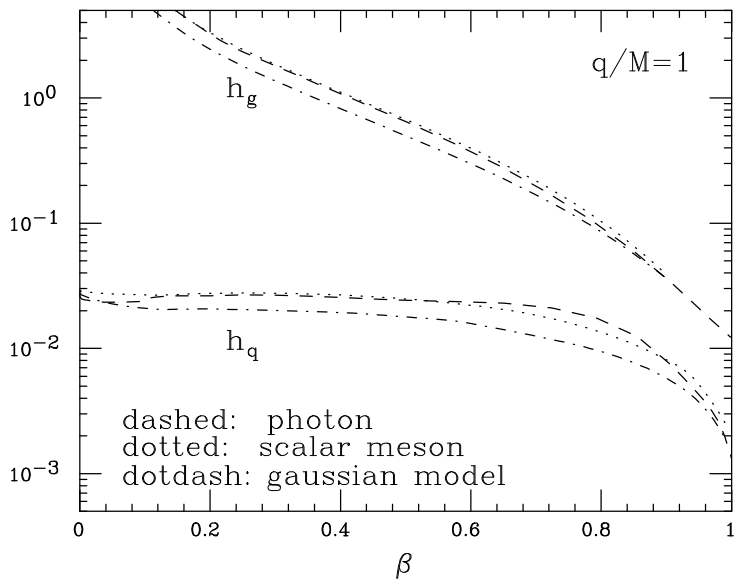} 
\caption{ The gluon ($ h_g $) and quark ($  h_q $) diffractive 
distributions for different incoming hadronic states. The dashed curves 
(photon case) are as in Fig.~\ref{fighlinbeta}. The 
dotted curves (scalar meson case) and the dotdashed curves (gaussian 
case) have been rescaled by an overall normalization factor independent 
of $\beta$ (the same factor for gluons and quarks). }
\label{figvarwv}
\end{figure}

The results described above have been obtained for the 
diffractive scattering of a model vector meson 
made of a photon that couples to heavy quarks. 
 To what extent do they depend on the assumed 
incoming state?  
To address empirically this question, we first consider the case of a 
scalar meson that couples to scalar quarks with an interaction 
$\lambda \, \phi \, {\bar q} \, q$ ($\lambda$ being a coupling 
with dimension of  mass).   A  
 calculation analogous to  the one described in Sec.~3.3 
 for the photon shows that in the scalar case  
 the wave function  (\ref{spinphi}) gets replaced 
by 
\begin{equation}
\label{scalphi} 
\Phi^{({\rm {scalar}})} (z, {\mbox{\bf k}} ,M) 
=  \lambda \, { { \sqrt{z \, (1-z)} } \over 
{  \left(  {\mbox{\bf k}}^2 + M^2 \right) } } \, 
\hspace*{0.3 cm} .  
\end{equation} 
Then, we also consider a gaussian 
model in which the incoming bound state is described by the 
 wave function 
\begin{equation}
\label{gauphi} 
\Phi^{({\rm {gauss}})} (z, {\mbox{\bf k}} ,M)  
=  { 1 \over M } \, \exp \left( - {\mbox{\bf k}}^2 /  M^2 \right)  
\hspace*{0.3 cm} .  
\end{equation}  
We compute the resulting diffractive distributions for these cases. 
In Fig.~\ref{figvarwv} we plot the results. In this figure, 
the normalization for each of the  two models (\ref{scalphi}) and 
(\ref{gauphi}) has been fixed so that it includes 
coupling and charge factors as well as an overall 
arbitrary numerical constant. This 
arbitrary normalization factor is 
independent of $\beta$ and is the same for 
quarks and gluons. 
Aside from this absolute normalization, we see that the results 
for the diffractive distributions 
are remarkably stable against variation of the hadronic 
wave function. In particular, the shape in $\beta$ of both the gluon 
and the quark distribution as well as the relative size between 
the gluon and the quark is qualitatively very similar for all 
initial states. 
We take this as an indication that such features 
of the diffractive parton distributions may hold 
with more generality. 
They do not so much depend on the specific form of the 
incoming hadron wave function, but rather they result 
 from the color and $x^-$ ordering constraints 
that give rise to the convolution formula (\ref{conv}) 
and the expressions (\ref{comb})-(\ref{psiq}) for the Green 
functions $U_a$.

\begin{figure}[htb]
\vspace{95mm}
\includegraphics{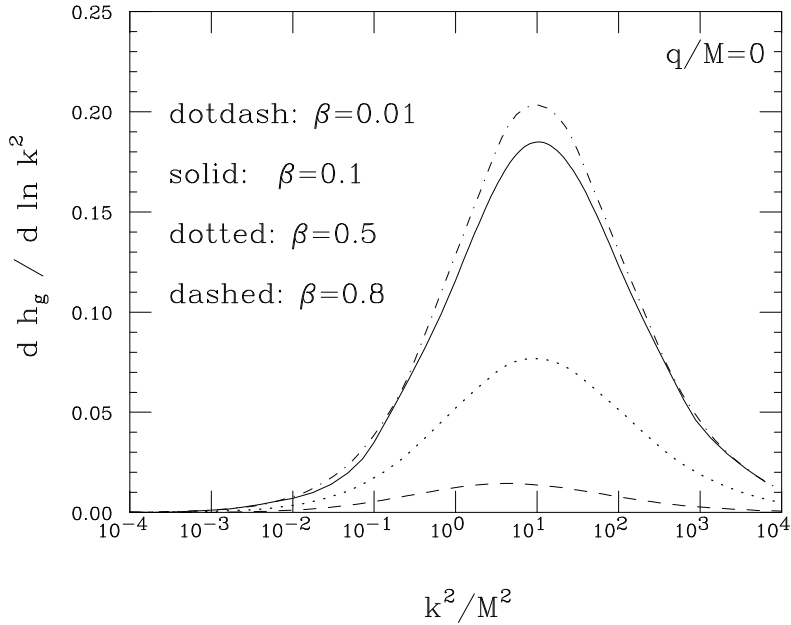} 
\caption{ The distribution in the 
relative transverse momentum $ {\mbox{\bf k}} $ for 
the gluon case at various fixed values of $\beta$ and 
$ {\mbox{\bf q}} = 0$. }
\label{figkmatrel}
\end{figure}

 Having given results for 
 the diffractive parton distributions, 
it is of interest to  ask 
 what are the typical 
transverse sizes that dominate the answer. 
A measure of the  transverse separations between the two 
outgoing particles in  the upper subgraph (see Fig.~2) 
may be obtained by looking at the relative transverse momentum  
in the final state, that is, the momentum  
$ {\mbox{\bf k}} $ in Eq.~(\ref{capu}).  
In Fig.~\ref{figkmatrel} we plot the distribution in 
$ {\mbox{\bf k}}^2 $ for the gluon 
operator matrix element at some fixed values of 
$\beta$ with $ {\mbox{\bf q}} = 0$. We observe that there is quite a 
large spread in $ {\mbox{\bf k}}^2 $ values. In the range of 
$\beta$ values considered, the typical value of $ | {\mbox{\bf k}} |$ 
does not depend much on $\beta$ and 
is of order $M$, about $ | {\mbox{\bf k}} | \sim 3 \, M$.  
This plot provides a numerical illustration of the remark made 
below Eq.~(\ref{gluonkminus}) about the dominant integration 
regions in the upper subgraph. 
A qualitatively similar behavior holds for the quark matrix element. 

\vskip 1.5 true cm 

\setcounter{equation}{0}
\setcounter{sect}{5}

\noindent  {\Large \bf 5.   Ultraviolet behavior  
and renormalization group  } 
\vskip .1 true cm

To the order in $\alpha_s$  at which we have worked so far,   the matrix
elements do not have ultraviolet divergences.  Correspondingly, the
calculation that we have discussed does not describe scaling violation. When
additional gluons are emitted from the top subgraph in 
Fig.~1, on the other hand,
ultraviolet divergences  arise. The renormalization of these divergences leads
to the dependence of the diffractive parton distributions on a renormalization
scale $\mu$. (The scale at which parton distributions are renormalized is often
called the factorization scale in applications).  Because the factorization
theorem applies, this scale dependence is governed by the same renormalization
group  evolution equations as in the case of the inclusive parton 
distributions~\cite{proo,bere,kust}. 

The higher order, ultraviolet divergent graphs are suppressed compared to
the graphs considered so far by a factor $\alpha_s\,\log(\mu^2/M^2)$.
When $\log(\mu^2/M^2)$ is large, these contributions are important, and thus
evolution is important. On the other hand, when $\mu$ is of the same order as
the heavy quark mass $M$, the higher order contributions are small corrections 
to the graphs considered so far. Thus one may interpret the result given in 
the previous sections as a result for the diffractive  parton distributions 
at a fixed scale of order $\mu^2 \approx M^2$.   Then the diffractive parton
distributions at higher values of $\mu^2$ are given by solving the evolution
equations with the results of Eq.~(\ref{conv}) as a boundary condition.

We will explore the results of evolution in the next section.

\vskip 1.5 true cm 

\setcounter{equation}{0}
\setcounter{sect}{6}

\noindent  {\Large \bf 6. Evolution and 
diffractive deeply inelastic scattering  } 
\vskip .1 true cm

 Until now, we have considered diffractive parton distributions 
in a model hadron with a transverse size $1/M$ that is as small as one
likes. Then lowest order perturbation theory is applicable. We have found 
that in this context we can evaluate the diffractive parton distributions 
as exactly as numerical integration permits. In Sec.~4, we have
investigated some of the properties of the results. Among other things, 
we have found that the results are rather insensitive to the precise wave 
function of the small-size hadron; if we change wave functions we get 
diffractive parton distribution functions that are qualitatively similar.

Now suppose that one had available a hadron of adjustable size 
$R$. For a very
small size, $R \ll 1/\Lambda_{\rm {QCD}}$, 
the diffractive parton distributions 
at any $\mu^2 \greatersim 1 / R^2$ must be similar to those
obtained by evolution beginning with Eq.~(\ref{conv}) with $M = 1 / R$. 
Let   the size now increase.  
Longer and longer distances are now allowed to contribute to the 
diffraction process. What would the answer look like 
when $R \approx 1 / ( 300 \, {\mbox {MeV}})$?    
In a perturbation expansion, 
 the result    
would be completely dominated by the soft region   
$ k_\perp \sim 300 \, {\mbox {MeV}}$. Then one 
possible scenario is  that 
the diffractive parton distributions are radically 
different from those for a small hadron. 
A different, but conceptually related, scenario 
is that, even in the soft region,  perturbation theory 
 gives a  description that is not too far off. 
In this case the diffractive parton distributions 
would be similar 
to what one gets from the calculation presented above using evolution 
from the scale $M \approx 300\ {\rm MeV}$ to the multi-GeV scale 
relevant for experiments. 
On the other hand, 
as the size of the hadronic system 
increases, 
we may hypothesize that 
nonperturbative dynamics sets in that   
reduces the infrared sensitivity 
suggested by the perturbative 
power counting. 
As we go to larger and larger sizes, then, 
the distance 
scales that dominate the diffraction process, rather than 
continuing to grow,  
stay of the order of some intermediate, semihard scale $1 / M_{\rm SH}$.  
This suggests a conceptually different scenario for the 
diffractive parton distributions, in which  
the contribution from hard physics is  enhanced 
with respect to the contribution from soft physics. 
Under this hypothesis, the diffractive parton 
distributions would be similar to what one gets
from the calculation presented above using evolution from the scale $M
\approx M_{\rm SH}$ to the multi-GeV scale.

Although this hypothesis does not have 
 a firm theoretical justification at present, 
 there are some indications in its favor. A set of 
 indications  comes from  lattice QCD.  
Lattice investigations of glueballs~\cite{glueball} suggest  
that the
correlation length for a color singlet pair of 
gluons is on the order of $1\ {\rm GeV}$, not
$300\ {\rm MeV}$. Another set of  indications comes from recent experimental 
measurements on the $x_\pom$ dependence in diffraction. 
Recall that in our model calculation the leading 
dependence is $x_\pom^{-2}$, corresponding to a pomeron intercept 
$\alpha_\pom (0) = 1$. The inclusion of higher order corrections 
of the type $  \alpha_s^k \, \ln^k (1 / x_\pom) $ would,   
roughly speaking, modify this into 
a structure of the form 
$\alpha_\pom (0) \sim  1 + {\mbox {const.}}  \times \alpha_s$.  
The value measured in diffractive deeply inelastic 
scattering~\cite{h1scalviol,zeusf2d2} is $\alpha_\pom (0) \approx 1.15$.  
The first, very general, observation is that 
this result is  not inconsistent with 
evaluating $\alpha_s$ at a relatively short distance scale 
in the above  formula for $\alpha_\pom (0)$. 
More specifically, it has been 
stressed~\cite{h1scalviol,zeusf2d2,muedis98} that the 
experimental value of $\alpha_\pom (0) - 1$  given above 
differs by a factor of $2$ from the corresponding value measured in  
soft hadron-hadron cross sections. 
This  may be taken to suggest 
that semihard physics dominates the diffractive parton distributions.

In this section, we explore the hypothesis of a semihard scale  by
comparing its predictions to results from diffractive deeply
inelastic scattering from protons at HERA. 
To carry out this study, we 
choose a value for $M$ in Eq.~(\ref{conv}) and  
 take the scale dependence of the diffractive parton distributions to
be that given by the (two-loop) evolution equations (\ref{eveq}) 
with the results (\ref{conv}) as a boundary condition at $\mu = M$. 
We expect a semihard scale  to be of the order 
of one or two GeV. In what follows we set 
$M =  1.5\ {\rm GeV}$. 
Note however that 
 in this study the value of $M$ is to be regarded as a free parameter 
to be adjusted phenomenologically. 
If, for instance,  the data prefer  
a value $M \ll 1\ {\rm GeV}$, this 
would indicate that the diffractive parton distributions 
are completely dominated
by soft physics but that nevertheless the perturbative result is not 
far off.  If the data do not agree with the predictions for
any $M$, then we would learn that soft physics dominates and completely
transforms the answer. We will comment later on what 
happens when we vary the value of $M$.

\begin{figure}[htb]
\vspace{115mm}
\includegraphics{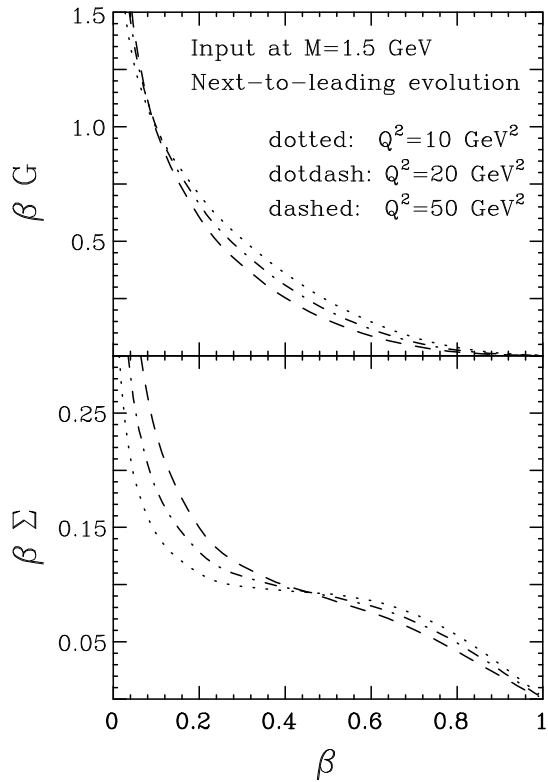}
\caption{  Evolution 
of the  gluon (above) and  singlet quark (below) diffractive 
distributions. 
 The integrated distributions $G$ and $\Sigma$ are defined 
 in Eq.~(\ref{intt}). We assume the initial scale to be 
 $ M = 1.5 \, {\mbox {GeV}}$ and we use 
  evolution equations in next-to-leading order (two-loop). } 
\label{figevgsigma} 
\end{figure}

The study of evolution 
may be done for the distributions at $ | t | \simeq  
{\mbox{\bf q}}^2$ fixed or for the distributions integrated over $| t |$. 
Here we consider 
the integrated case.  
We determine 
the gluon distribution 
 and the 
(flavor-singlet) quark distribution 
at the initial scale $\mu = M$   
by integrating  
the result (\ref{conv}) according to the definition  
(\ref{intt}).  
 We then  compute the 
evolution  of
these initial distributions up to 
$\mu = Q$ for
different values of $Q$. 
In Fig.~\ref{figevgsigma} we show the results.

\begin{figure}[htb]
\vspace{70mm}
\includegraphics{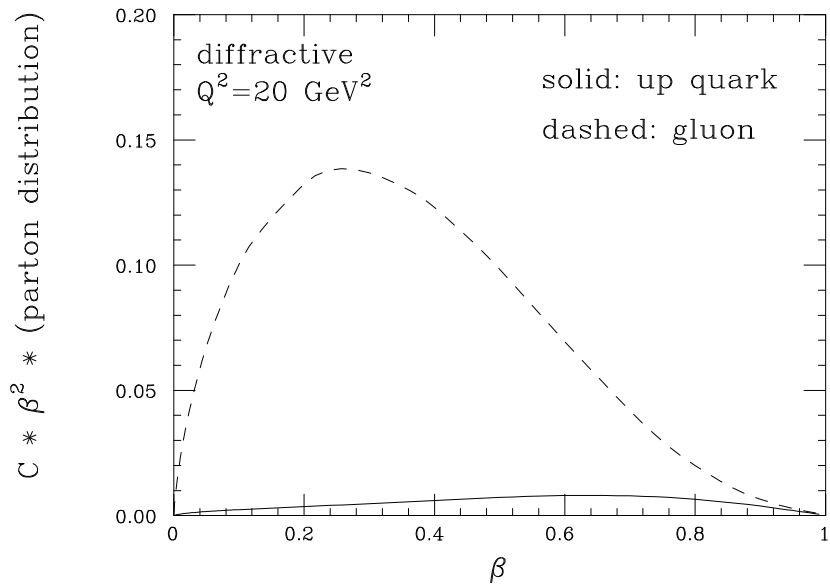} 
\caption{ The $\beta$ dependence of the gluon and up quark diffractive 
distributions (multiplied by $\beta^2$)
at $Q^2 = 20 \, {\mbox {GeV}}^2$. The overall normalization constant 
$C$ is arbitrary.}
\label{figcompdiff}
\end{figure}

\begin{figure}[htb]
\vspace{70mm}
\includegraphics{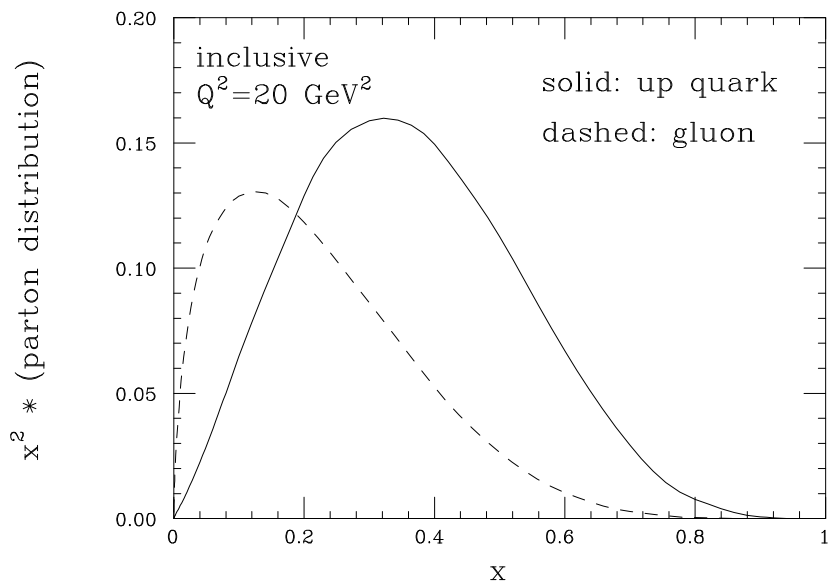} 
\caption{ The $x$ dependence of the gluon and up quark inclusive  
distributions (multiplied by $x^2$)
at $Q^2 = 20 \, {\mbox {GeV}}^2$ (from the set CTEQ4M).  }
\label{figcompincl}
\end{figure}

It is interesting to compare directly the distributions that we obtain 
from this calculation 
with the ordinary (inclusive) parton distributions. 
In Figs.~\ref{figcompdiff} and \ref{figcompincl} 
we report the gluon and up quark distributions at a certain scale 
($Q^2 = 20 \, {\mbox {GeV}}^2$) 
for, respectively, the diffractive case and the inclusive case. 
The diffractive result 
 (Fig.~\ref{figcompdiff}) is from our calculation. 
The inclusive result   (Fig.~\ref{figcompincl}) is 
from the standard set CTEQ4M of parton 
distributions in a proton~\cite{cteq4}.
In the diffractive case 
we look at the $\beta$ dependence and 
plot 
$\beta^2$  times the distributions. In the inclusive case 
we look at the 
$x$ dependence and plot 
 $x^2$ times the distributions.

Recall that the absolute normalization 
of the diffractive distributions 
has been obtained by dividing out numerical factors and coupling 
factors associated with the incoming state (see Eq.~(\ref{intt})). 
Therefore the absolute scale of Fig.~\ref{figcompdiff} 
compared to Fig.~\ref{figcompincl} has to be regarded as arbitrary. 
In contrast, the relative normalization 
between gluon and quark as well as 
the shape in $\beta$ 
is determined by our calculation. 
Fig.~\ref{figcompincl} shows 
 that, for ordinary 
\vspace{20mm}
\vspace{20mm}
{\noindent partons} in 
 a proton at the $Q^2$ scale 
considered, the gluon is dominant for small momentum fractions, 
 but, as 
the momentum fraction increases to about $x \approx 0.2$, the up quark 
starts to dominate 
over the gluon. The behavior changes 
dramatically in the 
diffractive case (Fig.~\ref{figcompdiff}). 
The gluon distribution is broader and 
stays large compared to the quark  throughout 
the  range of momentum fractions. 
The origin of this is in the behavior 
observed at a fixed scale  $M$ (see 
Figs.~\ref{fighlogbeta}, \ref{fighlinbeta}).
Fig.~\ref{figcompdiff} illustrates  that 
the feature found in the fixed scale 
calculation  persists qualitatively after the inclusion of  loop 
corrections through perturbative evolution. 

\begin{figure}[htb]
\vspace{115mm}
\includegraphics{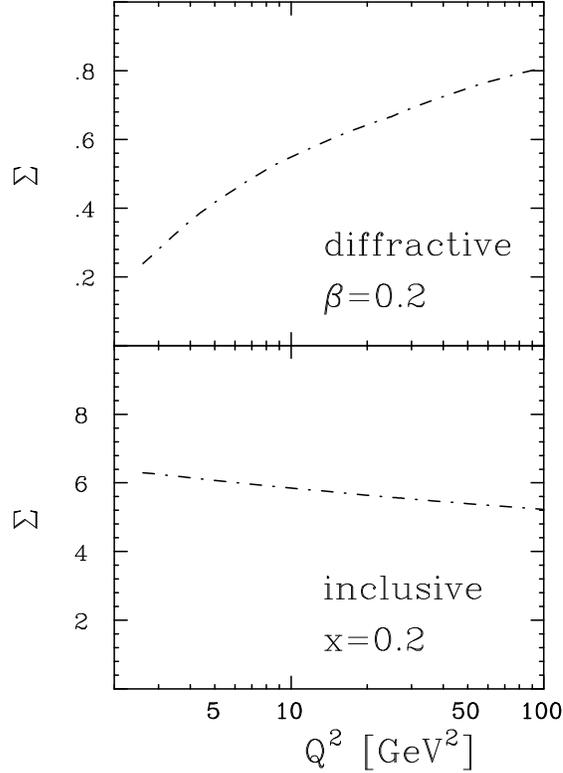}
\caption{ Scaling violation in the (flavor-singlet) quark 
distribution $\Sigma$ at moderate values of momentum fractions.  
Above is the case of the  diffractive  distribution, below is the 
case  of the inclusive distribution (from the set CTEQ4M).  }
\label{figcompqdep}
\end{figure}

This behavior has consequences on the pattern of the scaling 
violation in $Q^2$. As can be seen from Fig.~\ref{figevgsigma}, 
 the diffractive distributions grow with $Q^2$ at
low $\beta$ and decrease with $Q^2$ at high $\beta$. 
 In particular, for the  quark  
the stability point 
at which the behavior changes is about $\beta \approx 0.5$. 
 This is to be contrasted with  
 the  case of the ordinary (inclusive) quark 
 distribution in a proton,
for which the stability point is at $x \approx 0.08$. There is a large, 
 important range of moderate values of momentum fractions, 
 say,  approximately, $ 0.1 $ to $0.5$, 
 in which
  the ordinary quark distribution is flat or 
 weakly decreasing with $Q^2$, while the 
diffractive 
 distribution is rising with $Q^2$. This
is illustrated    in Fig.~\ref{figcompqdep}.       
The explanation for the rise 
 in the diffractive case 
lies with the 
 gluon distribution being dominant even at
  large 
momentum fractions (Figs.~\ref{fighlinbeta},\ref{figcompdiff}). 
As $Q^2$ increases, 
gluons splitting into $  q {\bar q}$ pairs feed 
the quark distribution and cause it to grow  
in the region of moderately large  $\beta$.

\begin{figure}[htb]
\vspace{95mm}
\includegraphics{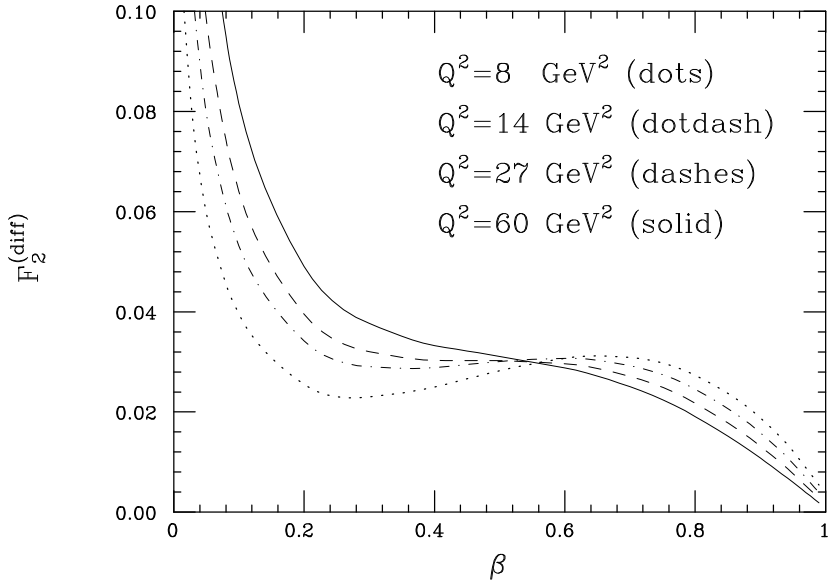} 
\caption{ The $\beta$ dependence 
of the  diffractive structure function $F_2^{\rm {diff}}$ 
for different values of $Q^2$. We compute $F_2^{\rm {diff}}$ 
in next-to-leading order.   } 
\label{figevf2nl}
\end{figure}

\begin{figure}[htb]
\vspace{115mm}
\includegraphics{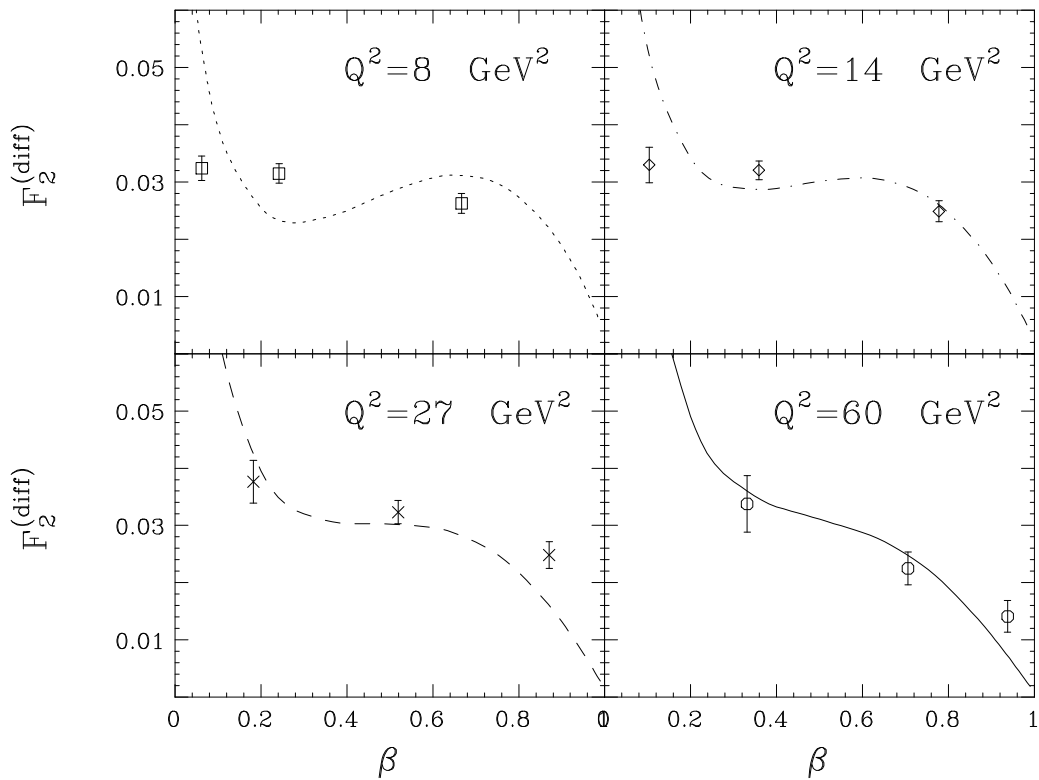}
\caption{ Same as in Fig.~\ref{figevf2nl}. Also shown are the ZEUS data 
from Ref.~\cite{zeusf2d2}. }
\label{figfourpanel}
\end{figure}

Let us  turn to the calculation of 
the deeply inelastic structure function $F_2^{\rm diff}$.
Having determined the
diffractive parton distributions and their 
evolution, we may 
 use the factorization formula (\ref{fact}) 
 to compute 
results for $F_2^{\rm diff}$. 
We evaluate $F_2^{\rm diff}$ in next-to-leading order 
 by using  the one-loop expressions~\cite{cfp}  for 
 the hard scattering functions ${\hat F}_{a} $. 
Note that the use of the factorization theorem 
allows us  to systematically take into account 
 corrections to the diffractive scattering 
beyond the leading logarithms 
for both  the quark and the gluon  contributions, much as 
in the case of inclusive deeply inelastic scattering. 
We set the factorization scale $\mu^2$ in 
Eq.~(\ref{fact})   equal to $Q^2$.
In Fig.~\ref{figevf2nl}  we plot the results for $F_2^{\rm diff}$ 
versus $\beta$  at different values of $Q^2$. Here $F_2^{\rm diff}$ 
is related to the differential structure function 
of Eq.~(\ref{fact}) by  
\begin{equation}
\label{deff2d2}
F_2^{\rm diff} = a \, 
{{x_\pom^2} \over { \alpha^2 \, e_Q^4 \, \alpha_s^4}} \, 
\int_0^{M^2} d | t | \, 
{ { d F_2^{\rm diff}  } \over { dx_\pom\, dt}} \hspace*{0.3 cm} , 
\end{equation}
where $a$ is an arbitrary numerical normalization. 
In Fig.~\ref{figfourpanel} we show these results along with 
the ZEUS data for the proton diffractive structure function at the same  
values of $Q^2$~\cite{zeusf2d2}. 
These data are obtained from the structure function integrated over $t$
by fitting the $x_\pom$ dependence to a power $x_\pom^{-2\alpha_\pom}$ and
extracting the coefficient of $x_\pom^{-2\alpha_\pom}$.

Notice the main qualitative features of the curves in 
Fig.~\ref{figevf2nl}. For $0.2 < \beta < 0.8$  the diffractive
structure function is rather flat in $\beta$. As $Q^2$ increases,
$F_2^{\rm diff}$ increases for $\beta < 0.5$, reflecting the behavior
already observed for the quark distribution $\Sigma$. For $0.2 < \beta <
0.8$ the $Q^2$ dependence is quite mild. Similar rather flat
dependences on both $\beta$ and $Q^2$ are striking features of the
data for $F_2^{\rm diff}$ and distinguish the diffractive $F_2^{\rm
diff}$ sharply from the inclusive $F_2$.

  We observe from Fig.~\ref{figfourpanel} that these  features  are 
 in qualitative agreement with what is seen 
in the HERA data on proton diffraction (except  for 
the  two data points at the smallest values of 
$\beta$: see below). 
Recall that, given the value of $M$, both the 
$ Q^2 $ dependence and the $\beta $ dependence of $F_2^{\rm diff}$ are
determined from theory. Only the overall normalization, associated with
the absolute normalization of the diffractive parton distributions, is
free. (The $t$ dependence is also determined in principle. In
the results presented here $t$ is integrated over.)  The agreement
represented in Fig.~\ref{figfourpanel} between the predictions and the
data is evidently not perfect, but given the simple nature of the
theoretical calculation, one may suspect that this agreement is telling
us something.
 
Recall that the predictions represented in Fig.~\ref{figfourpanel} are
based on the choice $M = 1.5\ {\rm GeV}$. If we decrease $M$ to 1 GeV,
we find that the agreement between the predictions and experiment is also not
too bad, but if we choose $M$ much below 1 GeV then 
we find that there is too strong a 
dependence on $Q^2$ and too much
of a slope in $\beta$. Thus a semihard scale for $M$ seems to be preferred
by the data.

In the region of small $\beta$,  the curves of 
Figs.~\ref{figevf2nl},\ref{figfourpanel}  
have a different behavior from that suggested by the 
two data points at the lowest values of $\beta$ and lowest values of 
$ Q^2 $ ($ Q^2 = 8 \, {\mbox {GeV}}^2$ and $ Q^2 = 14 \, {\mbox {GeV}}^2$). 
If further data were to  confirm this difference, this 
could point to  interesting effects.     
Here we limit ourselves to a few qualitative remarks. 
As far as the theoretical curves are concerned, 
we note that the 
diffractive distributions that  
serve as a starting point for 
the evolution are  fairly mild  as $\beta \to 0$. The gluon 
distribution goes like $1 / \beta$, while the quark distribution 
goes like a constant (see Eq.~(\ref{hb0})). The 
small-$\beta$ rise 
of the 
structure function $F_2^{\rm diff}$ 
in the curves of Figs.~\ref{figevf2nl},\ref{figfourpanel}    
is essentially due to the 
form of the perturbative evolution kernels.  
As regards the data, it has been 
observed~\cite{ellisross} that 
for small $\beta$ the experimental identification of 
the rapidity gap signal may be complicated 
by the presence of low $p_\perp$ particles in the final state.  
If the current data hold up and especially if the same features are  
observed at lower values of $\beta$, it would be interesting to see 
whether detailed models for the 
saturation of the unitarity 
bound~\cite{satura}, which might bring in a new kind of physics for very
small $x_\pom$,  
could accommodate this small $\beta$ behavior.

\vskip 1.5 true cm 

\setcounter{equation}{0}
\setcounter{sect}{7}

\noindent  {\Large \bf 7. Conclusions} 
\vskip .1 true cm 

We have analyzed  diffractive deeply inelastic scattering  in QCD,  
 drawing on two main ideas. 
One is the notion of factorization of the hard scattering. 
For processes with only one hadron in the initial state, 
this allows us to introduce diffractive parton distributions, 
defined through certain measurement operators 
 in terms of the fundamental quark and gluon fields.   
The other is the notion, widely used in small $x$ physics, that the
spacetime structure of the diffraction process looks simple in a
reference frame in which the struck hadron is at rest (or, at least,
has small momentum). 
That is, the process is dominated by 
 configurations in  which the parton 
probed by the measurement operator is created at light cone 
times $x^-$ far in the past 
 and much later, following a slow 
 evolution in $x^-$, interacts with the 
color field of the incoming hadron. 

We find that  this physics can be described most transparently 
by using a hamiltonian formulation in which the theory 
is quantized on planes of equal light cone coordinates.  
We develop this description 
in detail along the lines of Ref.~\cite{bks}. 
 
Using this method, we examine the problem of diffractive deeply 
inelastic scattering from a color source with small transverse size 
$1/M$. Because the size is small, perturbation theory is applicable. 
Using perturbation theory at lowest order (order $\alpha_s^4$) for the 
diffractive parton distributions, we find a solution in closed form. 
This solution has the property that (except for the overall 
normalization factor) it is independent of the size of the source. It
is also scale invariant, since the diagrams are ultraviolet convergent.
The scale dependence that arises from diagrams with more loops is to be
included by solving the renormalization group evolution equations 
with the results of
the lowest order  calculation as input at $\mu = M$.

We find that, although the diffractive distributions are generally
softened by evolution, certain distinctive qualitative features
survive. Both the diffractive quark distribution and the diffractive
gluon distributions fall off less quickly as $\beta$ increases than the
comparable inclusive distributions fall off as $x$ increases.
Furthermore, the diffractive gluon distribution is much larger than the
diffractive quark distributions. This has implications for the
diffractive structure function: $F_2^{\rm diff}$ is rather flat as a
function of both $\beta$ and $Q^2$ as long as $Q^2$ is not too much
larger than $M^2$. 

We observe that these qualitative features can be seen in the data from
HERA on the proton diffractive structure function $F_2^{\rm diff}$ in
the range $0.2<\beta<0.8$ and $8\ {\rm GeV}^2 < Q^2 < 60\ {\rm GeV}^2$.
If the small hadron calculation is regarded as a model with an
adjustable parameter $M$, then we find rough agreement between the
model and the experimental results if we take $M$ to be $M_{\rm SH}
\approx 1.5\ {\rm GeV}$. We conclude that the model may not be too far
from reality provided that some essentially nonperturbative effect
intervenes to provide a semihard scale $M_{\rm SH}$ in diffractive
deeply inelastic scattering.

\vskip 2.5 true cm 

\bigskip
\noindent {\large \bf Acknowledgements.} We are grateful to J.~Collins 
for a number of discussions. We thank H.~Abramowicz and 
J.~Whitmore for providing us with the 
data shown in Fig.~\ref{figfourpanel}.  
This research is supported 
in part by the U.S. Department of Energy grants No. DE-FG02-90ER40577 
and No. DE-FG03-96ER40969. 

\vskip 2.5 true cm 

\renewcommand{\theequation}{A.\arabic{equation}}
\setcounter{equation}{0}

\noindent  {\Large \bf Appendix A. Wave function for quarks} 
\vskip .1 true cm 

In this appendix, we evaluate the wave function 
$\psi_{\bar s s}({\bf k},{\bf p})$
for the antiquark state created by the quark measurement operator. We follow
the gluon case closely, merely noting what needs to be changed to deal with the
quark measurement operator instead of the gluon measurement operator.

In Eq.~(\ref{opq}) we have the operator
\begin{eqnarray}
{\beta x_\pom p_A^+ \over 2}\
{\overline \Psi}(0)\, E^\dagger(0) \cdots
& \gamma^+ &
\cdots E(0,y^-,{\bf 0})\,\Psi(0,y^-,{\bf 0})
\nonumber\\
={\beta x_\pom p_A^+ \over \sqrt 2}\ 
\Psi^\dagger(0)\, E^\dagger(0)\cdots
&{1 \over 2}\,\gamma^-\gamma^+&
\cdots E(0,y^-,{\bf 0})\,\Psi(0,y^-,{\bf 0}) \hspace*{0.2 cm} . 
\end{eqnarray}
We use a representation of the $\gamma$ matrices with
\begin{equation}
\gamma^+ = \sqrt 2 
\left(\matrix{0&0\cr 1&0}\right),\
\gamma^- = \sqrt 2 
\left(\matrix{0&1\cr 0&0}\right),\
\gamma^j =
\left(\matrix{i \sigma_j&0\cr 0&-i\sigma_j}\right) \hspace*{0.2 cm} .
\label{dirac}
\end{equation}
Then
\begin{equation}
{\scriptstyle{1 \over 2}}\,\gamma^-\gamma^+ =  
\left(\matrix{1&0\cr 0&0}\right) \hspace*{0.2 cm} .
\end{equation}
Let us break the Dirac field into two parts:
\begin{equation}
\Psi = \left(\matrix{\Psi_U \cr \Psi_L}\right) \hspace*{0.2 cm} .
\end{equation}
Then the operator in Eq.~(\ref{opq}) is
\begin{equation}
{\beta x_\pom p_A^+ \over \sqrt 2}\ 
\Psi^\dagger_U(0)\, E^\dagger(0)
\cdots E(0,y^-,{\bf 0})\,\Psi_U(0,y^-,{\bf 0}) \hspace*{0.2 cm} .
\end{equation}

Thus in order to compute the upper subgraph for quarks, we need to compute a
matrix element analogous to that in Eq.~(\ref{calMdef}):
\begin{equation}
{\cal M} = \int dy^- e^{i\beta x_\pom p_A^+ y^-}
[{\beta x_\pom p_A^+ / \sqrt 2}]^{1/2}
\langle k, s| 
E(0,y^-,{\bf 0})\,\Psi_U^{\bar s}(0,y^-,{\bf 0})
|0\rangle_{\cal A} \hspace*{0.2 cm} .
\end{equation}
The Dirac field operator $\psi$ creates the antiquark state that, after
interaction with the external field ${\cal A}$, becomes the antiquark state
$\langle k, s|$. Proceeding as in the gluon case, we obtain the analogue of
Eq.~(\ref{Mfourth}),
\begin{equation}
{\cal M} =
{ -1 \over (2\pi)^{3}} \int { d p^- \over 2 p^-}\int d^2 {\bf p}\sum_{s'}\
\langle k^-,{\bf k};s;{\cal E}|
[{\bf F}- 1]_{g^2}
| p^-,{\bf p};s';{\cal E}\rangle\
\psi_{\bar s s'}({\bf k},{\bf p}) \hspace*{0.2 cm} ,
\label{usepsiencore}
\end{equation}
where the wave function analogous to that in Eq.~(\ref{wavefctndef}) is
\begin{equation}
\label{wavefctndefq}
\psi_{\bar s s'}({\bf k},{\bf p}) = 
i\ 
\left({\beta x_\pom p_A^+ \over \sqrt 2}\right)^{1/2}
{ \langle k^-,{\bf p};s'|
\Psi_U^{\bar s}(0)
|0\rangle 
\over \beta x_\pom p_A^+ + {\bf p}^2/(2k^-)} \hspace*{0.3 cm} 
\end{equation}
with $k^- = {\bf k}^2/[2(1-\beta)x_\pom p_A^+]$ as in Eq.~(\ref{kminus}). 
Here $s$, $s'$ and $\bar s$ are two component spinor indices.
We evaluate the matrix element in free field theory, so the result is
simple:
\begin{equation}
\langle k^-,{\bf p},s'|
\Psi_U^{\bar s}(0)
|0\rangle 
=
{\cal V}_U^{\bar s}(k^-,{\bf p},s') \hspace*{0.2 cm} , 
\end{equation}
where ${\cal V}_U(k^-,{\bf p},s)$ is a two-component spinor consisting of the
top two components of the four-component Dirac spinor for antiquarks, ${\cal
V}(k^-,{\bf p},s)$:
\begin{equation}
{\cal V}(k^-,{\bf p},s)
=
\left(\matrix{
{\cal V}_U(k^-,{\bf p},s)\cr 
{\cal V}_L(k^-,{\bf p},s)}
\right) \hspace*{0.2 cm} .
\end{equation}

It remains to specify precisely the spin, and here we come to an important
technical point. As we have seen, the operator that we need is expressed very
simply in terms of the upper two components of the Dirac field. These are the
components that have a simple partonic interpretation for a system of partons
with large momentum in the plus direction, the direction in which the hadron
is moving. However, our derivation has been based on null planes with fixed
$x^-$, which is the natural formulation of the theory for a system of partons
with large momentum in the minus direction. In this formulation of QCD, it is
the lower two components of the Dirac field that are the independent degrees
of freedom for the quarks. At any given $x^-$, the upper two
components are given in terms of the lower two at the same $x^-$ by an
equation of constraint~\cite{KogutSoper,bks}. The lower part of the Dirac  
field has two components, which create the two antiquark states with the 
corresponding null-plane helicities:
\begin{equation}
\langle k^-,{\bf p},s_2|
\Psi_L^{s_1}(0)
|0\rangle 
\propto \delta_{s_1 s_2} \hspace*{0.2 cm} .
\end{equation}
With this definition of helicity, the antiquark helicity is preserved as the
fast antiquark passes through the external field.

We conclude that with the appropriate definition of the antiquark spin the
lower two components of the spinors $\cal V$ are simple:
\begin{equation}
{\cal V}_L^{\bar s}(k^-,{\bf p},s')
= [\sqrt 2 k^-]^{1/2}\,\delta_{\bar s s'} \hspace*{0.2 cm} .
\end{equation}
Here the normalization is fixed to give the conventional normalization for the
four-component spinors, $\overline {\cal V}(k,s) \gamma^\mu {\cal V}(k,s) = 2
k^\mu$. The upper two components of the spinor are related to the lower two
components by the free Dirac equation
\begin{equation}
{\cal V}_U(k^-,{\bf p},s)
=
{-i \over \sqrt 2 k^-}\ {\bf p}\cdot \sigma\
{\cal V}_L(k^-,{\bf p},s) \hspace*{0.2 cm} .
\end{equation}
Thus
\begin{equation}
\langle k^-,{\bf p},s'|
{\cal O}^{\bar s}(0)
|0\rangle
=
{-i  \over \left[ \sqrt{2} k^-\right]^{1/2}}\
{\bf p}\cdot \sigma_{\bar s s'}\ \hspace*{0.2 cm} .
\end{equation}

We now insert this result into Eq.~(\ref{wavefctnmid}) in order to obtain the
wave function for the antiquark state created by the  measurement operator. 
For our application in Eq.~(\ref{usepsiencore}), we want $s' = s$ because of
null-plane helicity conservation and we want $ k^- \equiv {\bf k}^2
/[2(1-\beta)x_\pom p_A^+]$. With these
replacements, we find
\begin{equation}
\psi_{\bar s s}({\bf k},{\bf p}) =
{\left[\beta(1-\beta) {\bf k}^2\right]^{1/2} \over
\beta{\bf k}^2 + (1 - \beta){\bf p}^2}\
{\bf p}\cdot\sigma_{\bar s s} \hspace*{0.2 cm} .
\end{equation}
This expression for $\psi_{\bar s s}$ is ready to be inserted 
into Eq.~(\ref{comb}) for the upper subgraph $U$.

\vskip 1.5 true cm 

\renewcommand{\theequation}{B.\arabic{equation}}
\setcounter{equation}{0}

\newpage

\noindent  {\Large \bf Appendix B. Wave function for the 
incoming heavy-quark system  } 
\vskip .1 true cm 

In this appendix, we evaluate the function $\Phi$ used in 
Eqs.~(\ref{Phiuse1}),(\ref{Phiuse2}). We define $\Phi$ as 
\begin{eqnarray}
\lefteqn{
{\langle
z p_A^+,{\bf r}_1,s_1; (1-z) p_A^+,{\bf r}_2,s_2
|\varepsilon_\mu\overline\Psi(0)\gamma^\mu\Psi(0) |0\rangle
\over
({\bf r}_1 + {\bf r}_2) ^2
- [{\bf r}_1^2 + M^2]/ z
- [{\bf r}_2^2 + M^2]/ (1-z) }
=}
\nonumber\\
&&\hskip 1cm
\sqrt{z(1-z)}\
\Phi(z,(1-z){\bf r}_1 - z {\bf r}_2,M,\varepsilon)_{s_1 s_2} 
\hspace*{0.3 cm} .
\label{phidef}
\end{eqnarray}
It is not immediately obvious that $\Phi$ depends on the combination
$(1-z){\bf r}_1 - z {\bf r}_2$ of the transverse momenta and not on
${\bf r}_1$ and ${\bf r}_2$ separately. This is a consequence of the
invariance of the wave function under the subgroup of the Lorentz group in
which ${\bf r}_1 \to {\bf r}_1 + z {\bf v}$ and ${\bf r}_2 \to  {\bf r}_2 + 
(1-z) {\bf v}$ \cite{bks}. We shall find this property by  explicit
calculation.

First, the denominator is
\begin{equation}
({\bf r}_1 + {\bf r}_2) ^2
- {{\bf r}_1^2 + M^2 \over  z}
- {{\bf r}_2^2 + M^2\over (1-z)}
=
-{[ (1-z){\bf r}_1 - z {\bf r}_2 ]^2 + M^2\over z(1-z)} 
\hspace*{0.2 cm} .
\end{equation}
The numerator is
\begin{equation}
\varepsilon(p_A^+, {\bf r}_1 + {\bf r}_2)_\mu\
\overline{\cal U}((1-z) p_A^+,{\bf r}_2,s_2)\,
\gamma^\mu\,
{\cal V}(z p_A^+,{\bf r}_1,s_1) \hspace*{0.2 cm} .
\end{equation}
For the Dirac algebra, we use the representation (\ref{dirac}) of the
Dirac matrices. Then, using $\varepsilon^+ = 0$ gauge for the polarization
vector, the numerator has the form
\begin{equation}
\varepsilon_\mu\
\overline{\cal U}\,
\gamma^\mu\,
{\cal V}
=
({\cal U}^\dagger_U,{\cal U}^\dagger_L)
\left(\matrix{
\sqrt{2} \varepsilon^- & i\varepsilon_\perp\cdot\sigma_\perp \cr 
- i\varepsilon_\perp\cdot\sigma_\perp & 0}
\right)
\left(\matrix{{\cal V}_U\cr {\cal V}_L}
\right) \hspace*{0.2 cm} .
\end{equation}

We need the quark and antiquark spinors $\cal U$ and $\cal V$. In our
application, the quark and antiquark move through an external field with a
large momentum in the plus direction. The external field leaves the spin
unchanged provided that we use null-plane helicity adapted to the plus
direction. Then the upper two components of $\cal U$ and $\cal V$ are
simple:
\begin{eqnarray}
{\cal U}_U^s((1-z) p_A^+,{\bf r}_2,s_2) 
&=& [\sqrt 2\, (1-z) p_A^+]^{1/2}\,\delta_{s s_2} \hspace*{0.2 cm} , 
\nonumber\\
{\cal V}_U^s(z p_A^+,{\bf r}_1,s_2) 
&=& [\sqrt 2\, z p_A^+]^{1/2}\,\delta_{s s_1} \hspace*{0.2 cm} .
\end{eqnarray}
The lower two components are then given by the free Dirac equation:
\begin{eqnarray}
{\cal U}_L^s((1-z) p_A^+,{\bf r}_2,s_2) 
&=& 
{ i {\bf r}_2\cdot \sigma + M \over \sqrt{2} (1-z) p_A^+}\
{\cal U}_U^s((1-z) p_A^+,{\bf r}_2,s_2) \hspace*{0.2 cm} ,
\nonumber\\
{\cal V}_L^s(z p_A^+,{\bf r}_1,s_2) 
&=&
{ i {\bf r}_1\cdot \sigma - M \over \sqrt{2} z p_A^+}\
{\cal V}_U^s(z
p_A^+,{\bf r}_1,s_2) \hspace*{0.2 cm} .
\end{eqnarray}

Finally, we need the polarization vector $\varepsilon(p_A^+, {\bf r}_1 + {\bf
r}_2)^\mu$ of the photon. Since $p_\mu\varepsilon(p)^\mu = 0$ in
$\varepsilon^+ = 0$ gauge, we have
\begin{equation}
\varepsilon^- = 
{ ({\bf r}_1 + {\bf r}_2)\cdot \varepsilon \over  p_A^+} \hspace*{0.3 cm} .
\end{equation}

With these ingredients, we obtain Eq.~(\ref{phidef}) with
\begin{equation}
\label{spinphiapp} 
\Phi (z, {\mbox{\bf k}} ,M, \varepsilon) 
=  { 1 \over 
{  \left(  {\mbox{\bf k}}^2 + M^2 \right) } } \, 
 \left[  
(1 -  z) \, \varepsilon \cdot \sigma \, {\mbox{\bf k}} \cdot \sigma 
- z \, {\mbox{\bf k}} \cdot \sigma \, \varepsilon \cdot \sigma 
 + i \, M \, \varepsilon \cdot  \sigma   \right]
\hspace*{0.2 cm} .
\end{equation}

\vskip 1.5 true cm 

\renewcommand{\theequation}{C.\arabic{equation}}
\setcounter{equation}{0}

\noindent  {\Large \bf Appendix C. Covariant formulation
  } 
\vskip .1 true cm 

In the main text   
we have  derived 
the  results 
for the diffractive parton distributions 
collected in Sec.~3.4 
by using the null-plane formulation of perturbation theory and 
emphasizing the picture of diffraction scattering in configuration space. 
Here we outline the main aspects of the alternative 
derivation based on the covariant formalism in momentum space.   
We limit ourselves to highlighting the most 
important ingredients of the calculation.   

We first discuss the gauge for the gluon field. The conceptually 
simplest choice
would be to use $A^- = 0$ gauge. However, some calculational simplicity 
can be
achieved by using Feynman gauge. It is necessary only to replace the sum,
$-g^{\mu\nu}$, over four polarizations for the gluon 
that enters the final state by a sum
\begin{equation}
\sum_{\lambda = 1}^2 
\varepsilon^\mu(k,\lambda)\,
\varepsilon^\nu(k,\lambda)
\label{gaugechoice}
\end{equation}
over physical polarizations, with the choice $\varepsilon^-(k,\lambda) = 0$
for the polarization vectors. Recall that
\begin{equation}
-g^{\mu\nu} =
\sum_{\lambda = 1}^2 
\varepsilon^\mu(k,\lambda)\,
\varepsilon^\nu(k,\lambda)
-
{k^\mu n^\nu + n^\mu k^\nu
\over
k\cdot n} \;\;\; ,
\label{gtoepsilon}
\end{equation}
where $n^\mu$ is the gauge fixing vector defined by $n\cdot A = A^-$. The
unphysical polarizations, represented by the second term, do not contribute 
to the result after summing over graphs. Thus we can drop them. 
Effectively, then,
we have $A^- = 0$ gauge for the final state gluon. We can retain the simple 
Feynman gauge propagator, $- i\,g^{\mu\nu}/ (k^2 + i\varepsilon)$ for virtual
gluons.

\begin{figure}[htb]
\centerline{ \desepsf(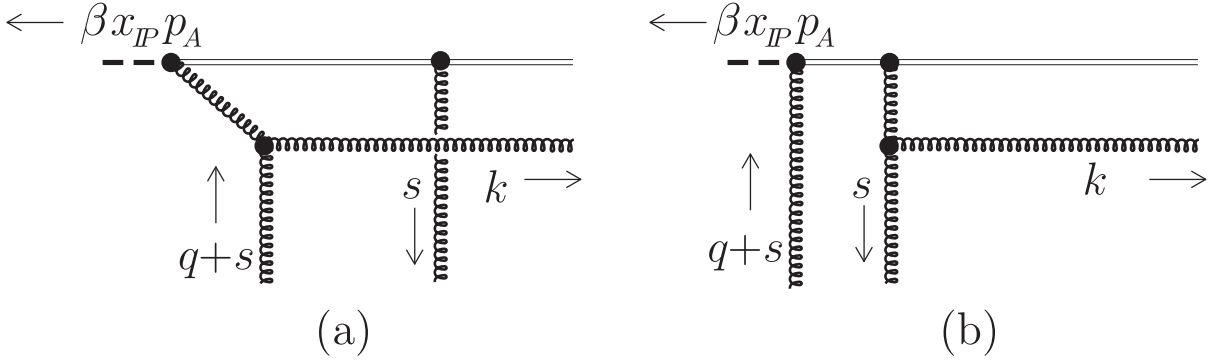 width 16 cm) }
\caption{Two examples of Feynman graphs contributing to 
the amplitude for the gluon Green function $U_g$. }
\label{figapp1}
\end{figure}

\begin{figure}[htb]
\centerline{ \desepsf(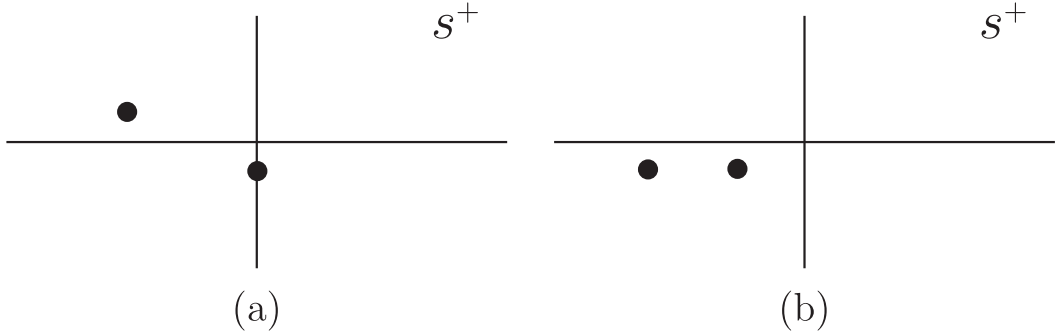 width 14 cm) }
\caption{Poles  in the complex $s^+$ plane from 
the graphs of Fig.~\ref{figapp1}.}
\label{figapp2}
\end{figure}

Let us now see which graphs contribute to the result in the 
$1 / x_\pom \to \infty $ limit. Consider  the upper 
subgraph in Fig.~2. 
  The 
definition of the Green function $U_a$ contains  integrations  
over the plus momenta $s^+$,$s^{\prime +}$ 
(see Eqs.~(\ref{udef}),(\ref{capu})).   
We recognize that only certain topologies give rise to 
pinch singularities in the complex $s^+$ plane. Consider 
for instance the two contributions in Fig.~\ref{figapp1}. 
From the eikonal propagator and the $t$-channel propagator 
we get, 
for graph (a), 
\begin{equation}
\label{intplusa}
{\cal I}_+^{(a)} = 
\int_{- \infty}^{+ \infty} \, { { d s^+ } \over {2 \, \pi} } \, 
{ {N^{(a)}} \over  { \left[ s^+ + i \, \varepsilon \right] \, 
\left[ (k-q -s)^2   + i \, \varepsilon \right] \, } } 
\hspace*{0.4 cm}  
\end{equation} 
and for graph (b) 
\begin{equation}
\label{intplusb}
{\cal I}_+^{(b)} = 
\int_{- \infty}^{+ \infty} \, { { d s^+ } \over {2 \, \pi} } \, 
{ {N^{(b)}} \over  { \left[ s^+ + k^+ + i \, \varepsilon \right] \, 
\left[ (k+s)^2   + i \, \varepsilon \right] \, } } 
\hspace*{0.4 cm} ,  
\end{equation} 
where ${N^{(a)}}$ and ${N^{(b)}}$
denote spin numerators.  
Using the kinematic relation (\ref{gluonkminus}) 
and the fact that 
in the limit $1 / x_\pom \to \infty$ we only need to evaluate
the upper subgraph  at  $s^- = q^- = 0$ (see Sec.~2.2), 
the position of the poles in the two cases is 
as in Fig.~\ref{figapp2}. 
 In the case of graph (a) the poles 
pinch the contour and one gets a leading 
$1 / x_\pom \to \infty$ contribution, 
while in the case of graph (b) one may deform the contour 
away from the poles and neglect this graph as $1 / x_\pom \to \infty$.

The gauge choice $\varepsilon^- = 0$ for the final state gluon is
crucial for this  argument to go through in the form described above. 
With this choice,  the spin numerators behave like $(s^+ )^0$.  Then the
integrals (\ref{intplusa}),(\ref{intplusb})  are sufficiently convergent
at large $s^+$ to allow the contour deformation.

\begin{figure}[htb]
\centerline{ \desepsf(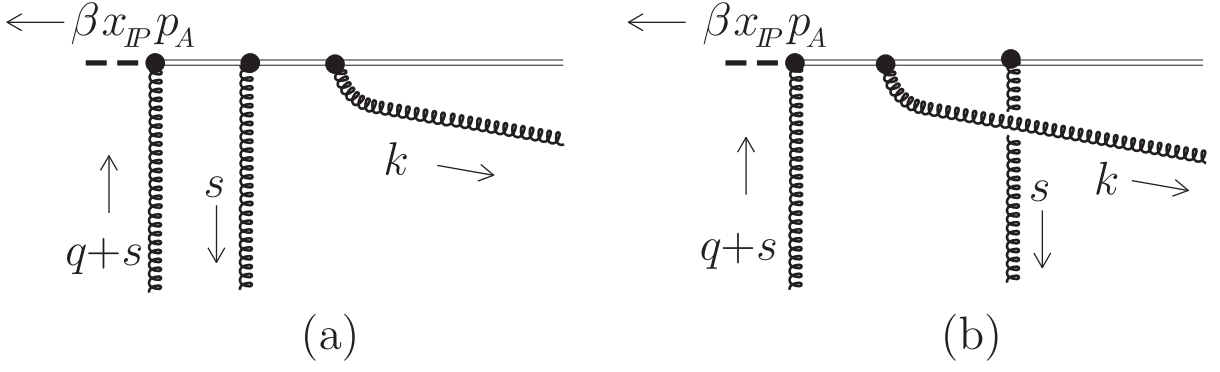 width 16 cm) }
\caption{Gluon emission from the eikonal line.}
\label{figapp3}
\end{figure}

Consider now the two graphs shown in Fig.~\ref{figapp3}, in which a
gluon is emitted into the final state from the eikonal line. 
The contribution in Fig.~\ref{figapp3}(a) is disallowed 
by color conservation, because the $t$-channel gluons 
are in a color-singlet configuration while the eikonal line 
carries color octet charge. 
The contribution in Fig.~\ref{figapp3}(b) may be 
dealt with by $s^+$ contour deformation, 
by an argument similar to that given above. 
  
Having seen why certain graphs do not contribute in the 
$1 / x_\pom \to \infty$
limit, we turn to the calculation of the result for a graph that
does contribute. Consider the contribution to the gluon  Green function
$U_g$ from the product of the amplitude in Fig.~\ref{figapp1}(a) with the
complex conjugate amplitude of the same form. 
This contribution   can be written as  
\begin{eqnarray}
\label{ug0}
U_g^{(0)}( x_\pom , \beta , {\mbox{\bf q}},{\mbox{\bf s}},
{\mbox{\bf s}}^\prime   ) &=& 
{ { g_s^4 \, C_A^2 \, (N_c^2 - 1) \, p_A^+ } \over 
{ 2 \, \pi \,  \beta \,  
x_{\pom} }}  
\int \, { { d s^+ } \over {2 \, \pi} } \,
\int \, { { d s^{\prime +} } \over {2 \, \pi} } 
\int \,{{ d^4  k}  \over 
{( 2 \, \pi )^4} } \,
2 \, \pi \, \delta_+ (k^2) \, 2 \, \pi \, \delta (k^{\prime +} ) 
\nonumber\\
&\times& { { - F^{\mu j} \, M_{\mu \nu} \, F^{\nu j} } \over 
{ \left[ s^+ + i \, \varepsilon \right] \, 
\left[ (k-q -s)^2   + i \, \varepsilon \right] \, 
\left[ (k-q -s^\prime)^2   - i \, \varepsilon \right] \, 
\left[ s^{\prime +} - i \, \varepsilon \right] } } 
\hspace*{0.1 cm} ,  
\end{eqnarray} 
where  $F^{\mu j}$ is the gluon operator vertex, $M_{\mu \nu}$ is 
the polarization tensor associated with the squared amplitude, 
and the resulting expression for the spin numerator is 
\begin{eqnarray}
\label{numer}
 - F^{\mu j} \, M_{\mu \nu} \, F^{\nu j} 
&=& 
4 \, ( k^- )^2 \left\{ \left( \beta \, x_\pom \, p_A^+ \right)^2 - 
{{\beta \, x_\pom \, p_A^+ } \over {k^-} } \, 
\left[ ({\bf k} - {\bf q} - {\bf s})^2 + 
({\bf k} - {\bf q} - {\bf s}^\prime )^2 \right] \right. 
\nonumber\\
&& \left. 
+ {{ \left[ ({\bf k} - {\bf q} - {\bf s}) \cdot 
({\bf k} - {\bf q} - {\bf s}^\prime ) \right]^2 } 
\over {(k^-)^2} } 
\right\} 
\hspace*{0.4 cm} . 
\end{eqnarray} 
We may  perform 
the integrations over the plus and minus components of the momentum $k$ 
by using Eq.~(\ref{longint}). The $s^+$,$s^{\prime +}$  integrations  
are of the form (\ref{intplusa}). By carrying out this integral we get 
\begin{equation}
\label{intplusbis}
{\cal I}_+^{(a)} = {1 \over { \beta \, {\bf k}^2 / (1 - \beta) + 
({\bf k} - {\bf q} - {\bf s})^2 } } 
\hspace*{0.4 cm} . 
\end{equation} 
Then we obtain 
\begin{eqnarray}
\label{ug0bis}
&& U_g^{(0)}( x_\pom , \beta , {\mbox{\bf q}},{\mbox{\bf s}},
{\mbox{\bf s}}^\prime   ) = 
{ { g_s^4 \, C_A^2 \, (N_c^2 - 1)  } \over 
{ 4 \, \pi \,  \beta \, (1 - \beta) \,  
x_{\pom}^2 }} \, \int \,{{ d^2 {\mbox{\bf k}} } \over 
{( 2 \, \pi )^2}} \, 
 \left\{ 
   2 \, ( \beta \, {\mbox{\bf k}}^2 )^2 \,   + 2 \, \beta \, 
 ( 1 - \beta) \, {\mbox{\bf k}}^2 \, \left[ 
({\bf k} - {\bf q} - {\bf s})^2 
\right. \right. 
\nonumber\\
&& \left. \left. 
+ 
({\bf k} - {\bf q} - {\bf s}^\prime)^2 \right] 
+ 4 \, (1-\beta)^2 \, \left[  
({\bf k} - {\bf q} - {\bf s}) \cdot 
({\bf k} - {\bf q} - {\bf s}^\prime) \right]^2 \right\} 
\nonumber\\
&&    
/ \left\{  
\left[ \beta \, {\mbox{\bf k}}^2 + (1-\beta) \, 
({\bf k} - {\bf q} - {\bf s})^2 \right] 
\, \left[ \beta \, {\mbox{\bf k}}^2 + (1-\beta) \, 
({\bf k} - {\bf q} - {\bf s}^\prime)^2 \right] \right\}   
\hspace*{0.4 cm} . 
\end{eqnarray}
We recognize in  this expression the contribution to $U_g$ 
from the last term in Eq.~(\ref{comb}). Indeed, the expression 
(\ref{ug0bis}) can be recast in the form 
\begin{equation}
\label{ug0tris}
U_g^{(0)}( x_\pom , \beta , {\mbox{\bf q}},{\mbox{\bf s}},
{\mbox{\bf s}}^\prime   ) = 
{ { g_s^4 \, C_A^2 \, (N_c^2 - 1) } \over { 4 \, \pi \,  \beta \, 
(1- \beta) \, 
x_{\pom}^2 }} \, \int \,{{ d^2 {\mbox{\bf k}} } \over 
{( 2 \, \pi )^2}} \, {\mbox{Tr}} 
\left( 
\psi_g^\dagger ( {\mbox{\bf k}} , {\mbox{\bf k}} - 
{\mbox{\bf q}} - {\mbox{\bf s}}^\prime ) \, 
\psi_g ( {\mbox{\bf k}} , {\mbox{\bf k}} - 
{\mbox{\bf q}} - {\mbox{\bf s}} ) \right) 
\hspace*{0.2 cm} ,  
\end{equation} 
with $\psi$ being the wave function given in Eq.~(\ref{psig}). 

The other terms in Eq.~(\ref{comb}) are obtained in a similar way 
by adding the graphs in which the exchanged gluons attach to the 
same line and using the color singlet projection 
(\ref{colproj}).

\vskip 1.5 true cm 

\renewcommand{\theequation}{D.\arabic{equation}}
\setcounter{equation}{0}

\noindent  {\Large \bf \boldmath Appendix D. An integral 
representation for the Green functions $U_a$  } 
\vskip .1 true cm 

In this appendix we give an integral representation 
for the Green functions $U_a$ 
alternative to the one given in the text, Eq.~(\ref{capu}). 
It is obtained from Eq.~(\ref{capu}) by performing the trace, 
introducing a Feynman parameterization to combine the denominators  
and  carrying out the integral over ${\bf k}$. The resulting 
formulas are lengthy and not as compact as the formulas used in 
the text. They have the advantages, though,  that the spin structure 
is explicitly worked out and that the cancellation of the singular 
terms at large ${\bf k}^2$ is also  explicitly worked out. They may 
be useful for numerical evaluation.  

The functions $U_a$ are written as 
\begin{eqnarray}
\label{ufey}
U_a( x_\pom , \beta , {\mbox{\bf q}},{\mbox{\bf s}},
{\mbox{\bf s}}^\prime   ) &=& 
{ { g_s^4 \, c_a (1- \beta)} \over { 16 \, \pi^2 \,  \beta \, 
x_{\pom}^2 }} \, \sum_{i = 0}^{3} \, \sum_{j = 0}^{3} \, (-1)^{i+j} 
\, \int_0^1 d x 
\nonumber\\
&& \times \, \left( {\cal F}_{a, i j} \ln {\cal M}_{a, i j}^2 + 
{ { {\cal G}_{a, i j} } \over { {\cal M}_{a, i j}^2  } } \right) 
\; \; ,    
\end{eqnarray}
where the arguments $(\bq,\bs,\bsp,\beta,x)$ of the
auxiliary functions 
${\cal F}$, ${\cal G}$, ${\cal M}$  
are suppressed. The color factors $c_a$ are 
given in Eqs.~(\ref{colg}),(\ref{colq}). 
The summation labels $i,j$ run over the four momentum configurations 
appearing in the definition of $u_a$ (see Eq.(\ref{comb})) 
$$ 
u_{a} ( \beta,\bk,\bq,\bs ) = 
\sum_{i=0}^3 (-1)^i\psi_a (\bk,\bk+\bri1 ) \;\;,  
$$ 
where
 for $i,j=0,1,2,3 \;$  \bri1,\brj2
have the values
$$
\bri1=
\{\bnul,\bs,-\bq,-\bq-\bs\} \;, \; 
 \qquad 
\brj2=\{\bnul,\bsp,-\bq,-\bq-\bsp\} \;\;\; . 
$$ 
The expressions for ${\cal F}_{a,i j },$ $ {\cal G}_{a,i j }$ 
and  ${\cal M}^2_{i j }$ are 
\begin{eqnarray}
\label{FGMij}
\nonumber
 {\cal F}_{a,i j } &=&
f_{a,1}(x,\gamma ) \bri1 \cdot \bri1 +  f_{a,2}(x,\gamma )\bri1 \cdot \brj2 + 
\left(x\to (1-x),\bri1\to\brj2\right)
\\ \nonumber
 {\cal G}_{a,i j } &=&
g_{a,1}(x,\gamma )(\bri1 \cdot \bri1)^2
   + g_{a,2}(x,\gamma )(\bri1 \cdot \bri1)(\brj2 \cdot \brj2)
 + g_{a,3}(x,\gamma ) 
(\bri1\cdot\brj2)^2
\\ \nonumber
&& + g_{a,4}(x,\gamma ) (\bri1 \cdot \brj2) (\bri1 \cdot \bri1) +
\left(x\to (1-x),\bri1\to\brj2\right)
\\ \nonumber
 {\cal M}^2_{i j }  &=&
\gamma  x(1-\gamma  x)\bri1 \cdot \bri1 -\,\gamma ^2x(1-x) \bri1 \cdot \brj2+
\left(x\to (1-x),\bri1\to\brj2\right) \hspace*{0.2 cm}  ,   
\end{eqnarray}
where $\gamma=1-\beta$. The new  auxiliary functions $f$ and $g$ 
are simple polynomials in $x$: 
\begin{eqnarray}
\label{qfi}
\nonumber 
f_{q,1}&=& 4 (1 - \gamma) x (2 - 3 \gamma x)
\\ \nonumber
f_{q,2}&=& ((1 - \gamma) (-1 + 2 \gamma - 12 \gamma^2 (1 - x) x))/\gamma
\\ \nonumber
g_{q,1}&=&-2 (1 - \gamma) \gamma^2 x^3 (1 - \gamma x)
\\ \nonumber
g_{q,2}&=&
-(1 - \gamma) \gamma^2 (1 - x) x [1 - 2 \gamma (1 - x) x]
\\ \nonumber
g_{q,2}&=&
 2 (1 - \gamma) \gamma (1 - x) x (1 - \gamma + 2 \gamma^2 (1 - x) x)
\\ \nonumber
g_{q,4}&=&
2 (1 - \gamma) \gamma x^2 (1 - \gamma (3 - 2 x) + 4 \gamma^2 (1 - x) x)
\end{eqnarray}
in the case of quarks and
\begin{eqnarray}
\label{gfi}
\nonumber
f_{g,1}&=&
 (-2 (1 - 2 (1 + 2 \gamma + 3 \gamma^2) x + 6 (\gamma + \gamma^3) x^2))/\gamma 
\\ \nonumber 
f_{g,2}&=&2 (-1 + 2 \gamma - 6 (1+\gamma^2) x (1-x) ) 
\\ \nonumber
g_{g,1}&=&2 \gamma x^2 (1 + \gamma (-1 + x)^2 - 2 \gamma^2 x + \gamma^3 x^2)
\\ \nonumber
g_{g,2}&=& \gamma \Bigl(\Big[1 - 2 (1+\gamma^2) x (1-x)   + 2 \gamma^3 (1 - x)^2 x^2 + 
\\ \nonumber
&&     \gamma \big\{-1 + 2 x (1-x) (2+x (1-x)\big\}\Big]\Bigr)
\\ \nonumber
g_{g,3}&=& 2 \Bigl(1 - 2 \gamma - 2 \gamma^3 (1 - x) x + 2 \gamma^4 (1 - x)^2 x^2 + 
\\ \nonumber
&&     \gamma^2 \big[1 + 2 x (1-x) \bigl\{1+x (1-x)\bigr\}\big]\Bigr)
\\ \nonumber
g_{g,4}&=&  -4 \gamma x \Bigl[1 + x + \gamma^2 (3 - 2 x) x - 
\\ \nonumber
&& 2 \gamma^3 (1 - x) x^2 - 
     \gamma (1 + 4 x^2 - 2 x^3)\Bigr]
\end{eqnarray}
in the case of gluons.
The $x$ integral can  easily be carried out  
but the result  is simple only  for the limiting
values $\bq=0,\beta\to 0$ or $\beta\to 1$. The 
expressions for these cases are given 
 in Sec.~4.2.

\vskip 3 true cm


\noindent{\Large \bf References}
\begin{enumerate} 
\bibitem{ingelschl}  
    G.~Ingelman and P.~Schlein,     
    Phys.\ Lett.\  {\bf 152B}, 256 (1985).    
\bibitem{hhdiffcern}      
    UA8 Collaboration (A.~Brandt {\it et al.}), 
    Phys.\ Lett.\ B {\bf 297}, 417 (1992). 
\bibitem{hhdifffnal}   
    CDF Collaboration (F.~Abe {\it et al.}), 
    Phys.\ Rev.\ Lett.\ 
    {\bf 74}, 855 (1995);
    {\bf 78}, 2698 (1997); {\bf 79}, 2636 (1997); 
    D0 Collaboration (S.~Abachi {\it et al.}), 
    Phys.\ Rev.\ Lett.\ 
    {\bf 72}, 2332 (1994); {\bf 76}, 734 (1996).  
\bibitem{disdiff1} 
    ZEUS Collaboration (M.~Derrick {\it et al.}),
    Phys.\ Lett.\  B {\bf 315}, 481 (1993); 
    {\bf 332}, 228 (1994); 
    {\bf 338}, 483 (1994); 
     {\bf 356}, 129 (1995);
    Z.\ Phys.\ C {\bf 68}, 569 (1995); 
     {\bf 70}, 391 (1996);    
    H1 Collaboration (T.~Ahmed {\it et al.}),
    Nucl.\ Phys.\  {\bf B429}, 477 (1994);
    Phys.\ Lett.\  B {\bf 348}, 681 (1995). 
\bibitem{h1scalviol}
    H1 Collaboration, (C.~Adloff {\it et al.}),
    Z.\ Phys.\ C {\bf 76}, 613 (1997); 
    ZEUS Collaboration 
    (J.~Breitweg {\it et al.}), Eur.\ Phys.\ J.\ C{\bf 1}, 81 (1998).         
\bibitem{zeusf2d2}
    ZEUS Collaboration 
    (J.~Breitweg {\it et al.}), Eur.\ Phys.\ J.\ C{\bf 6}, 43 (1999).      
\bibitem{css} 
    See, for instance, the review in J.C.~Collins, D.E.~Soper 
    and G.~Sterman, in {\em Perturbative Quantum Chromodynamics}, 
    edited by A.H.~Mueller (World Scientific, Singapore, 1989).    
\bibitem{spoil}
    J.C.~Collins, L.~Frankfurt and M.~Strikman,  
    Phys.\ Lett.\ B {\bf 307}, 161 (1993); 
    A.~Berera and D.E.~Soper, Phys.\ Rev.\ D  {\bf 50}, 
    4328 (1994).       
\bibitem{proo}
    J.C.~Collins, Phys.\ Rev.\ D  {\bf 57}, 3051 (1998). 
\bibitem{bere}
    A.~Berera and D.E.~Soper, Phys.\ Rev.\ D  {\bf 53}, 
    6162 (1996).     
\bibitem{fracture}    
    L.~Trentadue and G.~Veneziano, 
    Phys.\ Lett.\ B {\bf 323}, 201 (1994). 
\bibitem{hkslett}
    F.~Hautmann, Z.~Kunszt and D.E.~Soper, Phys.\ Rev.\ Lett.\ 
    {\bf 81}, 3333 (1998).  
\bibitem{alignedjet}
    J.D.~Bjorken, AIP Conference Proceedings No.~6, Particles 
    and Fields subseries No.~2, edited by M.~Bander, G.~Shaw and 
    D.~Wong (AIP, New York, 1972); 
    J.D.~Bjorken and J.~Kogut, Phys.\ Rev.\  D  {\bf 8}, 
    1341 (1973); J.D.~Bjorken,  preprint SLAC-PUB-7096 (1996), 
    hep-ph/9601363.  
\bibitem{cs82}
    J.C.~Collins and D.E.~Soper, Nucl.\  Phys.\ {\bf B194}, 
    445 (1982). 
\bibitem{cfp}    
    G.~Curci, W.~Furmanski and R.~Petronzio,  
    Nucl.\  Phys.\ {\bf B175}, 27 (1980).        
\bibitem{KogutSoper}
    J.~Kogut and D.E.~Soper, 
    Phys.\ Rev.\  D  {\bf 1},     2901 (1970).            
\bibitem{nikoplus}
    N.N.~Nikolaev and B.G.~Zakharov, Z.\ Phys.\ C {\bf 49}, 
    607  (1991); {\bf 53}, 331 (1992); 
    A.H.~Mueller,  Nucl.\  Phys.\ {\bf B335}, 115 (1990);             
    L.~Frankfurt and M.~Strikman, Phys.\ Rep.\ {\bf 160}, 235 (1988). 
\bibitem{buch}
    W.~Buchm{\" u}ller and  A.~Hebecker,  
    Nucl.\  Phys.\ {\bf B476}, 203 (1996);   
    A.~Hebecker,  
    Nucl.\  Phys.\ {\bf B505}, 349 (1997); 
    W.~Buchm{\" u}ller, T.~Gehrmann and  A.~Hebecker, 
    Nucl.\  Phys.\ {\bf B537}, 477 (1999).               
\bibitem{wuest}
    M.~W{\" u}sthoff, Phys.\ Rev.\ D  {\bf 56 }, 4311 (1997); 
%\bibitem{bartetal} 
    J.~Bartels, J.~Ellis, H.~Kowalski and M.~W{\" u}sthoff, 
    Eur.\ Phys.\ J.\ C{\bf 7}, 443 (1999).     
\bibitem{bks}
    J.D.~Bjorken, J.~Kogut and D.E.~Soper, 
    Phys.\ Rev.\  D  {\bf 3},     1382 (1971).   
\bibitem{kust} 
    T.~Gehrmann and W.J.~Stirling, 
    Z.\ Phys.\ C {\bf 70}, 89 (1996);  
    K.~Golec-Biernat and J.~Kwiecinski,  
    Phys.\ Lett.\ B {\bf 353}, 329 (1995);                            
    Z.~Kunszt and W.J.~Stirling, in {\it Proceedings of the 
    International Workshop on Deep Inelastic Scattering DIS96},
    Rome, Italy, 1996,
    edited by G.~D'Agostini and A.~Nigro 
    (World Scientific, Singapore, 1997), p.240;     
    L.~Alvero, J.C.~Collins, J.~Terron and J.~Whitmore, 
    Phys.\ Rev.\ D  {\bf 59}, 074022 (1999).  
\bibitem{glueball} 
    See, for instance, M.J.~Teper, hep-th/9812187; S.~Dalley 
    and B.~van de     Sande, Phys.\ Rev.\ Lett.\ 
    {\bf 82}, 1088 (1999); C.J.~Morningstar and M.~Peardon, 
    hep-lat/9901004.      
\bibitem{muedis98}
    A.H.~Mueller, in {\it Proceedings of the 
    International Workshop on Deep Inelastic Scattering DIS98},
    Brussels, Belgium,  1998, edited by G.~Coremans and  R.~Roosen 
    (World Scientific, Singapore, 1998), p.3.     
\bibitem{cteq4} 
    H.L.~Lai et al., Phys.\ Rev.\ D {\bf 55}, 1280 (1997). 
\bibitem{ellisross}
    J.~Ellis and G.G.~Ross,  
    Phys.\ Lett.\ B {\bf 384}, 293 (1996);    
    J.~Ellis, G.G.~Ross  and J.~Williams, hep-ph/9812385.    
\bibitem{satura}       
    E.~Gotsman, E.~Levin and U.~Maor, 
    Nucl.\  Phys.\ {\bf B493}, 354 (1997); 
    J.~Jalilian-Marian, A.~Kovner, A.~Leonidov and H.~Weigert, 
    Phys.\ Rev.\ D  {\bf 59}, 034007 (1999);
    K.~Golec-Biernat and  M.~W{\" u}sthoff, hep-ph/9903358.                  

\end{enumerate}
\end{document}